\documentclass[aps,floats,superscriptaddress,showpacs]{revtex4}
\usepackage{amsmath}
\usepackage{amsfonts}
\usepackage{amssymb}

\usepackage{epsfig}
\usepackage{graphicx}
\usepackage{dcolumn}
\usepackage{bm}

\renewcommand{\dag}{^{\dagger}}
\newcommand{\beq}{\begin{equation}}
\newcommand{\eeq}{\end{equation}}
\newcommand{\beqa}{\begin{eqnarray}}
\newcommand{\eeqa}{\end{eqnarray}}
\def\non{\nonumber }

\def\go{\vec g_1}
\def\gt{\vec g_2}

\def\gapp{\lower.35em\hbox{$\stackrel{\textstyle>}{\sim}$}}
\def\lapp{\lower.35em\hbox{$\stackrel{\textstyle<}{\sim}$}}

\begin{document}
\bibliographystyle{apsrev}
%

\title{Gauge fields in graphene}

\author{M. A. H. Vozmediano}
\affiliation{Instituto de Ciencia de Materiales de Madrid,\\
CSIC, Cantoblanco, E-28049 Madrid, Spain.}

\author{M. I. Katsnelson}
\affiliation{Radboud University Nijmegen, \\ Institute for
Molecules and Materials, Heijendaalseweg 135, 6525 AJ, Nijmegen,
The Netherlands}

\author{F. Guinea}
\affiliation{Instituto de Ciencia de Materiales de Madrid, CSIC,
Cantoblanco, E-28049 Madrid, Spain.}

\date{\today}
\begin{abstract}
The physics of graphene is acting as a bridge between quantum
field theory and condensed matter physics due to the special
quality of the graphene quasiparticles behaving as massless two
dimensional Dirac fermions. Moreover, the particular structure of
the 2D crystal lattice sets the arena to study and unify concepts
from elasticity, topology and cosmology. In this paper we analyze
these connections combining a pedagogical, intuitive approach with
a more rigorous formalism when required.

\end{abstract}
%
%
%
%

\maketitle

\tableofcontents

 \section{Introduction}
 \label{sec_intro}

One of the sad consequences of the very fast development of
physics during the last half of a century is that it has started
to loose its unity. Especially, the gap between condensed matter
physics which is a key ingredient of our scientific understanding
a ``world around us'' and microphysics investigating the most
fundamental laws of nature (high-energy physics, quantum field
theory, gravity and cosmology, etc.) has become wider. That is why
examples of fruitful exchange of ideas and methods between these
two branches of physics are very important now. A prototype
example of the relevance of macrophysics for microphysics is the
idea of spontaneously broken symmetry which first appeared in
Landau's theory of second-order phase transitions \cite{L37} and
then turned out to be one of the most important concepts in both
condensed matter \cite{pwa} and fundamental physics
\cite{perkins,linde}. Conversely, the idea of the renormalization
group, being developed first to study the problem of infinities in
quantum electrodynamics (QED) has been of crucial importance to
solve  problems of condensed matter physics as difficult as
critical behavior and the Kondo problem
\cite{C84,WK74,M76,wilson2,hewson}. The use of M\"{o}ssbauer
effect to check the general relativity theory \cite{pound} gives
an amazing example of an entanglement of condensed matter physics,
nuclear physics, and gravity.

One of the fields where condensed matter physics meets quantum
field theory and cosmology is that of the superfluidity of helium
3 \cite{V03}. A recent development in material science provides a
new and unexpected bridge between condensed matter and high-energy
physics. The experimental discovery of graphene,  a
two-dimensional allotrope of carbon formed by a single carbon atom
sheet, was made in 2004 when a technique called micromechanical
cleavage was employed to obtain the first graphene crystals
\cite{Netal05a,Netal04}. The observation of a peculiar
``Dirac-like'' spectrum of charge carriers and an anomalous
quantum Hall effect in graphene \cite{Netal05,ZTSK05} has ignited
an enormously growing interest to this field (for a review, see
\cite{GN07,K07,r3,RMP08,geim2009}). One of the most interesting
aspects of the graphene physics from a theoretical point of view
is the deep and fruitful relation that it has with QED and other
quantum field theory ideas
\cite{S84,H88,GGV94,KA06b,KNG06,falko,KN07,SKL07b,beenakker}.
Probably the most clear example is the Klein paradox
\cite{klein,dombey}, a property of relativistic quantum particles
of being able to penetrate with a probability of the order of
unity through very high and broad potential barriers. Previously
it was discussed only for experimentally unattainable (or very
hard to reach) situations such as particle-antiparticle pair
creation at the black hole evaporation, or vacuum breakdown at
collisions of super-heavy nuclei (a rather complete reference list
can be found in \cite{greiner}). At the same time, it appeared to
be relevant for graphene-based electronics and for electronic
transport in graphene \cite{KNG06,SKL07b,GMSHG08,gordon,YK09}.

Here we focus on a particular aspect of the physics of graphene
that establishes its connection with QED, field theory and
gravity, namely, the appearance  of gauge fields in graphene and
their effect on its properties. The concept of a gauge field in
general has been extensively discussed in condensed matter
physics, especially, in relation with modeling different types of
topological defects, phase transitions, and properties of glasses
\cite{K89,sadoc,nelson,toulouse,KV92}. In the modern context of
graphene the gauge fields were first introduced in
\cite{Metal06,MG06} in relation with the problem of weak
localization. It is well-established by now, both experimentally
and theoretically
\cite{Metal06,Metal07,Metal07b,Ietal07,Setal07,FLK07}, that
graphene is always corrugated and covered by ripples which can be
either intrinsic \cite{Metal07,Metal07b,FLK07,nelson} or induced
by a roughness of substrate \cite{Ietal07,Setal07}. In general,
the departure from flatness of graphene leads to the appearance of
an inhomogeneous pseudomagnetic gauge field \cite{Metal06,MG06}
acting on the charge carriers. This has important consequences
affecting the character of quantum Hall effect in graphene, it
produces an effective source of charge carrier scattering, and it
can provide a mechanism for an observed charge inhomogeneity, as
discussed in detail below. Whereas a smooth deformation of the
graphene sheets produces a gauge field similar to the
electromagnetic one, different topological defects in graphene
inducing inter-valley (Umklapp) processes can be considered
sources of a non-Abelian gauge field \cite{GGV92,LC04}. On the
other hand the description of some topological defects as cosmic
strings \cite{CV07a} creates an interesting connection with
general relativity. This will be also described in the present
review.

The work is organized as follows. Section ~\ref{sec_lattice}
presents the electronic features of the model that are most
relevant for the physics discussed. Section~ \ref{sec_dirac} gives
a short description of the way in which the Dirac equation
approximates the electronic states of graphene in the long
wavelength limit emphasizing the need for two component
wavefunctions. In subsection ~\ref{sec_stability} we show the
topological stability of the Fermi points in the honeycomb lattice
towards small perturbations like lattice deformations or disorder.
Then, in subsections~\ref{sec_gaugegeneral} and ~\ref{sec_gauge}
we describe the way in which gauge fields were defined in physics
and in geometry and we show the different types of effective gauge
fields that appear in graphene associated to various disorder
types. Different physical mechanisms which can contribute to these
gauge fields are discussed next, in sections ~\ref{sec_defects}
(topological defects) and ~\ref{sec_elastic} (smooth
deformations). Observable effects related to the gauge fields are
described in sections~\ref{sec_obs} (microscale gauge fields) and
~\ref{sec_mesos} (macroscale gauge fields). Section
~\ref{sec_conclusions} presents a summary of the work, open
questions, and some possible future developments. Technical
aspects related to general features of spinors in a curved two
dimensional surface are explained in
Appendix~\ref{sec_curvedspace} and some auxiliary material from
the elasticity theory is presented in Appendix~\ref{sec_radial}.

\section{The honeycomb lattice: Spinors and geometry in two
dimensions} \label{sec_lattice}

\subsection{The low energy electronic excitations of graphene. Continuum model}
\label{sec_dirac} 
\begin{figure}[!t]
\begin{center}
\includegraphics[width=5cm]{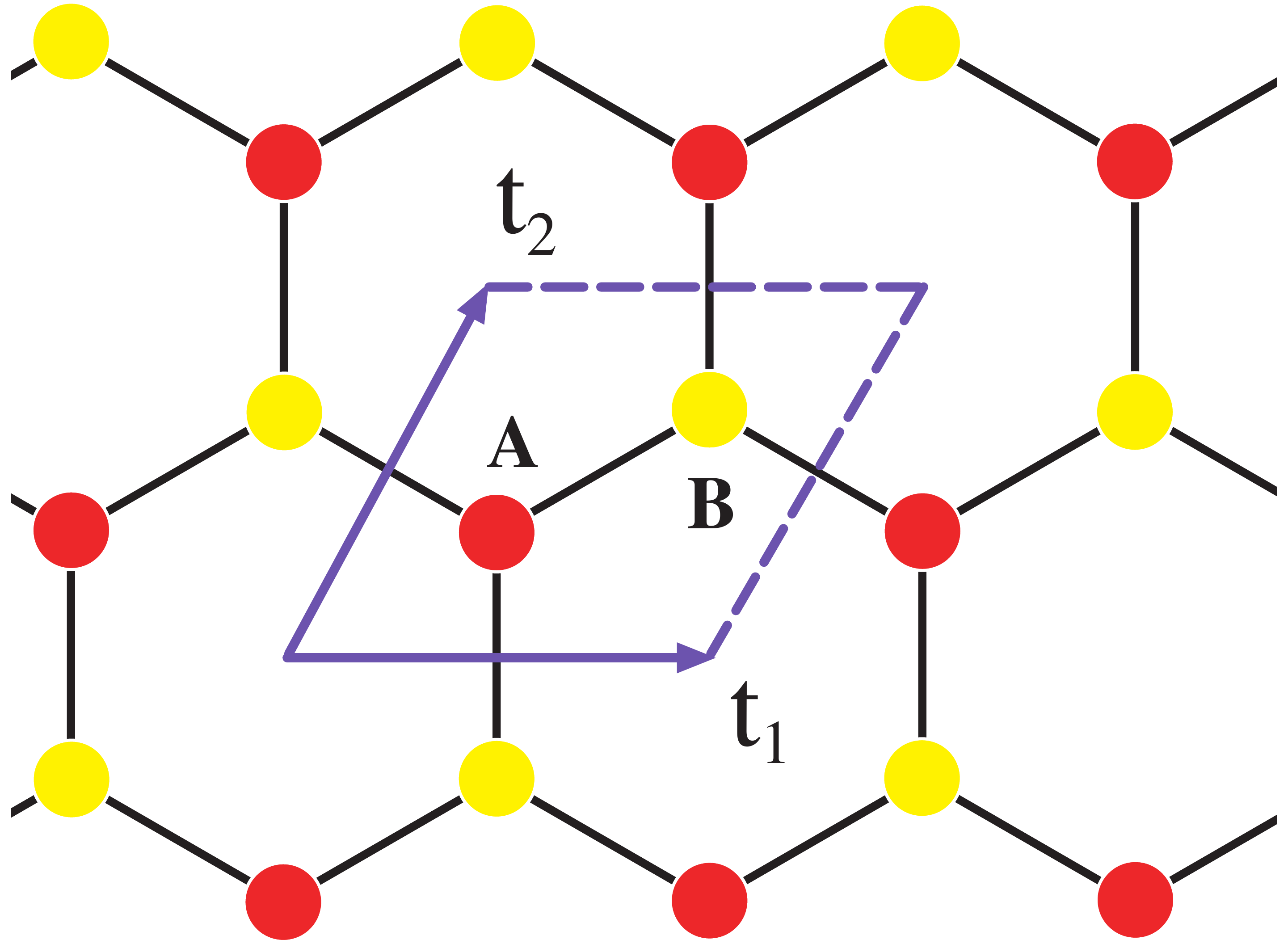}
\caption[fig]{\label{DL}(Color online) Lattice and unit
cell for monolayer graphene in real space. }
\end{center}
\end{figure}

Monolayer graphite - graphene - consists of a
planar honeycomb lattice of carbon atoms shown in Fig. \ref{DL}.
In the graphene structure the in-plane $\sigma$ bonds are formed
from 2$s$, 2$p_x$ and 2$p_y$ orbitals hybridized in a $sp^2$
configuration, while the 2$p_z$ orbital, perpendicular to the
layer, builds up covalent bonds, similar to the ones in the
benzene molecule. The $\sigma$ bonds give rigidity to the
structure, while the $\pi$ bonds give rise to the valence and
conduction bands.
The electronic properties around the Fermi energy of a graphene
sheet can be described by a tight binding model with only one
orbital per atom, the so-called $\pi$-electron approximation
because there is no significant mixing between states belonging to
$\sigma$ and $\pi$ bands in 2D graphite. Within this approximation
a basis set is provided by the Bloch functions made up of the
2$p_z$ orbitals from the two inequivalent carbon atoms A and B
which form the unit cell of the honeycomb lattice. One may define
two Bloch wave functions to be used in a variational
(tight-binding) computation of the spectrum
\beq
\label{bloch}
\Phi_i (\vec K)=\sum_{\vec t} e^{i \vec
K\cdot(\vec r_i+ \vec t)} \Phi (\vec r-\vec r_i-\vec t) \; \;\; ,
\; \;\; i=A,B
\eeq
where the sum runs over all the points in the direct lattice,
i.e., $\vec t = n_1 \vec t_1+n_2 \vec t_2$, $(\vec r_A,\vec r_B)$
are the positions of the atoms in the unit cell, and $\Phi(\vec
r)$ is a real ($\pi$-type) atomic   orbital. As it is well known,
a simple tight-binding computation~\cite{W47,SW58} yields at the
neutrality point a Fermi surface made of the six single points
located at the corners of the hexagonal Brillouin zone. Only two of them
are inequivalent and can be chosen to be located at $\vec K_1 =
-2\go/3- \gt/3$ and $\vec K_2= -\vec K_1$, where $\vec
t_i\cdot\vec g_j=2\pi \delta_{ij}$. A low energy expansion around
any of the two Fermi points $K_{+,-}$ gives an effective
hamiltonian linear in momentum which reduces to the massless Dirac
equation in two dimensions derived from the Hamiltonian:
\begin{align}
{\cal H}_{0} = - \hbar v_{\rm F} \int d^2 {\bf r} \bar{\Psi}({\bf
r}) ( i \sigma_x \partial_x + i \sigma_y \partial_y ) \Psi ({\bf
r})\;, \label{hamil}
\end{align}
where $\sigma_i$ are the Pauli matrices, $v_{\rm F} = (3 t a )/2$,
$t$ is the nearest-neighbor hopping parameter, and $a$ = 1.42
$\AA$  is the distance between nearest carbon atoms.

The components of the two-dimensional spinor:
\begin{equation}
\Psi( {\bf \vec{r}} )= \left( \begin{array}{c} \Psi_1 ( {\bf r} )
\\  \Psi_2 ( {\bf r }) \end{array} \right) \label{spinor}
\end{equation}
correspond to the amplitude of the wavefunction in each of the two
sublattices which build up the honeycomb structure. The two Fermi
points of the honeycomb lattice give rise to two similar equations
related by time reversal symmetry.

We can combine the two spinors attached to each Fermi point into a
four component Dirac spinor and write the four-dimensional
Hamiltonian:
\begin{eqnarray}
H_{D}=-iv_{F}\hbar(1 \otimes \sigma_{1} \partial_{x}+\tau_{3}
\otimes \sigma_{2} \partial_{y}) ,\label{hamil4}
\end{eqnarray}
where $\sigma$ and  $\tau$ matrices are Pauli matrices acting on
the sublattice and valley degree of freedom respectively.

Summarizing, the internal degrees of freedom of charge carriers in graphene
are: Sublattice index (pseudospin), valley index (flavor) and
real spin, each taking two values. The real spin is irrelevant
for the issues discussed in this review and will not be taken
into account except for an additional degeneracy factor 2 in some quantities.

\subsection{Topological stability of the Fermi points}
\label{sec_stability} The properties of graphene presented in this
work as well as most of the exotic properties of the material lie
on the special character of its low energy excitations obeying a
massless two dimensional Dirac equation. Hence it is important to
establish the robustness of the low energy description under small
lattice deformations, disorder  and other possible low energy
perturbations. In this section we will analyze the existence and
stability of the Fermi points in the graphene lattice. A related
issue is the possibility to open and control a gap in the
material, crucial for electronic applications. We will analyze
under which conditions a gap can open in neutral graphene. We will
see that the Dirac points in graphene are topologically preserved
under rather general circumstances and Coulomb interactions,
stress, phonons, and other moderate perturbations of the lattice will not
open a gap. The analysis of this section follows closely ref.
\cite{MGV07}.

The discrete symmetries of the system playing a very important
role in the analysis we will summarize them first. As it was
described in Section \ref{sec_dirac}, the Fermi surface of
neutral graphene consists of two independent Fermi points that can
be chosen in the Brillouin zone to be located at {\it K} and {\it -K}. The
effective low energy Hamiltonians around these points are
\beq\label{lin} H(\vec K_1+\vec k)\sim\left(
\begin{array}{cc} 0 & k^*
\\ k & 0 \end{array} \right)=k_x\sigma_x+k_y\sigma_y ,
\eeq and \beq
H(-\vec K_1+\vec k)\sim\left( \begin{array}{cc} 0 &- k \\
-k^* & 0 \end{array} \right) \sim - k_x \sigma_x + k_y \sigma_y.
\eeq
where $k\equiv k_x+ik_y$ and $\sigma_i$ are the Pauli matrices.

A gap will be opened by a  $k$-independent, translationally
invariant  perturbation of the type
\beqa
\label{const} & H(\vec K_1+\vec k)\to\left( \begin{array}{cc}
a_z & k^*+a^*
\\ k+a & -a_z \end{array} \right)
\eeqa where $a\equiv a_x+i a_y$.  The spectrum becomes
$E=\pm\sqrt{a_z^2+|k+a|^2}$ and a gap $2| a_z|$is generated.
However as we will see, such a perturbation is not allowed if the
discrete symmetries of the system are to be respected. The two
relevant discrete symmetries in graphene are time-reversal $T:t\to
-t$ and spatial inversion $I:(x,y)\to(-x,-y)$. The reality of the
$\pi$~orbitals implies that time reversal  merely reverses $\vec
K$ \beq\label{compl} T\Phi_i(\vec K) =\Phi^*_i(\vec K)=\Phi_i
(-\vec K) \eeq whereas the spatial inversion  also exchanges the
two types of atoms \beq\label{I} I\Phi_A (\vec K)=\Phi_B (-\vec
K)\;\;\; , \;\;\; I\Phi_B (\vec K)=\Phi_A (-\vec K). \eeq
Invariance under these symmetries imposes the following
constraints on the Hamiltonians: \beqa\label{trev}
T:\;  H (\vec K)&=&H^* (-\vec K)\non\\
I:\;H (\vec K)&=& \sigma_x H(-\vec K) \sigma_x. \eeqa From eqs.
(\ref{I}) and (\ref{trev}) we see that none of the two symmetries
are obeyed for the effective description around a single Fermi
point. The graphene system is only invariant under $T$ and $I$
individually if both Fermi points are considered simultaneously.
Nevertheless the product $TI$ imposes a constraint on the form of
$H(\vec K)$ at each given Fermi point: \beq\label{TI} TI:\; H(\vec
K)=\sigma_x H^*(\vec K) \sigma_x. \eeq If the allowed perturbation
respects $TI$ this implies that $H_{11}(\vec K)=H_{22}(\vec K)$,
what enforces
  $a_z\!=\!0$ in~(\ref{const}) and   no gap opens.

This has an interesting topological interpretation, which extends
  the previous arguments to $k$-dependent ---but translationally invariant---  perturbations.
  The low energy hamiltonian $H(\vec K_1+\vec k)$ in~(\ref{lin})
  defines a  map from the circle $k_x^2+k_y^2=R^2$ to the space
  of $2\times 2$ Hamiltonians
 $H=\vec h \cdot\vec \sigma$:
 \beq\label{map1}
 k=R e^{i\theta} \to (h_x,h_y, h_z) =R (\cos\theta ,\sin\theta,0).
 \eeq
  Since Fermi points correspond to zeroes of the determinant
$ -Det(H)=h_x^2+h_y^2+h_z^2$,
  a perturbation will be able to create  a gap only if the loop
  represented by the map~(\ref{map1}) is contractible in the space of
  Hamiltonians with non-vanishing determinants, which 
  is just $R^3-\{ 0\}$. This is clearly the case, since
$ \pi_1(R^3-\{ 0\})=\pi_1(S^2)=0$.
On the other hand, Hamiltonians invariant under $TI$ are
represented
 by points in $R^2$, and   we have
 \beq\label{hom2}
 \pi_1(R^2-\{ 0\})=\pi_1(S^1)=Z
 \eeq
This means that non-trivial maps such as the ones implied
by~(\ref{lin}) can only be extended to the interior of the circle
by going through the origin, i.e., by having at least one zero.
This precludes the creation of a gap. The non--trivial topological
charge associated to the topological stability of each individual
Fermi point in graphene is simply the winding number which can be
computed from
\beq\label{wind}
 N={1\over 4\pi i}\int_0^{2\pi} d\theta\,
 \mathrm{Tr} (\sigma_z H^{-1} \partial_\theta H).
 \eeq
It is important to realize that the two Fermi points have opposite
winding numbers implying that they could annihilate mutually if
brought together by a perturbation.

The nontrivial topological charge associated to the topological
stability of individual Fermi point in very general 2D systems has
been discussed in \cite{V06} and in \cite{FK07,M09,Xetal09}  in the context of
topological insulators.

\subsection{Gauge fields in physics and geometry}
\label{sec_gaugegeneral} Gauge invariance plays a key role in the
quantum field theory (QFT) description of fundamental forces between
elementary particles.  The seemingly abstract concept of a
non-Abelian gauge field introduced first in QFT to describe 
the electroweak interaction followed by  the
experimental discovery of the $W$ and $Z$ bosons is one of the
most impressive achievements of theoretical physics. Before
introducing the various gauge fields appearing associated to the
physics of graphene and in order to clarify their specific nature
we will make a brief description of the classical concept of gauge
invariance and of the associated gauge fields.

The concept of gauge invariance emerged from classical
electrodynamics. In particular the electromagnetic field (${\bf
E}, {\bf B}$) is expressed in terms of potentials ($\Phi$, ${\bf
A}$) through: \beq {\bf E}=-(\nabla\Phi+\partial_t{\bf
A})\;\;,\;\; {\bf B}=\nabla \wedge {\bf A}. \eeq The fields do not
change under the transformation \beq {\bf A}\to {\bf A}+\nabla
\chi \;\;,\;\; \Phi\to\Phi-\partial_t\chi, \label{eq_gauge} \eeq
where $\chi$ is an arbitrary (smooth)  function of space. This
invariance was shown to survive to the Quantum Mechanics
description of a charged spinless particle in an electromagnetic
field provided that  the wave function was simultaneously
transformed  to \beq \Psi\to\Psi\exp(ie\chi). \eeq The
relativistic wave equation for a spinless particle with charge $e$
interacting with electromagnetic fields is derived by first
performing the substitution $p^\mu\to p^\mu-e A^\mu$, where
$A^\mu=(A^0=\Phi,{\bf A})$ is the 4-vector electromagnetic
potential and then performing the usual substitution $p_\mu\to i
{\hbar}\partial_\mu$. A formal solution for the wave function of a
particle interacting with the electromagnetic potential $ A^\mu$
can be written in terms of the solution without interaction
$\Psi_0$ as \beq \Psi=\exp\left[-ie\int A^\mu dx_\mu\right]\Psi_0.
\eeq The original discovery by V. Fock was that the quantum
dynamics, i. e.  the form of the quantum equation, remains
unchanged by the transformations (\ref{eq_gauge}) if the wave
function of the particle is multiplied by a local
(space-time-dependent) phase. The modern description and the term
`` gauge invariance"  was established by H. Weyl. An interesting
account of the historical development of the gauge concept is
given in \cite{JO01}.

The first example of a  QFT gauge model is four dimensional  QED.
A free spin 1/2 Dirac fermion with charge $e$ and mass $m$ is
described by the action \beq S_\Psi=\int d^4
x\bar\Psi\left[\gamma^\mu\partial_\mu+m\right]\Psi,
\label{eq_diracaction} \eeq which is invariant under the global
$U(1)$ group  transformations: \beq \Psi(x)\to U \Psi(x) \;,\;
\bar\Psi(x)\to U^*\bar\Psi(x) \;,\; U=\exp(ie\chi), \label{u1}
\eeq where $\chi$ is a constant. Gauge invariance requires
invariance of the action under the {\it local} group of
transformations obtained by replacing $\chi\to \chi(x)$. This can
be achieved by replacing  the derivative in (\ref{eq_diracaction})
by the covariant derivative  $ D_\mu=\partial_\mu+ieA_\mu$: Under
a local $U(1)$ transformation defined by  (\ref{u1}) with a
space-time dependent function $\chi(x)$,  $A_\mu(x)$ transforms
as $A_\mu\to A_\mu-\partial_\mu\chi$, a generalization of
(\ref{eq_gauge}).

The invariance of the Maxwell equations under special relativity
allows a formulation of (classical) electromagnetism in terms of
quadrivectors and tensors. The equations can be written in a
covariant way by introducing the electromagnetic tensor
$F_{\mu\nu}$ defined by \beq F_{0i}=E_i
\;\;,\;\;F_{ij}=-\epsilon_{ijk}B_k, \label{eq_emtensor} \eeq and
the quadricurrent $J^\mu=(\rho, {\bf J})$ made of the charge
density and the current. In terms of these geometric objects the
four Maxwell equations reduce to \beq
\partial_\lambda F_{\mu\nu}+\partial_\mu F_{\nu\lambda}+\partial_\nu F_{\lambda\mu}=0,
\eeq
\beq
\partial_\mu F^{\mu\nu}=J^\nu.
\eeq The conservation of the current $\partial_\nu J^\nu=0$
follows  from the antisymmetry of $F_{\mu\nu}$. The first equation
is identified as a Bianchi identity that can be integrated by
introducing a gauge field $A_\mu$ such that
$F_{\mu\nu}=\partial_\mu A_\nu-\partial_\nu A_\mu$. It is readily
verified that two gauge fields related by the gauge transformation
$A'_\mu=A_\mu-\partial_\mu \Omega$ give rise to the same
electromagnetic tensor field. Maxwell's equations can be derived
from the action \beq S(A, J)=\int d^4
x\left[F_{\mu\nu}F^{\mu\nu}+J_\mu A^\mu\right], \eeq which is
precisely the full QED action.

The concepts of gauge fields and covariant derivatives can be
translated into the language of differential geometry, based on
differential forms. In general, the gauge field  has a
mathematical interpretation as a Lie-valued connection and is used
to construct covariant derivatives acting on fields, whose form
depends on the representation of the group  under which the field
transforms (for global transformations). The  field tensor
$F_{\mu\nu}$ is a curvature 2--form given by the commutator of two
covariant derivatives. It is an element of the Lie algebra
associated to the gauge group. The gauge connection generates
parallel transport of the geometric objects under gauge
transformations. The generalization of $U(1)$ to non-Abelian
groups as $SU(N)$ is straightforward, the main modification arises
in the definition of the field strength (\ref{eq_emtensor}) that
becomes $F_{\mu\nu}=\partial_\mu A_\nu-\partial_\nu
A_\mu+[A_\mu,A_\nu]$. Since most of the gauge fields arising in
the graphene context will not have dynamics we will not discuss
this point further.

Einstein's General relativity can be also interpreted as a gauge
theory where gauge invariance is invariance under diffeomorphisms
(local smooth  changes of coordinates) in the space-time manifold.
The  connection which generates parallel transport plays the role
of the gauge field. The similarity is better appreciated in the
vielbein formalism introduced in Section \ref{curve} and Appendix
1 where a gauge transformation corresponds to a change of local
frame (a local Lorentz transformation). Gauge invariance
corresponds to the independence of field equations from the choice
of the local frame. The  spin connection plays the role of the
gauge field.

In any gauge theory physical observables are related to
gauge-invariant operators. The gauge invariance allows to fix some
conditions on the gauge potentials that will not affect the
physical properties. In quantum gauge theories gauge-fixing is a
delicate issue that will no concern us here. In classical
electromagnetism, the gauge-fixing problem is simply the problem
of choosing a representative in the class of equivalent
potentials, convenient for practical calculations or most suited
to physical intuition. In non-relativistic problems one of the
most popular choices is the Coulomb gauge: $\nabla . {\bf A}(t,
{\bf x})=0$ whose relativistic counterpart  $\partial_\mu A^\mu(t,
{\bf x})=0, \mu=(0, 1, 2, 3)$ is the Landau or Lorentz's gauge.
The freedom to choose a gauge condition is related with the full
gauge invariance of the action. When fictitious gauge fields are
generated by analogy with the gauge formalism but there is no
dynamics associated to them it can happen that the gauge
potentials are fixed by the physics involved and no extra
conditions can be imposed. A particular example is provided by the
strain fields discussed in Sect. \ref{sec_elasticgauge}. Gauge
fields were introduced in condensed matter in the early works of
refs. \cite{DV78,DV80}.

\subsection{Types of gauge fields} \label{sec_gauge}
Different physical mechanisms give rise to perturbations in the
Dirac equation described above which can be described mathematically
as effective gauge fields.  By definition, a gauge field appears in
the off diagonal elements of the Dirac equation, which correspond to
the hopping between sublattices. There are two ways in which this
change in the hopping between sublattices can take place:
\begin{itemize}
\item
The topology of the lattice itself can change, requiring a
redefinition of the two sublattices. This happens in the presence of
topological defects, such as pentagons and heptagons
(disclinations), or pentagon-heptagon pairs (dislocations). If the
carbon atoms remain with threefold coordination, the local
electronic structure is not altered very much, and an effective
Dirac equation can be defined locally. The defect changes the global
properties of the lattice. Modifications of the internal degrees of
freedom of the electrons when they propagate over long distances can
be described by gauge fields, as defined more precisely in the
following section.

The substitution of hexagons by rings with fewer or more sites,
when the bond lengths remain more or less unchanged, leads to the
curvature of the structure, according to Euler's theorem. As
described below, these defects induce, in the continuum limit, the
spin connection which modifies the Dirac equation of a spinorial
field, extensively discussed in quantum field theory in curved
spaces. Because these gauge fields arise from general topological
features of the system, {\it these gauge fields do not depend on
any material parameter}.

\item The hopping between $\pi$ orbitals in different sublattices
can also change because the distance between the orbitals is
modified, or because a change of symmetry allows for indirect
hoppings through the $\sigma$ orbitals \cite{KN07b}, also present
in the system. {\it These fields depend on microscopic details of
the material}. The same changes in the electronic spectrum are
induced when the atoms oscillate around their equilibrium
positions, so that the parameters which describe the strength of
these gauge fields contribute also to the electron-phonon
coupling.

Small static deformations of the lattice are described by the strain
tensor. A field with the symmetries of a vector can be obtained by
contracting the strain tensor with a third rank tensor. The trigonal
symmetry of the honeycomb lattice allows for such a third rank
tensor\cite{SA02b,M07}  (note that a vector gauge field cannot be
defined in a fully isotropic system). The curvature away from the
flat configuration can be described in a similar way by the
curvature tensor. The induced gauge field in the Dirac equation is
obtained in analogous way by contracting the curvature tensor and
the third rank trigonal tensor described above.
\end{itemize}

\section{Topological defects in
graphene} \label{sec_defects}

As described in Sect. \ref{sec_dirac},  graphene  consists of a
planar honeycomb lattice of carbon atoms determined by the $sp^2$
hybridization of the 2$s$, 2$p_x$ and 2$p_y$ orbitals of the
carbon system. These $\sigma$ bonds give rigidity to the
structure, while the $\pi$ bonds give rise to the valence and
conduction bands. The elastic properties of the material are
related with the $\sigma$ bonds and involve energies of the order
of 7-10 eV similar to  the bandwidth of graphene ($\sim 14 eV$).
The carbon bond in graphite is one of the strongest chemical bonds
occurring in nature. The Young's modulus of graphene is of the
order of 350 N$\cdot$ m$^{-1}$, one of the highest values known
for any material \cite{K81,LWetal09,ZKF09}.
\begin{figure}
\begin{center}
\includegraphics[height=4cm]{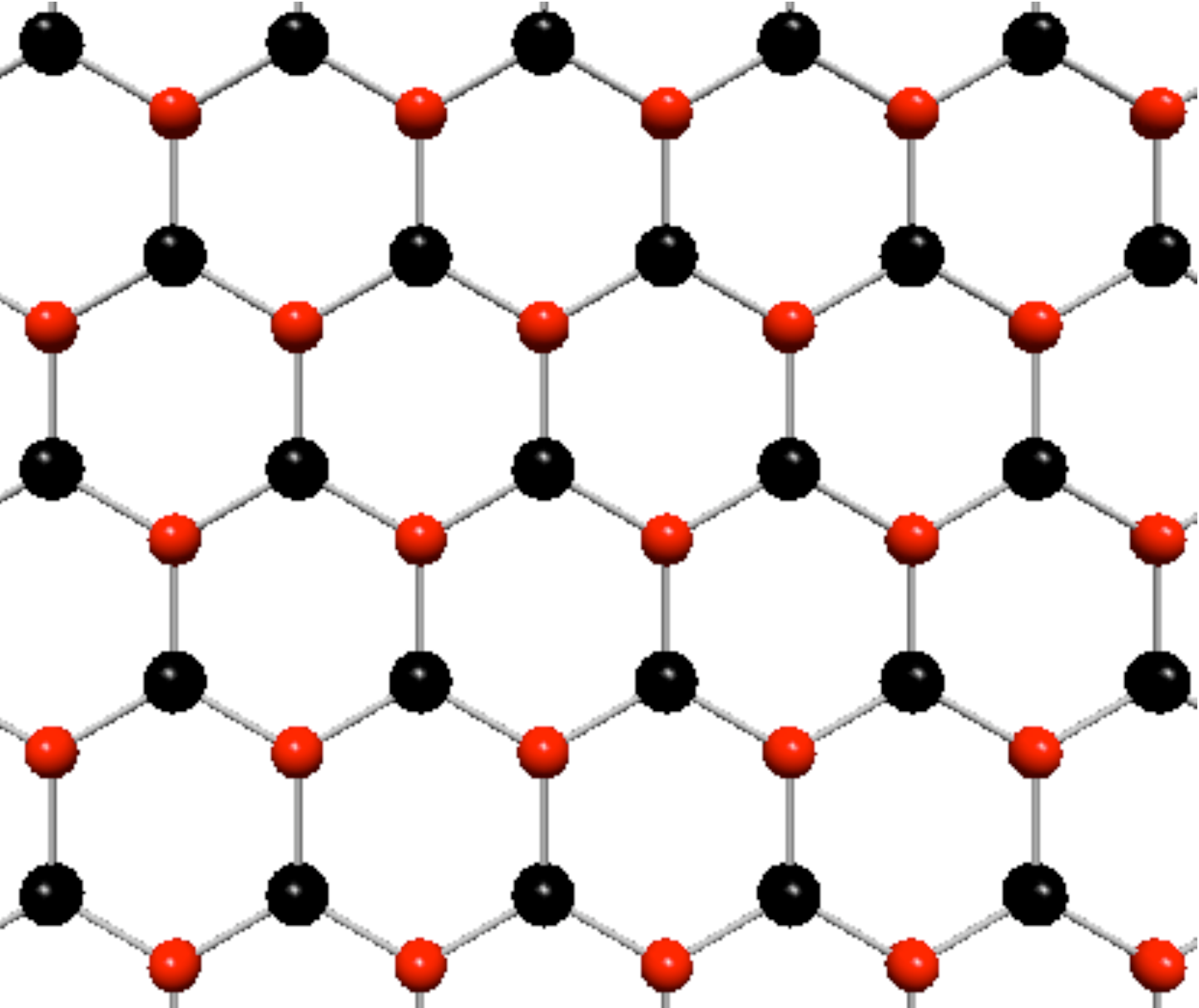}
\caption{(Color online) Honeycomb lattice of monolayer graphene. }
\label{lattice}
\end{center}
\end{figure}

Topological defects in solids are usually described by complicated
boundary conditions in elasticity theory.  It was very early
realized that they may be described more simply as sources of a
gravity-like deformation \cite{KV92}. In this approach the
boundary conditions imposed by defects in elastic media are
accounted for by introducing curvature in the given space. In the
continuum limit the crystal is described by a manifold where
curvature and torsion are associated to disclinations and
dislocations in the medium respectively. The Burgers vector of a
dislocation is associated to torsion, and the Frank angle of a
disclination to curvature.  The
combined effect of curvature and elasticity has been studied in
\cite{KO06} and  the scattering of phonons in the presence of a
disclination and of multiple disclinations in \cite{FCR06}.
The fact that elastic deformations of a crystal lattice induce
gauge fields that couple to the electronic degrees of freedom has
been known for a long time \cite{K89} and it has been applied to
nanotubes and graphene in \cite{FCR06,Metal06,MG06}.
The mechanics of defects of various classes in graphene has been
analyzed recently in \cite{ZJetal06}. The authors discuss the
failure of classical elasticity to describe divacancies and
Stone-Wales defects.
The defect formation under strain was studied in the context of
carbon nanotubes in \cite{LDetal05}.

\subsection{Topological defects. Formation
and naturalness} \label{sec_topological}

A very natural way of producing local curvature in the graphene
lattice is by substituting some of the hexagons by pentagons
(positive curvature) or heptagons (negative curvature). These
types of topological defects are present in all the previously
existing graphene structures (fullerenes and nanotubes), have been
observed experimentally \cite{Aetal01,Hetal04,DML04} and their
elastic and electronic properties have been studied at length
\cite{SDD98}. It is well known for instance from studies of
stability of carbon nanotubes that vacancies and similar defects
produced in the graphene structure by ion irradiation of the
samples are mended by forming higher membered rings. Subsequently
the unstable high-membered rings have been found to disappear by
Stone-Wales transformations \cite{SW86}, thus leading to very
stable structures  mainly constituted of five-, six-, and
seven-membered rings \cite{ARC98}. High-resolution Transmission
Electron Microscopy (TEM) with atomic sensitivity reported the
first direct imaging of pentagon--heptagon pair defects  in a
single wall carbon nanotube  heated at 2,273 K in \cite{SWetal07}.

The recent TEM experiments performed in suspended graphene sheets
\cite{Metal07,Metal07b} seem to indicate that  the observed
structure can not be explained by strain-free deformations of
graphene. Topological defects can be produced in the process of
mechanical cleavage which involves quite high energies. They would
induce strain-free curvature compatible with the experimental
data. Although the observed tilt reported is argued to produce a
strain too small to give rise to nucleations of defects
\cite{Metal07}, it can well be that the structure observed is the
final result of a fully relaxed structure where high energy
deformations have relaxed through formation of topological defects
and the lower energy elastic deformations remain. Topological
defects could also be preexisting in the underlying graphite
material. Even if the observed structure in graphene is not due to
these types of defects, it is worth to study how their presence
would affect the electronic and elastic properties of the
material.
\begin{figure}
  \begin{center}
    \includegraphics[height=4cm]{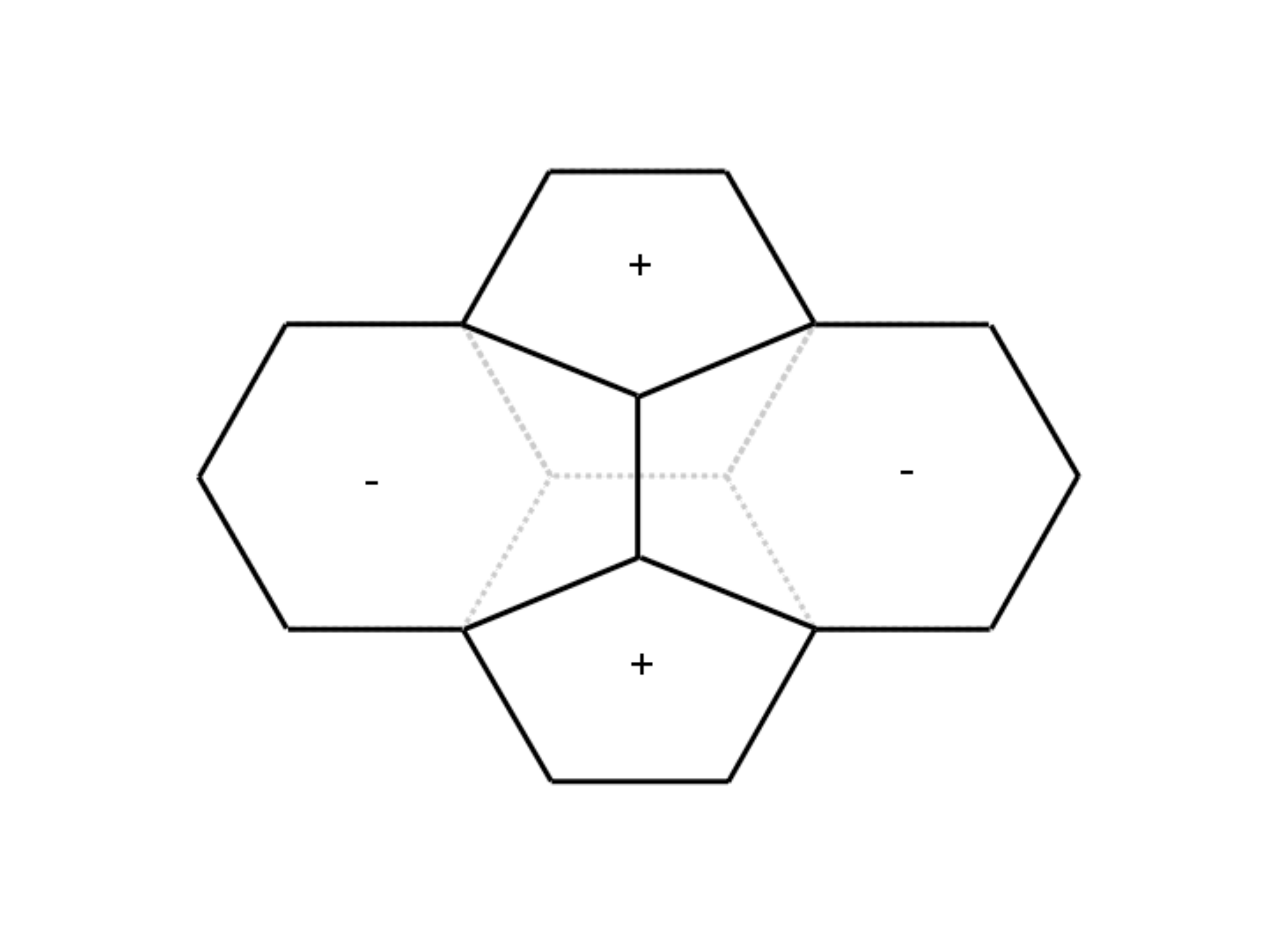}
    \caption{A Stone-Wales defect is made by  a 90 degrees rotation
    of a bond in the honeycomb lattice what produces two pentagon-heptagon
    rings.
    }
    \label{SW}
\end{center}
\end{figure}
In terms of the elasticity theory the pentagons and heptagons
represent disclinations of the lattice. Each disclination is made
by either removing or inserting a wedge of material as shown in
Fig. \ref{pentagon} for the particular case of a pentagon. Such a
procedure would create an enormous elastic strain in flatland,
i.e., if you insist in gluing the edges in flat space. The
structure is much more - or fully - relaxed if the crystal is
allowed to bend in the third direction forming a cone (or a saddle
point structure in the case of excess angle). This procedure is
not possible in three dimensional crystals where disclinations are
very rare. A dislocation can be formed by joining a pentagon and a
heptagon through a line which will represent the Burgess vector of
the dislocation. The Stone-Wales defects \cite{SW86} play a very
important role in the structural stability of the previously
existing graphene structures. They are formed by two
pentagon-heptagon pairs oriented as displayed in Fig. \ref{SW} and
are the result of a 90 degrees rotation of a bond in the honeycomb
lattice. In terms of the elasticity theory they can be represented
as two adjacent dislocations. The spontaneous formation of the
Stone-Wales defects is known to be the main mechanism to relief
strain in plastically deformed nanotubes \cite{NYB98}.

The electronic properties of topological defects in graphene have
been studied at length in the literature by applying various
numerical methods \cite{TT94,K00,CR01}. The main founding is that
they induce charge inhomogeneities in the samples with
characteristic patterns that can be observed in STM and EFM
experiments. Substitution of a hexagonal ring by odd-sided
polygons break the electron-hole symmetry which is preserved by
even-sided rings. Pentagonal rings have been suggested as
excellent field emitters \cite{Setal01}. In the recent field of
topological insulators \cite{M09,Xetal09} topological defects are
able to bind excitations with fractionalized quantum numbers
\cite{PST08,Retal09}. The tight-binding approach for graphene in
the presence of pentagonal and heptagonal rings has been worked
out in the context of nanotubes in the pioneer paper of ref.
\cite{MA98}. Analytical approaches give rise to the appearance of
various gauge fields which we will review. An interesting question
is whether these ``plastic'' deformations can be distinguished
from the elastic deformations which also give rise to gauge fields
in the continuum limit \cite{K89}. One of the main papers to
address this issue that appeared again in the nanotube literature
is \cite{KM97}. It has also been discussed very recently in
\cite{VO07}.

\begin{figure}
  \begin{center}
   \includegraphics[height=4cm]{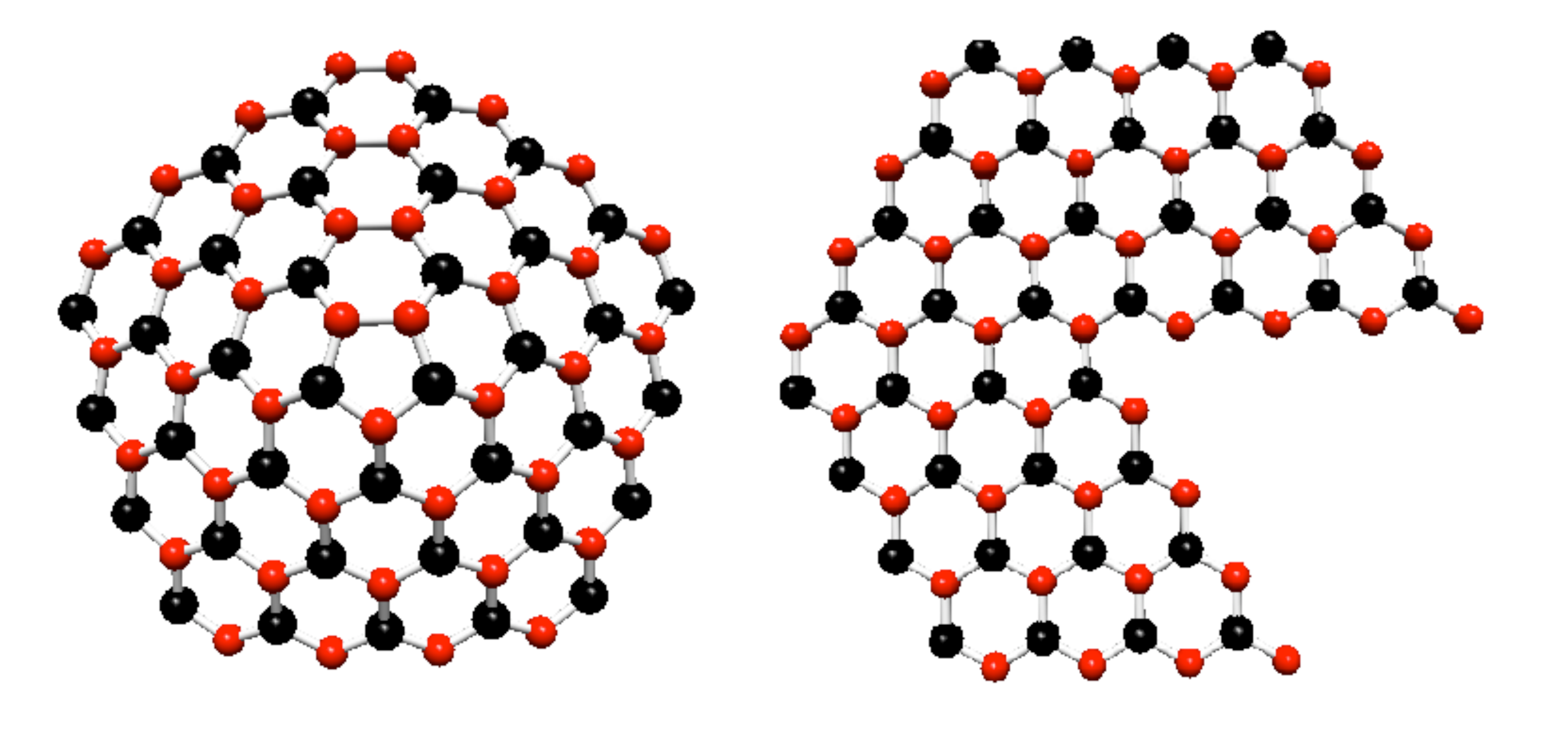}
    \caption{(Color online) Left: Effect of a pentagonal defect in a graphene layer.
    Right: Cut-and-paste procedure to form the pentagonal defect. The  points at the
    edges are connected by a link what induces the frustration of the bipartite character of the lattice at the seam (taken with permission from Ref. \cite{CV07b}).
    }
    \label{pentagon}
\end{center}
\end{figure}
We will first review the case of  disclinations where a hexagonal
ring of the honeycomb lattice is replaced by either a pentagon or
a heptagon. We will deal with the continuum model under the
assumption that  the electronic structure of graphene is well
described by (two) massless Dirac equations in two space
dimensions.

\subsection{The gauge approach to disclinations}
\label{singlediscl}

We will see that modelling disclinations in the continuum gives
rise to three different gauge fields coupled to the electronic
degrees of freedom. One is associated to the phase mismatch seen
by the electron when circling around the defect (holonomy). This
one is independent of the lattice structure and will take care of
the conical shape of the graphene electron energy surfaces
\cite{GGV92,GGV93,LC00,OK01,FMC06}. The second class has to do
with the frustration of the triangular sublattice ordering that
occurs in the seam (see Fig. \ref{pentagon}) and the exchange of
the Fermi points that occurs when going around a pentagon or an
heptagon. These effects where described in full detail in
\cite{GGV92,GGV93}
 and can be modelled by non-Abelian gauge fields.  Finally
the presence of various defects forces the introduction of a third
type of gauge field associated to the fact that the tight binding
phase does not commute with the holonomy associated with the
exchange of the Fermi points \cite{LC04,OK05,CV07d}.

\subsubsection{Two gauge fields for a single disclination}
\label{singlediscl2}

In relativistic quantum field theory where there is a very tight
connection between the spin  and the statistics, a spin 1/2
particle is described by a field that belongs to the so-called
spinor representation of the Lorentz group. A distinct
characteristic of spinor representations is that when they move
around a closed path they pick up a minus sign (upon a $2\pi$
rotation they acquire a phase of $\pi$). A spinor going around a
closed path encircling the pentagon in Fig. \ref{pentagon} will
acquire a phase proportional to one half the total angle of the
path $\phi=2\pi(1-1/6)$. When solving the Shr\"{o}dinger equation
$H\Psi = E\Psi$ with the Hamiltonian (\ref{hamil}) this condition
has to be imposed on the wave function as a boundary condition
which is often hard to deal with. A way to incorporate the
constraint attached to the boundary condition to the Hamiltonian
is to remind the Bohm-Aharonov (BA) effect and substitute the
pentagon by a fictitious magnetic field located at the same point.
The flux of the field can be adjusted so that the phase acquired
by the spinor when going around the vector potential is the same
as the one induced by going around the pentagon.
 This procedure has the advantage of being easily generalized to conical defects of
 arbitrary opening angle. In
the BA effect the phase is proportional to the circulation of the
vector potential along the closed path:
\begin{equation}
\oint {\bf A.dr}=2\pi(1-1/6). \label{vortex}
\end{equation}
 The simplest vector potential having the property (\ref{vortex})
 is a vortex:
 \begin{equation}
 {\bf
 A}(x,y)=\frac{5\pi}{3}\left(\frac{-y}{x^2+y^2},\frac{x}{x^2+y^2}\right)
 =-\frac{5\pi}{3}{\bf \nabla} \theta(r),
 \end{equation}
 which in polar coordinates reduces to the
gradient of the polar angle $\theta$. The presence of a pentagon
in the lattice has an additional consequence which was discussed
at length in \cite{GGV93}: the two sublattices of black and white
points are exchanged by going around a conical singularity, the
admissible wave functions are made of pairs of plane waves with
opposite momenta as described in Se. In momentum space the
inversion with respect to the origin exchanges the two independent
classes of Fermi points. It becomes clear that, for the mentioned
honeycomb lattices, the states of the theory have to accommodate
into the spectrum of two coupled Dirac spinors. This new condition
can be  fulfilled by attaching a quantum number (flavor) to the
Fermi points, and considering the two bi-spinors $\psi_i({\bf r}),
i=+,-$  as
 the two components of an SU(2)
flavor doublet. The vector field will now be a non-Abelian gauge
field able to rotate the spinors in this flavor space. The full
boundary condition to impose on the spinor when circling a
pentagon (or any conical singularity of arbitrary defect angle
$\varphi$) is
\begin{eqnarray}
\Psi(\theta=0)=T_{C}\Psi(\theta=2\pi)\Leftrightarrow \\ \nonumber
\Psi(\theta=0)=
\exp({\oint_{_{C}}\textbf{A}_{a}T^{a}d\textbf{r}})\Psi(\theta=2\pi),
\label{boundary}
\end{eqnarray}
$$\oint {\bf A.dr}=2\pi-\varphi.$$
where $\textbf{A}_{a}$ are a set of gauge fields and $T^{a}$ a set
of matrices related to the flavor SU(2) degree of freedom of the
system. In this way we end up with the Dirac equation coupled to a
gauge potential:
\begin{eqnarray}
H=-i\hbar v_{_{F}}{\vec \gamma}.{\vec\partial}+g\gamma^{q} {\vec
\gamma}.{\vec A}({\bf r}), \label{hamgauge}
\end{eqnarray}
where $v_{_{F}}$ is the Fermi velocity, $ \gamma^q$ is a
$4\times4$ matrix constructed from the Pauli matrices chosen so as
to accomplish the gauge transformation produced by the given
defect, the Latin indices run over the two spatial dimensions and
$g$ is a coupling parameter. The external field $A_{i}({\vec{r}})$
takes the form of a vortex
\begin{eqnarray}
A^{j}({\vec{r}})=
\frac{\Phi}{2\pi}\epsilon^{3ji}\frac{x_{i}}{r^{2}},\label{gaugefield}
\end{eqnarray}
and the constant $\Phi$ is a parameter that represents the
strength of the vortex: $\Phi=\oint\vec{A}d\vec{r} $ and is
related to the opening angle of the defect.

Since disclinations corresponding to even--sided polygons do not
frustrate the sublattice symmetry, the associated matrix
$\gamma^q$ in eq. (\ref{hamgauge}) for this case is
\begin{equation}
\gamma^q=\frac{\sigma_3}{2}\otimes I.
\end{equation}

\subsubsection{Generalization to several defects: gauge fields
arising from the holonomy} \label{severaldiscl}

A generalization of the gauge approach to include various
topological defects was presented in \cite{LC04,OK05,FMC06}. The
strategy consists of determining the phase of the gauge field by
parallel transporting the state in suitable form along a closed
curve surrounding all the defects.
\begin{eqnarray}
\Psi(\theta=0)=T_{C}\Psi(\theta=2\pi)\\ \nonumber
\Leftrightarrow\Psi(\theta=0)=
\exp({\oint_{_{C}}\textbf{A}_{a}T^{a}d\textbf{r}})\Psi(\theta=2\pi),\nonumber
\label{boundary2}
\end{eqnarray}
where $\textbf{A}_{a}$ are a set of gauge fields and $T^{a}$ a set
of matrices related to the pseudospin degrees of freedom of the
system.

\begin{figure}[h]
  \begin{center}
\includegraphics[height=4cm]{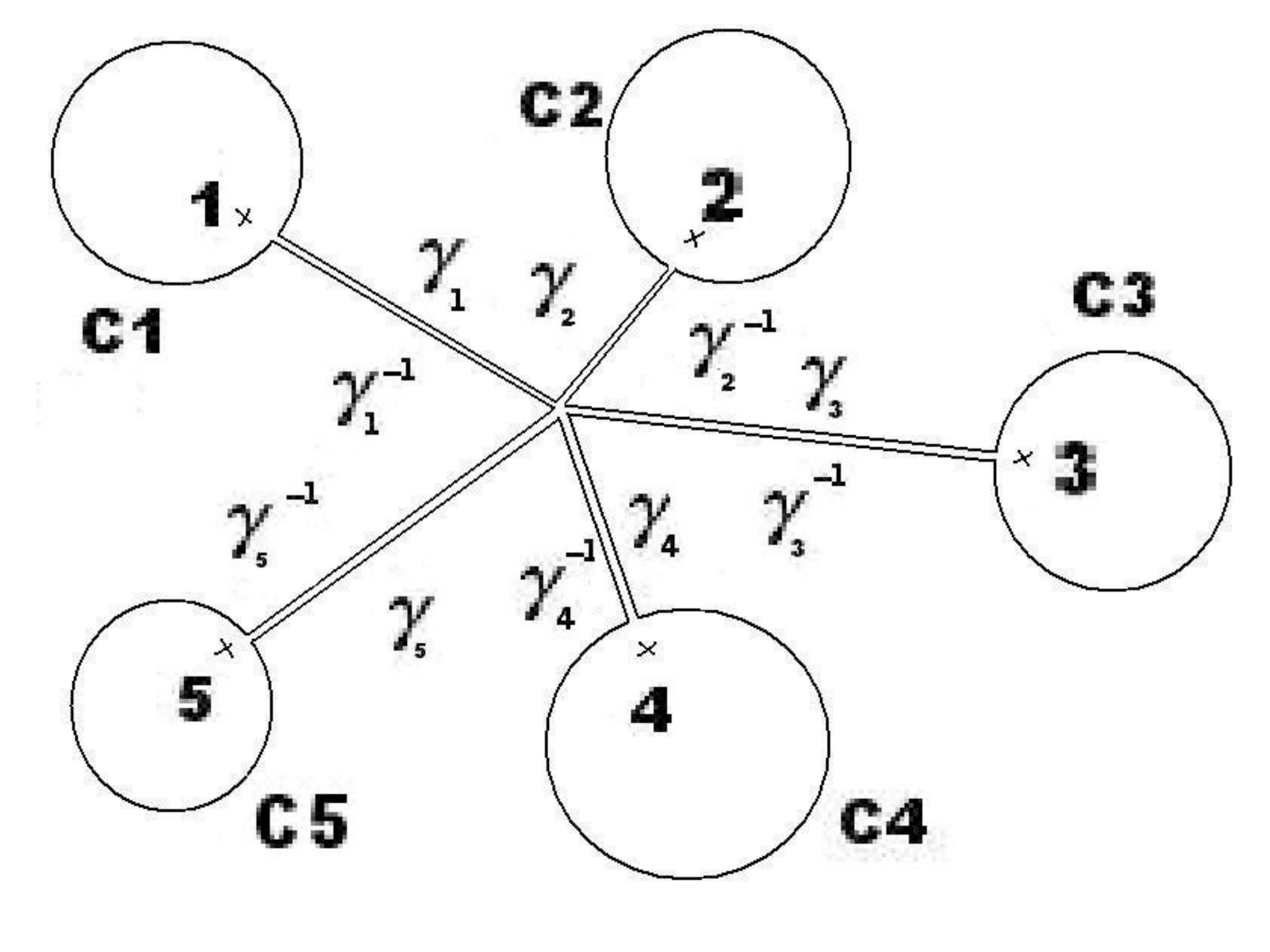}
\end{center}
\caption{Scheme of a prototypical curve enclosing multiple defects
in which the state will be parallel transported.}
\label{scheme}       
\end{figure}

When dealing with multiple defects, we must consider a curve
surrounding all of them, as the one sketched in Fig. \ref{scheme}.

The contour $C$ is made of closed circles enclosing each defect and
straight paths linking all the contours to a fixed origin. The
parallel transport operator $P_{C}$ associated to the closed path
is thus a composition of transport operators over each piece:
\begin{equation}
P=P_{\gamma1}\cdot P_{1}\cdot P^{-1}_{\gamma1}\cdot...\cdot
P_{\gamma N}\cdot P_{N}\cdot P^{-1}_{\gamma N},\label{composite}
\end{equation}
where $P_i$ is the phase associated with going around a single
defect discussed in the previous subsection and $P_{\gamma i}$ is
the chiral phase acquired by the wave function while travelling in
the straight paths and depends on the distances $(n, m)$ between
defects.

The total holonomy turns out to be
\begin{eqnarray}
P=(i)^{N}(\tau_{2})^{N}exp\left(\frac{2\pi
i}{6}(N^{+}-N^{-})\sigma_{3}\right). \nonumber \\
exp\left(\frac{2\pi
i}{3}\sum^{N}_{j=1}(n_{j}-m_{j})\tau_{3}\right),
\label{composite2}
\end{eqnarray}
where $N$ is the total number of defects equal to the sum of
pentagons ($N_+$) and heptagons $(N_-)$, and $(n, m)$ is the
position of the given defect in the usual chiral basis for the
graphene lattice. Then we see that in the modelling of
disclinations with  the gauge approach described in this section
there appear three different gauge fields to incorporate to the
Dirac equation, which couple to the matrices $\sigma_{3}$,
$\tau_{2}$ and $\tau_3$:

1. Gauge fields due to the holonomy: deal with the deficit or
surplus of graphene slices
$$\Lambda_\sigma=\exp(-i\pi\sigma_3/6).$$
This boundary condition only affects the AB pseudo-spin structure.

2. The second type deals with the identification of the two
sublattices
$$\Lambda_\tau=\exp(i\pi\tau_2/2).$$
This part of the boundary condition affects the valley ($K,K'$)
structure of the spinor: the two conical points are intertwined.

3. A third holonomy appears when more than one defect is
considered:
$$\Lambda_3=\exp\left( \frac{2\pi i}{3}\sum_{j=1}^N (n_j-m_j)\tau_3 \right).$$
This boundary condition comes from the fact that the tight binding
phase does not commute with the holonomy associated with $\tau_2$.

From (\ref{composite2}) it is easy to see that for a Stone-Wells
defect we have $N_+=N_-=2$, $n_i=m_i=1$ and $\sigma_2^4=1$ and hence
there is no gauge field associated to it. This casts some doubts
on the completeness of the gauge description as it is known that
such a defect alters the electronic structure of the graphene
sheet \cite{CV07a,CV07b}. In the next section we will see that
curvature effects account for the missing piece.

\subsection{Modelling curvature. Dirac fermions in curved space} \label{curve}

Many of the unusual properties of graphene arise from the fact
that its quasiparticles are described by Dirac spinors. The
special structure of the spinors is linked in particular to the
Klein paradox and to the absence of back scattering in the samples
\cite{KNG06}. When trying to study the possible consequences of
the ripples found in the samples on the electronic properties it
looks very natural to use the formalism of quantum field theory in
curved spaces \cite{BD82}. The main justification to use this
approach is the robustness of the Dirac description versus
deformations of the underlying lattice described in Section
\ref{sec_stability}. We must notice that, although the formalism
is similar to that used in general relativity, here the geometry
(curvature) of the space is fixed and the description is purely
geometrical: the aim is to construct a Hamiltonian with scalar
quantities made out of vectors or spinors in a curved background.
This approach was  applied to study the electronic spectrum of the
fullerenes in \cite{GGV92,GGV93,GGV93A,OK05,KO06}. A somehow
similar approach is used in  \cite{FM94}  in the framework of the
equivalence between the theory of defects in solids and the
three-dimensional gravity \cite{KV92} to study  the response of
electromagnetic charges to conical defects in planar graphene.
The geodesic around a dislocation has been treated in
\cite{M96}. The coupling of the electronic degrees of freedom of
planar graphene to conical defects within the geometric formalism has been explored in
\cite{CV07a,CV07b,CV07d,SV07}. There
it was found that a distribution of pentagons and heptagons
induces characteristic inhomogeneities in the density of states of
the graphene surface. It was shown that the charge density was
enhanced around the pentagonal defects and depressed near the
heptagons, a result that was also predicted earlier in numerical
calculations \cite{TT94}. The effect was very localized and
disappeared a few lattice constants away from the defect.

In order to investigate the effect of pure curvature on the
electronic properties of graphene, the conical defects studied
previously present two difficulties. First, they correspond to
surfaces with zero intrinsic curvature; moreover, the extrinsic
curvature is accumulated at the apex of the cone where the surface
has a singularity. It is then not clear if the results obtained --
which look similar to the ones got with vacancies in
\cite{WBetal06} -- are due to the singularity or to the curvature.
To disentangle the two effects  a flat graphene sheet with a
smooth curved portion with intrinsic curvature was analyzed in
\cite{JCV07}. In all cases the curvature gives rise to gauge
fields with a derivative coupling.  In what follows we will sketch
the formalism and give a summary of the results obtained so far
within the geometric approach. A derivation of the main
geometrical factors associated to the formalism is given in
Appendix \ref{sec_curvedspace}.

The dynamics of a massless Dirac spinor in a curved spacetime with
the metric tensor $g^{\mu\nu}$ is governed by the modified Dirac
equation:
\begin{equation}
i\gamma^{\mu}({\bf r})\nabla_{\mu}\psi=0 \label{dircurv}.
\end{equation}
The two geometric objects entering the Dirac equation are the
Pauli (gamma) matrices and the derivative of the spinor field.
Both are vectors that need to be properly defined in a curved
space. The  curved space $\gamma$ matrices must obey  the
generalized anticommutation relations:
\begin{equation}
\{\gamma^{\mu}({\bf r}),\gamma^{\nu}({\bf r})\}=2g^{\mu\nu}({\bf
r}),\nonumber
\end{equation}
and become functions dependent on the point of the space. They are
related to the constant flat matrices through the ``fielbeins''
described in the Appendix \ref{sec_curvedspace}.  The covariant
derivative operator is defined as
$$
\nabla_{\mu}=\partial_{\mu}-\Gamma_{\mu}
$$
where $\Gamma_{\mu}$ is the spin connection of the spinor field
whose detailed construction is also given in the Appendix
\ref{sec_curvedspace}.

\begin{figure}
  \begin{center}
    \includegraphics[height=4cm]{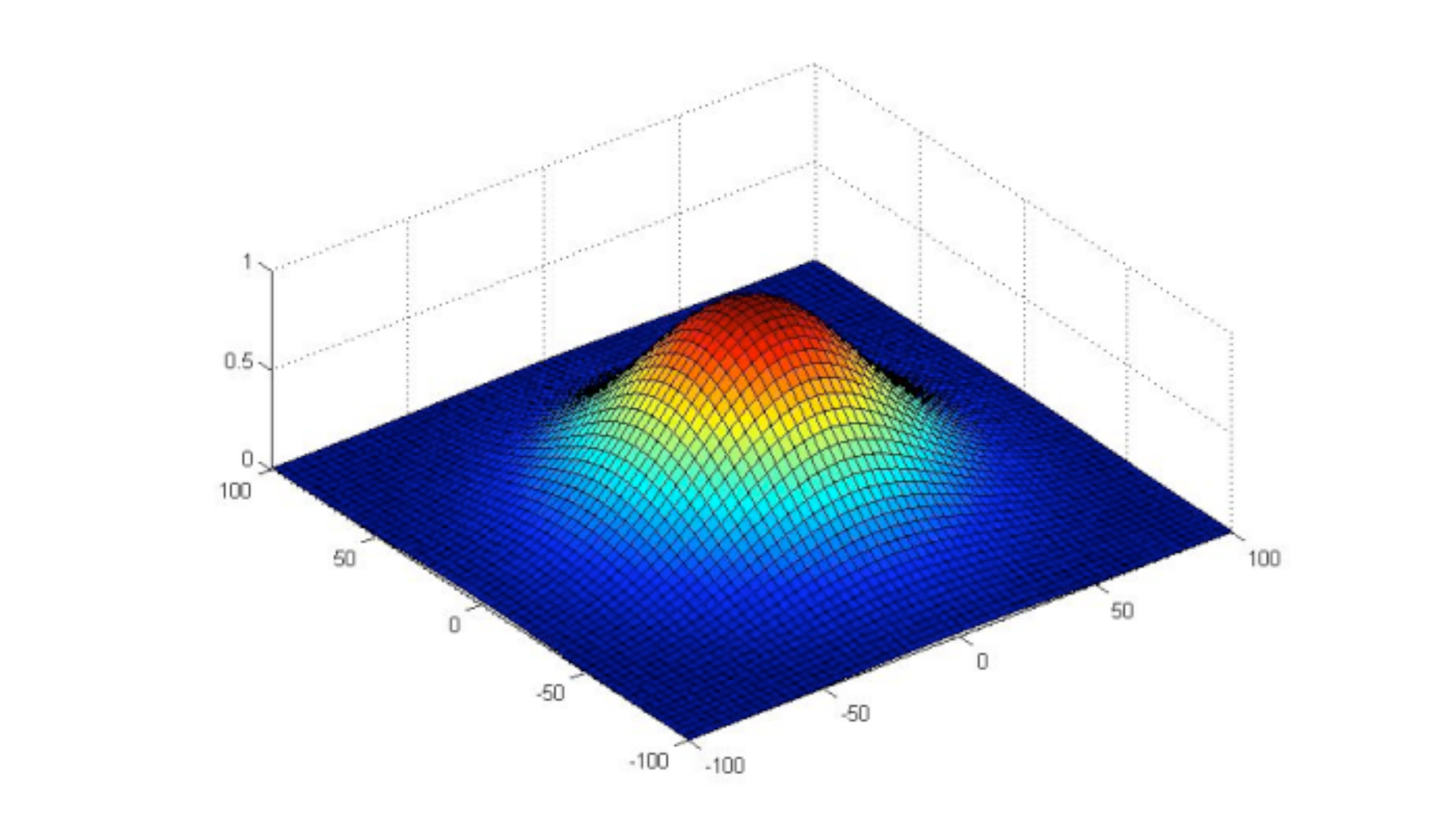}
    \caption{(Color online) A smooth curved bump  in the graphene sheet (taken with permission from Ref. \cite{JCV07}). }
    \label{gaussian}
\end{center}
\end{figure}

Once the metric of the curved space is known there is a standard
procedure to get the geometric factors that enter into the Dirac
equation. In the modelling of the graphene ripples, the metric can
be treated as a smooth perturbation of the flat surface and
physical results are obtained by a kind of perturbation theory.
Very often, the final result can be casted in the form of the flat
Dirac problem in the presence of a potential induced by the
curvature. In ref. \cite{JCV07} a smooth general metric was
considered of which an example is given by the gaussian shape of
Fig. \ref{gaussian}.
A simple way to derive the metric of a smooth shape (spherically
symmetric for convenience) is as follows: Start by embedding a
two-dimensional surface with polar symmetry in three-dimensional
space (described in cylindrical coordinates). The surface is
defined by a function $z(r)$ giving the height with respect to the
flat surface $z=0$, and parametrized by the polar coordinates of its
projection onto the $z=0$ plane. To obtain the metric we compute

\begin{equation}
dz^{2}=\left(\frac{dz}{dr}\right)^{2}dr^{2}\equiv \alpha
f(r)dr^{2}, \label{surface}
\end{equation}

\noindent
and substitute for the line element:

\begin{equation}
ds^{2}=dr^{2}+r^{2}d\theta^{2}+dz^{2}=\left(1+\alpha
f(r)\right)dr^{2}+r^{2}d\theta^{2}. \label{generalmetric2}
\end{equation}

For the particular example of the gaussian bump given in Fig.
\ref{gaussian}: $z=A\exp(-r^{2}/b^{2})$, the line element reads:
\begin{equation}
ds^{2}=-\left(1+\alpha f\right)dr^{2}-r^{2}d\theta^{2},
\label{fullmetric}
\end{equation}
where $\alpha$ is the ratio of the height to the mean width of the
gaussian that can be used as a perturbative parameter over the
flat case. The metric (\ref{fullmetric}) can be written in a more
usual form:
\begin{equation}
g_{\mu\nu}=
\left(%
\begin{array}{cc}

   -(1+\alpha f(r)) & 0 \\
   0 & -r^2 \\
\end{array}%
\right). \label{metricmatr}
\end{equation}
Comparing  the flat Hamiltonian written in polar coordinates:

\begin{equation}
H_{flat}=\hbar v_F
\left(%
\begin{array}{cc}
  0 & \partial_r+i\frac{\partial_\theta}{r}+\frac{1}{2r} \\
  \partial_r-i\frac{\partial_\theta}{r}+\frac{1}{2r} & 0  \\
\end{array}%
\right), \label{Hflat}
\end{equation}
with the curved Hamiltonian
\begin{equation}
H_{curved}=\hbar v_F
\left(%
\begin{array}{cc}
  0 & (1+\alpha f(r))^{-1/2}\partial_r+i\frac{\partial_\theta}{r}+A_\theta \\
  (1+\alpha f(r))^{-1/2}\partial_r-i\frac{\partial_\theta}{r}+A_\theta & 0  \\
\end{array}
\right), \label{Hcurved}
\end{equation}
it can be seen that  the curved bump induces an effective Fermi
velocity $\tilde{v}_{r}$ in the radial direction given by
\begin{equation}
\tilde{v}_r(r,\theta)=v_F(1+\alpha f(r))^{-1/2},
\end{equation}
and an effective magnetic field perpendicular to the graphene
sheet given by
\begin{equation}
B_z=-\frac{1}{r}\partial_r(rA_\theta)=\frac{1}{4r}\frac{\alpha
f'}{(1+\alpha f)^{3/2}}. \label{B}
\end{equation}
The magnitude of this effective magnetic field is estimated to be
of the order of 0.5 to 2-3 Tesla in the region spanned by the
bump, compatible with the estimations given in \cite{Metal06}, and
it  plays the same  role  in the issue of the weak localization of
graphene  as the effective magnetic  fields discussed there and in
\cite{MG06}.

To first order in the perturbative parameter $\alpha$ the
effective potential that appears in the Dirac equation in polar
coordinates is
\begin{equation}\label{potential1}
V(r,\theta)=i\Gamma(\theta)\left[\frac{1}{2}\alpha
f(r)\right](\frac{1}{2r}-\partial_{r}).
\end{equation}
The physical properties of the system are obtained from the Green's function in the curved space as explained in Section \ref{inhom}.

The metric approach was used to describe topological defects in
\cite{CV07a}. The proposed metric was as a generalization to  the
one used to model cosmic strings \cite{Vilenkin}:
\begin{equation}
ds^{2}=-dt^{2}+e^{-2\Lambda(x,y)}(dx^{2}+dy^{2}),\label{genmetric}
\end{equation}
where $$\Lambda(\textbf{r})=\sum^{N}_{i=1}4\mu_{i}\log(r_{i}),$$
$\mu_i$ is related to the defect angle of the disclination. In the
case of having heptagonal defects with an ``excess'' angle, the
sign of $\mu_i$ is negative. Here
$$r_{i}=[(x-a_{i})^{2}+(y-b_{i})^{2}]^{1/2},$$
where $(a_i, b_i)$ are the positions of the defects.  The curved
space formalism applied to this geometry  induces an effective
potential:
\begin{equation}
V(\omega,\textbf{r})=2i\Lambda\gamma^{0}\partial_{0}+i\Lambda\gamma^{j}
\partial_{j}+\frac{i}{2}\gamma^{j}(\partial_{j}\Lambda),\label{effV}
\end{equation}
which gives rise to spacial inhomogeneities in the local density
of states as discussed in   Section \ref{sec_obs}. The fictitious
gauge field coming from the spin connection is similar to the one
encountered in elasticity formalisms. A distinctive feature of the
covariant approach is the space dependent coefficient in the
kinetic term of the effective Hamiltonian that can be interpreted
as a space-dependent Fermi velocity that could eventually be used
to test the validity of the model. The influence of this extra
term on the minimal conductivity has been analyzed in \cite{CV09}.

\subsection{The electronic spectrum of fullerenes}
\label{sec_fullerenes} The ideas of modelling conical defects by
gauge fields and adding geometric curvature were established in
the early publication \cite{GGV93} where they were applied to
compute the electronic spectrum of the spherical fullerenes. The
most popular of them, the ``buckyball'' $C_{60}$ has the shape of
a soccer ball, consisting of 12 pentagons and 20 hexagons.  The
structure is that of a truncated icosahedron. There is a full set
of fullerenes of the same symmetry consisting of adding crowns of
hexagons around the pentagons. Following the analysis described
before,  modelling the $C_{60}$ molecule can be done with 12
non-Abelian vortices that would  account for both the singular
curvature, and the exchange of the Fermi points associated to the
pentagons. Given the compact nature of the system, the model
proposed in \cite{GGV93} was to solve the Dirac equation on the
surface of a sphere with a fictitious magnetic monopole at its
center. The sphere smears the  curvature accumulated at each
pentagon, and the charge $g$ of the fictitious magnetic monopole
is adjusted by adding up the individual fluxes of all the lines:
\begin{equation}
g=\frac{1}{4\pi}\sum_{i=1}^N\frac{\pi}{2}=\frac{N}{8},
\end{equation}
$N$ being the number of conical singularities on the surface. It
is interesting to note that the value of $g$ required for the
icosahedron ($N=12$), $g= 3/2$, is compatible with the standard
quantization condition of the monopole charge.

The given model  reproduces the low-energy spectra and correct
numbers of zero modes for the $C_{60}$ family of fullerenes and
works better in the limit of large number of points. The spectrum
is obtained by solving the eigenvalue problem for the covariant
Dirac operator
\begin{equation}
i\sigma^ae_a^\mu(\nabla_\mu-iA_\mu)\Phi_n=\epsilon_n\Phi_n, \;\;\;
a, \mu=1,2,
\end{equation}
where $e_\mu^a$ is the zweibein for the sphere, and the
geometrical factors in spherical coordinates are
$$\nabla_\theta=\partial_\theta,$$
$$\nabla_\phi=\partial_\phi-\frac{1}{4}[\sigma^1\sigma^2]\cos\theta,$$
$$A_\theta=0,$$
$$A_\phi=g\cos\theta\tau^{(2)}.$$
The Dirac equation becomes
\begin{align}
\frac{v_F}{R} \left[ i \partial_\theta - \frac{1}{\sin ( \theta )} \partial_\phi + \frac{i ( 1 + g )}{2 \cos ( \theta )}{2 \sin ( \theta )} \right] \Psi_1 ( \theta , \phi ) &= E \Psi_2 ( \theta , \phi ) \nonumber \\
\frac{v_F}{R} \left[ i \partial_\theta + \frac{1}{\sin ( \theta )} \partial_\phi + \frac{i ( 1 - g )}{2 \cos ( \theta )}{2 \sin ( \theta )} \right] \Psi_2 ( \theta , \phi ) &= E \Psi_1 ( \theta , \phi )
\end{align}
where $R$ is the radius of the sphere.

The Dirac equation can be diagonalized by introducing a
generalized angular momentum operator ${\bf J}$ taking into
account spinor indices, magnetic field and curvature. We find
\begin{align}
J^+ &= e^{i \phi} \partial_\theta + i e^{i \phi} \frac{\cos ( \theta )}{\sin ( \theta )} \partial_\phi + e^{i \phi} \frac{g-1}{2 \sin ( \theta )} \nonumber \\
J^- &= - e^{-i \phi} \partial_\theta + i e^{- i \phi} \frac{\cos ( \theta )}{\sin ( \theta )} \partial_\phi + e^{- i \phi} \frac{g-1}{2 \sin ( \theta )} \nonumber \\
J_z &= - i \partial_\phi
\end{align}
By squaring the Dirac operator each of
the spinor components obeys the equation
\begin{equation}
E_J^2 = \left( \frac{v_F}{R} \right)^2 \left( J ( J + 1 ) +\frac{1}{4}-g^2\right)
\label{jequation}
\end{equation}
where $J$ is the eigenvalue of the angular momentum, which is an integer. The degeneracy of a state is $2J + 1$. From Eq. (\ref{jequation}) it can be
seen that there is minimum value of the angular momentum $J$
dictated by $g$, so that the number of zero modes in the spectrum
depends exclusively on the value of the monopole charge, i. e. on
the number of pentagons in the lattice. It is also interesting to
note that, since the sphere has a constant curvature, there are no
inhomogeneous geometrical factors and all the effect of the
curvature reduces to a rescaling of the energy. Substituting
$g=3/2$ in Eq. (\ref{jequation}) gives the spectrum of the
$C_{60}$ family of fullerenes: the ground state is a couple of
triplets lying at zero energy, and the first excited levels follow
the degeneracy of the sphere up to the point in which the highest
dimension of an irreducible representation of the icosahedron
symmetry group is reached. The model works very good given its
simplicity. The elastic properties of the fullerenes were analyzed
within this approach in \cite{GGV92}.

\subsection{Dislocations and torsion: a general relativity
approach} \label{torsion}

The gravity connection can be pushed forward and used to model a
density of dislocations in the space in the continuum limit by
adding torsion to the space spanned by the lattice. The connection
of torsion with the continuum theory of crystal dislocations goes
back to Kondo \cite{K52} and has been formalized in
\cite{KV92,K89}. Various aspects of the problem have been explored
in \cite{FM99}. A nice review on the relation of gravity with
topological defects in solids is \cite{M00}. The geometric
approach to defects in solids \cite{K89,SN88} relates the metric
of the curved crystalline surface with the deformation tensor and
establishes that disclinations are associated to the curvature
tensor and a finite density of dislocations generates a torsion
term.

Dislocations in graphene are made of pentagon--heptagon pairs and
they have been widely studied in connection with the properties of
carbon nanotubes \cite{SDD98} and, more recently, in the flat
graphene sheets or ribbons \cite{Cetal08,LJV09}. The
covariant approach described above was extended in \cite{JCV10} to
include a connection with torsion. As we discussed in Section
\ref{sec_gauge} the minimal coupling of any geometrical or
physical fields to the Dirac spinors adopts always the form of a
covariant derivative:
\begin{equation}
\partial_\mu \Rightarrow D_\mu=\partial_\mu + A_\mu ,
\end{equation}
where the given vector can be an electromagnetic potential induced
by a real electromagnetic field or any other real or fictitious
gauge field associated to deformations or to geometrical factors.
In the case of having a density of dislocations in the graphene
sheet modelled by torsion we can construct  two potential
candidates to gauge fields:
\begin{equation}
V_\mu=g^{\nu\rho}T_{\nu\rho\mu}\;\;,\;\;
S_\mu=\epsilon_{\mu\nu\rho\sigma}T_{\nu\rho\sigma},
\label{vectors}
\end{equation}
where $T_{\mu\nu\rho}$ is the rank three torsion tensor related to
the  the antisymmetric part of the connection:
\begin{equation}
T^{\lambda}_{\mu\nu}=\Gamma ^{\lambda}_{\mu\nu}-\Gamma
^{\lambda}_{\nu\mu}. \label{torsiontensor}
\end{equation}

The field $V_\mu$ is an ordinary vector and can be associated to the
density of edge dislocations while $S_\mu$ is an axial (pseudo)
vector associated to the density of screw dislocations.

\section{Strain fields}
\label{sec_elastic}

\subsection{Modulations of the tight binding parameters}
\label{sec_elasticgauge}

In this section we will consider gauge fields which arise in
defect-free graphene due to smooth elastic deformations. The
latter can be created by applying external stress to graphene
flakes. It is important to notice that these deformations and the
related gauge fields are also intrinsic for graphene at finite
temperatures due to thermal fluctuations (see  Section
\ref{thermal}). These gauge fields were discussed in a context of
electron-phonon interactions in nanotubes  prior to the synthesis
of graphene in \cite{SA02b,SKS05}.

Let us start with a tight-binding model of the electronic structure
of graphene in the nearest-neighbor approximation. In the deformed
graphene all bonds are, in general, nonequivalent and the three
nearest-neighbors hopping parameters $t_i$ can be all different. If
one repeats the derivation of the Dirac Hamiltonian in the
effective mass approximation but with nonequal hopping parameters
one finds for the electron states in the vicinity of the $K$ point
the effective Hamiltonian:
\begin{equation}
H=-i\hbar v_F \vec{\sigma}\left(\vec{\nabla} -i\vec{ A}%
\right) ,  \label{df1}
\end{equation}
where \cite{SA02b,SKS05,KN07}
\begin{eqnarray}
A_x &=&\frac{\sqrt{3}}2\left(t_3-t_2\right) , \nonumber
\\
A_y &=&\frac 12\left( t_2+t_3-2t_1\right), \label{df2}
\end{eqnarray}
(note that, following Ref. \cite{SA02b} and  most of the subsequent
works, we use a choice of the axes $x,y$ different from that made in Refs.
\cite{SKS05,KN07}).  In the weakly deformed lattice, assuming that the atomic displacements $%
\vec{u}$ are small in comparison with the lattice
constant $a$ one has
\begin{equation}
t_i=t+\frac{\beta t}{a^2}\vec{\rho }%
_i\left( \vec{u}_i-\vec{u}_0\right) ,  \label{df3}
\end{equation}
where  $\vec{\rho }_i$ are the nearest-neighbor vectors,  $%
\vec{u}_0$ is the displacement vector for the central atom, and
\begin{equation}
\beta =-\frac{\partial \ln t}{\partial \ln a}\simeq 2 \label{df4}
\end{equation}
is the electron Gr\"uneisen parameter (for more details, see
below). The continuum limit (elasticity theory)  with a
displacement field $\vec{u}\left( \vec{r}\right) $ is performed by making the substitution
\begin{equation}
\vec{u}_i-\vec{u}_0\propto \left( \vec{\rho }_i\nabla \right)
\vec{u}\left( \vec{r}\right),
\end{equation}
from where we obtain the effective gauge field  \cite{SA02b,M07}:
\begin{eqnarray}
A_x &=&c\frac{\beta t}a\left( u_{xx}-u_{yy}\right) ,  \nonumber
\\
A_y &=&-c\frac{2\beta t}au_{xy},  \label{df5}
\end{eqnarray}
where $c$ is a numerical factor depending on the detailed model of
chemical bonding; in what follows we will make our estimations with the value $c=1$.

For the other valley, $K^{\prime }$, the sign of the vector
potential (\ref {df5}) is opposite, in agreement with the requirement
of time-reversal invariance. Indeed, atomic displacements can
break effectively the time-reversal symmetry (involving inversion
of the wave vector) for a given valley but not for the crystal in
general.

If the deformation is not a pure shear and the  dilatation does
not obey the condition $\nabla \vec{u}=u_{xx}+u_{yy}=0,$ a scalar
potential proportional to $\nabla \vec{u}$ also arises
\cite{SA02b}. The gauge field (\ref{df5}) is proportional to the
deformation tensor which is directly involved in the density of
elastic energy (see the next Section). This means that the problem
as a whole is {\it not} gauge invariant. For example we could add
to the gauge potential the gradient of a scalar function and shift
the electronic wave function by a phase as discussed in Section
\ref{sec_gaugegeneral} leaving the electronic part of the
Lagrangian invariant but  these transformations will change the
elastic part. The simplest argument to understand this lack of
gauge invariance is to notice that the kinetic energy term for the
elastic part is isotropic (see Eq. (\ref{enner})) while the
effective vector field given by Eq. (\ref{df5}) is not.

On general symmetry grounds, strains can also lead to  scalar potential\cite{SA02b,M07},
\begin{align}
V ( \vec{\bf r} ) &= g ( u_{xx} + u_{yy} )
\end{align}
This term enters in the diagonal elements of the Dirac equation. The most recent calculations suggest that $g \approx 4$eV for
a neutral graphene single layer\cite{CJS10}, although higher values, $g \approx 20$eV have been cited in earlier literature\cite{OS66,S83,SA02b}. The parameters $\beta$ and $g$ describe the  coupling to acoustical in graphene, and they are usually considered simultaneously,  as they both contribute to the deformation potential, $D$. Combinations of the type  $D^2 =  c_1 \times ( v_F^2 \beta^2 ) / a^2 + c_2  \times g^2 $, where $c_1$ and $c_2$ are numerical constants determine the phonon contribution to transport coefficients, like the resistivity. Microscopic models for the deformation potential have considered only the effect of the gauge potential\cite{PSZ80,G81,WM00,YAHL99}, or both the scalar and gauge potentials\cite{VCKAC01}.

\subsection{Gauge fields as function of the in plane strains}
Long wavelength distortions within the graphene layers can be
described using the strain tensor\cite{LL70}, $u_{\alpha \beta}$.
This tensor defines the local deformation of the lattice. Given the
deformation field $\vec{u} ( x , y )$, the strain tensor in the
linear approximation is:
\begin{align}
u_{\alpha \beta} &= \frac{\partial_{\alpha} u_{\beta} +
\partial_{\beta} u_{\alpha}}{2}.
\end{align}
The elastic energy is:
\begin{align}
E_{elas} &= \frac{\lambda}{2} \int d^2 \vec{r} \left( u_{xx} +
u_{yy} \right)^2 + \mu \int d^2 \vec{r} \left( u_{xx}^2 + u_{yy}^2
+ 2 u_{xy}^2 \right), \label{enner}
\end{align}
where $\lambda$ and $\mu$ are the elastic Lam\'e coefficients. We
can also define the stress tensor, $\sigma_{\alpha \beta} =
\partial E_{elas} / \partial u_{\alpha \beta}$,
\begin{align}
\sigma_{xx} &= \lambda \left( u_{xx} + u_{yy} \right) + 2 \mu u_{xx}
\nonumber \\
\sigma_{yy} &= \lambda \left( u_{xx} + u_{yy} \right) + 2 \mu u_{yy}
\nonumber \\
\sigma_{xy} &= 2 \mu u_{xy}.
\end{align}

As discussed in Section \ref{sec_gauge}, strains modify the bond
lengths and the hopping between $\pi$ orbitals. The dependence of
the hopping on the interatomic distance is approximately the same
in the organic compounds with $sp^2$ coordination, and it is
parametrized by the dimensionless constant $\beta$ (\ref{df4}).
A change in the hopping between nearest neighbor atoms modifies
the terms in the Hamiltonian which couple the two sublattices. In
the previous section this was done explicitly, in the
nearest-neighbor approximation. Actually, the expression
(\ref{df5}) is more general. In the continuum limit, this coupling
must take the form of a gauge field  with  the matrix structure of the Pauli
matrices, $\sigma_1$ and $\sigma_2$. A vector can be constructed
from the strain tensor by contracting it with a rank three tensor.
This tensor should be invariant under the discrete set of
symmetries of the honeycomb lattice. The only possible tensor with
these properties has the non zero components:
\begin{align}
K_{xxx} &= 1 \nonumber \\
K_{yyx} &= K_{yxy} = K_{xyy} = - 1 ,\label{tensor3}
\end{align}
where the $x$ axis coincides with one of the unit vectors of the
lattice. The same tensor describes the first deviation from complete
isotropy in the bands of the electrons in the honeycomb lattice, the
so called trigonal warping \cite{RMP08}.

From Eq.~(\ref{tensor3}) t can be seen that the gauge field must take the form
(\ref{df5}).

The combination of strains which enter in the gauge field,
Eq.~(\ref{df5}), can also be written in terms of the stress
tensor, $\sigma_{ij}$:
\begin{align}
u_{xx}-u_{yy} &= \frac{\sigma_{xx} - \sigma_{yy}}{2 \mu} \nonumber
\\
2 u_{xy} &= \frac{2 \sigma_{xy}}{2 \mu},
\end{align}
where $\mu$ is a Lam\'e coefficient (shear modulus).

The calculation of the strains in a two dimensional system with
only in-plane displacements is drastically simplified in
comparison with a generic three dimensional case, as the stresses
can be written in terms of a function \cite{LL70}:
\begin{align}
f ( x , y ) &= {\rm Re} \left[ ( x - i y ) g ( x + i y ) + h ( x +
i y ) \right],
\end{align}
where $g(z)$ and $h(z)$ are analytic functions of $z=x+iy$.  The
stress tensor is:
\begin{align}
\sigma_{xx} &= \frac{\partial^2 f}{\partial y^2} \nonumber \\
\sigma_{yy} &= \frac{\partial^2 f}{\partial x^2} \nonumber \\
\sigma_{xy} &= - \frac{\partial^2 f}{\partial x \partial y}.
\label{conformal}
\end{align}

A pure shear deformation has $\sigma_{xx} = - \sigma_{yy}$, which
implies that $g(z)=0$. Then we get the effective gauge potential:
\begin{align}
A_x &\propto {\rm Re} \left[ h'' ( z ) \right] \nonumber \\
A_y &\propto {\rm Im} \left[ h'' ( z ) \right],
\end{align}
and the effective magnetic field is:
\begin{equation} B =-\partial_y
A_x + \partial_x A_y \propto {\rm Im} \left[ h''' ( z ) \right].
\end{equation}
A pure shear deformation that leads to a constant effective
magnetic field is given by $h(z)=A z^3$, where $A$ is a constant.
A general deformation which satisfies the equilibrium equations of
elasticity and which leads to a constant effective gauge field is
determined by the function $f ( z ) = A z^3 + B \bar{z} z^2$.

A constant effective magnetic field requires a strain tensor which
increases linearly with position, $u_{ij} = \bar{u} ( \vec{r} \,
\vec{n} ) / L$, where $\vec{n}$ is a given unit vector, and $L$ is
the size of the system. The maximum stresses are of order
$\bar{u}$. The effective magnetic field has an associated magnetic
length:
\begin{equation}
\frac{1}{l_B^2} \approx \beta \frac{\bar{u}}{a L},
\end{equation}
where $a$ is the lattice constant. For $\bar{u} = 0.01$ and
$L=1\mu$m we obtain $l_B \approx 10^2$nm, which corresponds to a
magnetic field of 0.5T. A description of the strains and effective
fields induced by displacements with radial symmetry is given in
the Appendix \ref{sec_radial}.

\subsection{Out of plane displacements}

\begin{figure}[!t]
\begin{center}
\includegraphics[height=4cm]{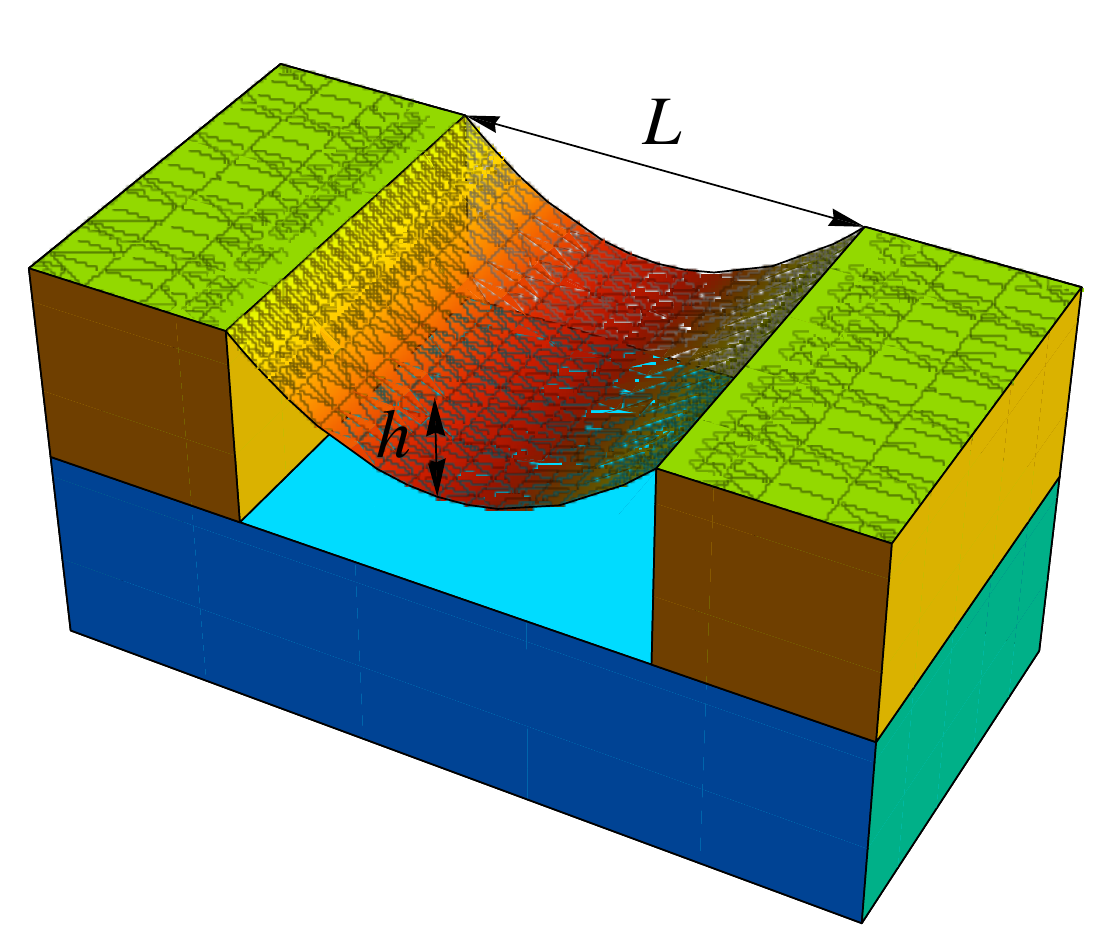}
\caption[fig]{Sketch of a suspended rectangular flake.}
\label{sketch}
\end{center}
\end{figure}

When the layer can be distorted along the out of plane direction,
the strain tensor becomes:
\begin{align}
u_{\alpha \beta} &= \frac{\partial_{\alpha} u_{\beta} +
\partial_{\beta} u_{\alpha}}{2} + \frac{\partial_{\alpha} h \partial_{\beta} h}{2},
\label{deffr}
\end{align}
where $h$ is the displacement along the third dimension. The
distortion of the bonds is proportional to the square of $h$.

The simplest and most ubiquitous deformation expected in suspended
graphene is due to the force induced by the electric field between
the gate and the flake \cite{FGK08}. The total  energy of a
monolayer flake under a constant vertical force, as function of the
maximum vertical deformation, $h$, can be written as \cite{LL70}:
\begin{equation}
E_{tot} = E_{bend} + E_{elas} + E_{field} \approx c_{bend} \kappa
\frac{h^2}{L^2} + \left( c_\lambda \lambda + c_\mu \mu \right)
\frac{h^4}{L^2} + c_{field} {\cal F} h L^2, \label{energy}
\end{equation}
where $c_{bend} , c_\lambda , c_\mu$ and $c_{field}$ are numerical
constants, $\kappa \approx 1$ eV is the bending rigidity of the
flake, $\lambda \approx 2.4$ eV \AA$^{-2}$ and $\mu \approx 9.9$
eV \AA$^{-2}$ are the in plane elastic Lam\'e coefficients
\cite{ZKF09}, and ${\cal F} = 2 \pi e^2 n^2$ is applied field, and
$n$ is the carrier density that it induces. The bending energy in
Eq.~(\ref{energy}) is negligible with respect to the in plane
elastic energy for $h \gtrsim \sqrt{\kappa / {\rm max} ( \lambda ,
\mu )} \approx 1-3$\AA. Neglecting it, we obtain:
\begin{equation}
h \approx C L \left(\frac{e^2 n^2 L}{{\rm max} ( \lambda , \mu ) }
\right)^{1/3}, \label{height}
\end{equation}
where $C$ is a numerical constant which depends on details of the
shape of the flake, and the ratio $\lambda / \mu$. The neglect of
the bending energy is justified for $n^2 L^4 \gtrsim
\sqrt{\kappa^3 / [ e^4 {\rm max} ( \lambda , \mu ) ]} \approx
0.1$. As $N = n L^2$ is the total number of carriers in the flake,
the bending energy can be neglected in all realistic situations.
For $n \sim 10^{12}$cm$^{-2}$ and $L \sim 1 \mu$m, we obtain $h
\sim 30$nm.

The strains associated to the deformation in Eq.~(\ref{height})
are of order:
\begin{equation}
u_{\alpha \beta} \sim \left( \nabla h \right)^2 \sim C
\left(\frac{e^2 n^2 L}{{\rm max} ( \lambda , \mu ) } \right)^{2/3},
\label{strains}
\end{equation}
leading to an effective magnetic field characterized by a magnetic
length, $l_B$:
\begin{equation}
\frac{1}{l_B^2} \sim \frac{C} \beta {a L} \left(\frac{e^2 n^2
L}{{\rm max} ( \lambda , \mu ) } \right)^{2/3}, \label{field}
\end{equation}
where $a$ is the lattice constant. For $n \approx
10^{12}$cm$^{-2}$ and $L \approx 1\mu$m, we obtain $l_B \sim
30-100$nm, which corresponds to effective fields of $2-10$T.

The effective magnetic field is determined by the full strain
tensor, which includes a contribution from the in plane
displacements \cite{GHL08}. If the height profile, $h ( \vec{r})
$, is known, we can define:
\begin{equation}
f_{i,j} ( \vec{r} ) = \frac{\partial h}{\partial r_i}
\frac{\partial h}{\partial r_j}.
\end{equation}
Then, the Fourier component of the effective field, including the
effect of the in plane strains can be written as:
\begin{equation}
B ( \vec{k} ) \propto i k_y \frac{ ( 3 k_x^2 - k_y^2 ) ( \lambda +
\mu )}{(\lambda + 2 \mu ) | \vec{k} |^4} \left[ k_y^2 f_{xx} (
\vec{k} ) + k_x^2 f_{yy} ( \vec{k} ) - 2 k_x k_y f_{xy} ( \vec{k}
) \right],
\end{equation}
where $\vec{k}$ is the wave vector. These corrections do not
change the earlier order of magnitude estimates for the effective
field, except in cases with high symmetry. For instance, when $h$
is constant along one direction, i. e. $y$, the relaxation of the
in plane displacements lead to a stress tensor such that
$\sigma_{xx} = {\rm constant} , \sigma_{yy} = \sigma_{xy} = 0$,
and to a constant effective gauge field.

\begin{figure}[!t]
\begin{center}
\includegraphics[height=4cm]{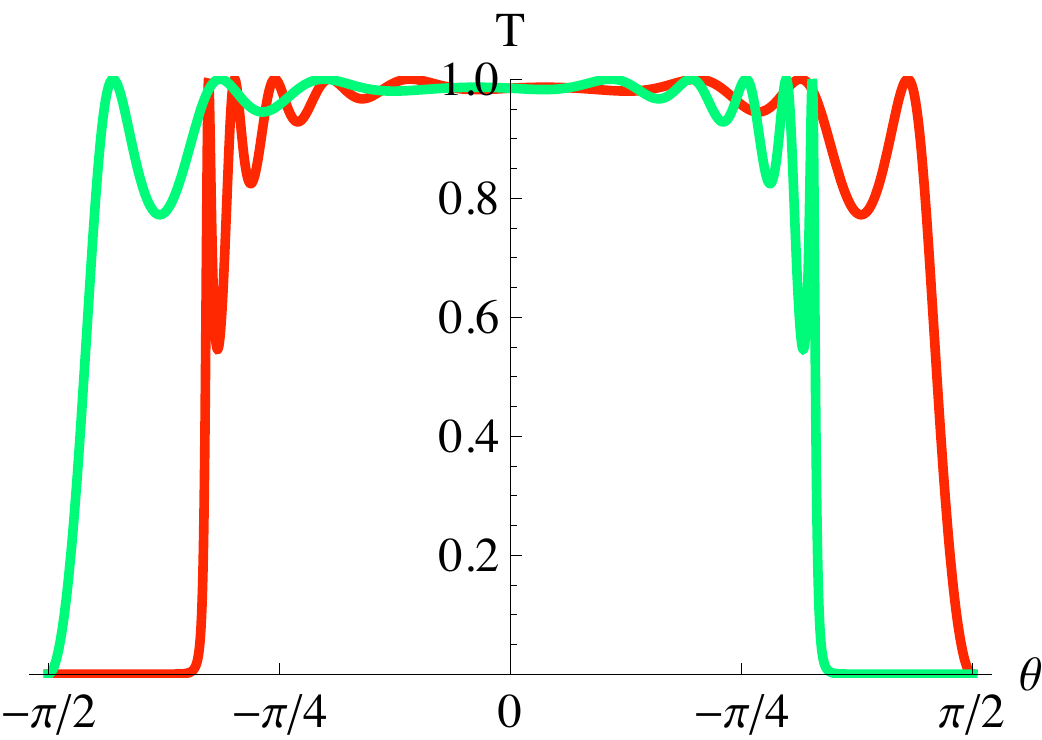}
\caption[fig]{Transmission through a suspended rectangular flake
as function of incident angle. The carrier density is $\rho =
10^{12} {\rm cm}^{-2}$, the length of the flake is $L=100$nm, and
the maximum deformation is $h_0 = 3$nm. Red and green lines
correspond to the two valleys in the Brillouin Zone (see text for
details).} \label{transmission_angle}
\end{center}
\end{figure}

\subsection{Thermal fluctuations and intrinsic ripples in graphene}
\label{thermal}  From the general analysis of the previous sections it follows that
corrugations of graphene with spatial scale much larger than the
interatomic distance $a$ lead to the formation of an Abelian gauge
field. It turns out that for two-dimensional crystals at finite
temperatures the corrugations are unavoidable, due to the instability of the crystal
with respect to the bending fluctuations. In this section we will
review the corresponding theoretical results.

The standard theory of lattice dynamics and thermodynamics
\cite{BH98} starts with the assumption that typical amplitudes of
atomic displacements $u$ due to thermal fluctuations are small in
comparison with the interatomic distances, $\left\langle
u^2\right\rangle \ll a^2$. It was understood as early as in 1930th
\cite{P34,P35,L37,LL80} that for two-dimensional crystals at
finite temperatures this cannot be the case, due to the
logarithmic divergence of $\left\langle u^2\right\rangle$. This
has given rise to the hypothesis that long-range crystal order
cannot exist in two dimensions which has been proven later, as a
particular case of Mermin-Wagner theorem \cite{M68}. Therefore the
experimental realization of graphene and other truly
two-dimensional crystals \cite{Netal04,Netal05a} was really
surprising. The point is that these objects are not
two-dimensional crystals in two-dimensional space but
two-dimensional crystals in three-dimensional space; it follows
from a general theory of flexible membranes \cite{N89} that such
systems can exist but cannot be really flat \cite{K07,Metal07}.
Very soon after the synthesis of graphene  corrugations of freely
suspended graphene membrane were observed experimentally
\cite{Metal07}; the existence of intrinsic ripples in graphene due
to thermal instability has also been confirmed  by atomistic Monte
Carlo simulations \cite{FJK07}.
\begin{figure}
  \begin{center}
    \includegraphics[height=4cm]{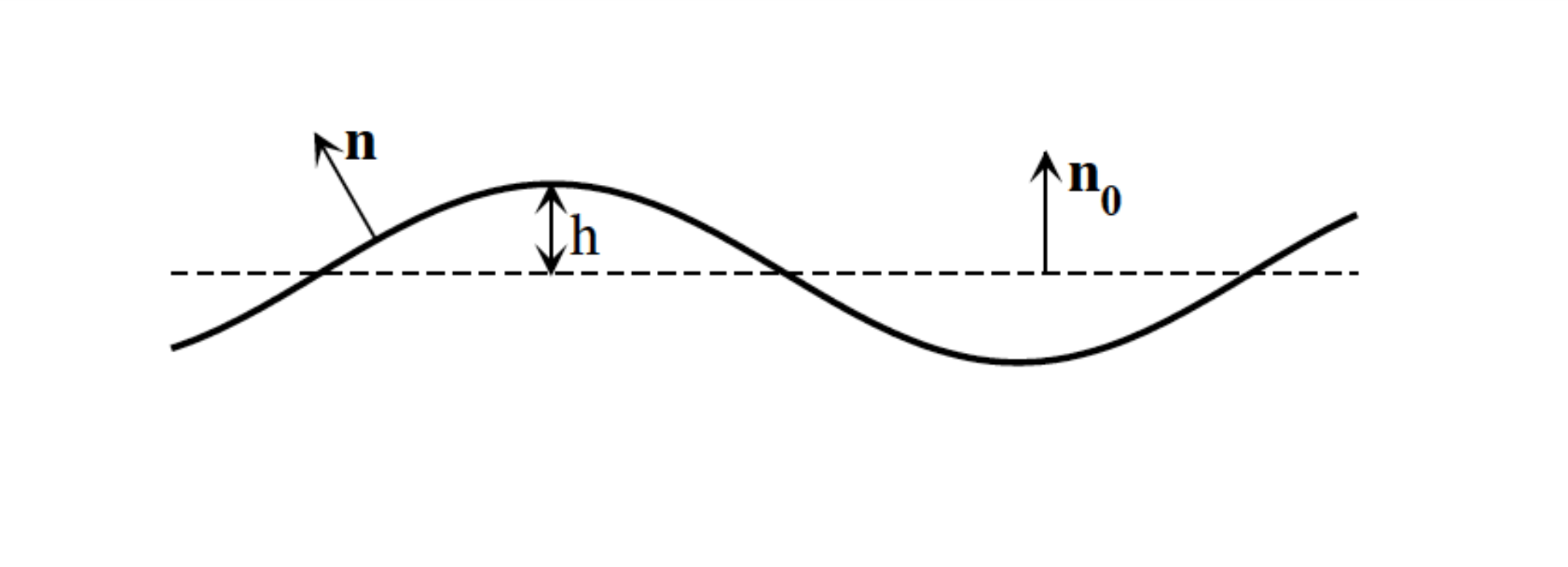}
    \caption{Sketch of a flexible membrane (solid line). $h$ is the out of plane deviation with respect to the z = 0 plane (dashed line) defined by the center of mass.
The unit vector {\bf n} and ${\bf n_0}$ are the normals to each
point in the membrane and in the reference plane respectively.}
\label{membrr}
\end{center}
\end{figure}

Let us start with the model of a continuum infinitely thin membrane
(Fig. \ref{membrr}). It is characterized by the dependence of
out-of-plane deformation on coordinates, $h\left( x,y\right) $.
The normal vector to the membrane at a given point has components
\cite{N89}
\begin{equation}
\mathbf{n}\left( x,y\right) =\frac{\left( -\frac{\partial h}{\partial x},-%
\frac{\partial h}{\partial y},1\right) }{\sqrt{1+\left| \nabla
h\right| ^2}}. \label{normal}
\end{equation}
For liquid membranes, according to the Helfrich model \cite{H73,S94}
the density of elastic energy contains two terms, proportional to
the square of mean curvature $H^2$ and to the Gaussian curvature
$K$ but the second one gives just a constant contribution to the
total energy (due to Gauss-Bonnet theorem) and is irrelevant for
our analysis. In the lowest-order (harmonic) approximation
$H\propto \nabla ^2h$ and, thus, the bending energy has the form
\begin{equation}
\mathcal{H}_b=\frac \kappa 2\int d^2\mathbf{r}\left( \nabla
^2h\right) ^2 ,\label{Eben}
\end{equation}
In the harmonic approximation, the correlation function of the
Fourier components of $h\left( \mathbf{r}\right) $, $\left\langle
\left| h_{\mathbf{q}}\right| ^2\right\rangle $ is calculated
immediately:
\begin{equation}
\left\langle \left| h_{\mathbf{q}}\right| ^2\right\rangle =\frac
T{\kappa q^4}.  \label{corr110}
\end{equation}
Thus, the mean square amplitude of out-of-plane displacement is
\begin{equation}
\left\langle h^2\right\rangle
=\sum\limits_{\mathbf{q}}\left\langle \left| h_{\mathbf{q}}\right|
^2\right\rangle \simeq \frac T\kappa L^2, \label{corr111}
\end{equation}
where $L$ is the sample size and we have introduced a cutoff of
divergent integral in Eq. (\ref{corr111}) at $q_{\min }\simeq
1/L$. Thus, typical amplitude of out-of-plane fluctuations is
proportional to the sample size. The correlation function of the
Fourier components of the normal vectors,
\begin{equation}
G\left( q\right) =q^2\left\langle \left| h_{\mathbf{q}}\right|
^2\right\rangle =\frac T{\kappa q^2}  \label{corr112}
\end{equation}
is also singular at small wave vectors and, thus, the correlator $%
\left\langle \vec{n}\left( \vec{R}\right) \vec{n}\left( 0%
\right) \right\rangle $ behaves logarithmically at $R\rightarrow
\infty $ whereas for a flat membrane it should tend to a constant
(normals are more or less parallel at large distances). As a
result, the model with the Hamiltonian (\ref{Eben}) describes not
a flat but a \textit{crumpled} membrane.

The higher-order terms in the expansion of the mean curvature $H$
in $\nabla h$ renormalize the effective bending rigidity
\cite{PL85},
\begin{equation}
\kappa _{eff}\left( T\right) =\kappa -\frac{3T}{4\pi }\ln \frac La,
\label{bendeff1}
\end{equation}
thus, the membrane becomes less rigid as the temperature
increases.
\begin{figure}[!t]
\begin{center}
\includegraphics[height=4cm]{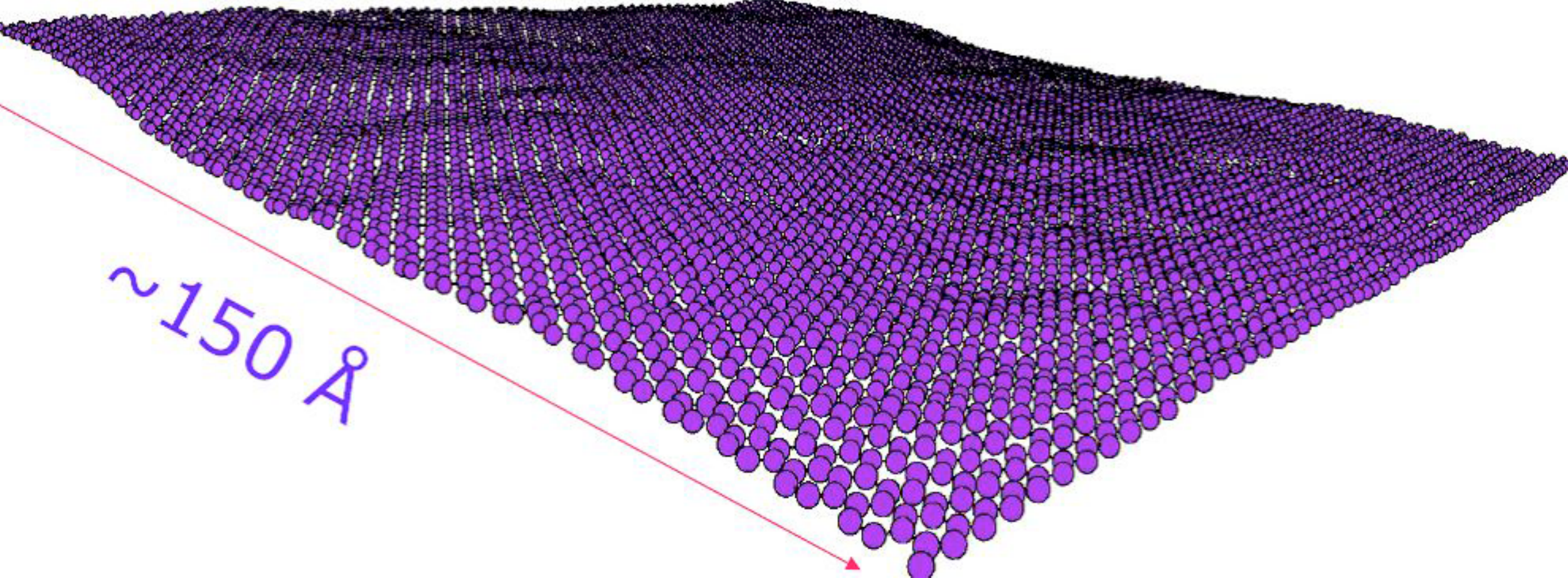}
\caption[fig]{A representative atomic configuration in Monte Carlo
simulations \cite{FJK07,LKYZF09} for graphene at room temperature
(courtesy A.Fasolino).}
\label{rip1}
\end{center}
\end{figure}

The situation is totally different for \textit{crystalline}
membranes where the coupling between out-of-plane (bending)
$h\left( \mathbf{r}\right) $ and in-plane $\mathbf{u}\left(
\mathbf{r}\right) $ atomic displacement becomes crucially
important \cite{NP87,N89}. It follows from the elasticity theory
\cite{LL70,TW59} that the total elastic energy in this case is
written as (cf. Eq. (\ref{enner}))
\begin{equation}
\mathcal{H}=\int d^2\mathbf{r}\left[ \frac \kappa 2\left( \nabla
^2h\right) ^2+\mu u_{\alpha \beta }^2+\frac \lambda 2 u_{\alpha
\alpha }^2\right]  \label{Ebencrys},
\end{equation}
$u_{\alpha \beta}$ is given by Eq. (\ref{deffr})
We keep there quadratic terms in the derivatives of out-of-plane
deformation but not in the derivatives of in-plane deformations
since the former are, at the average, much larger. The model
described by the Hamiltonian (\ref{Ebencrys}) is essentially
nonlinear. Since the Hamiltonian is quadratic in in-plane
components of the distortion tensor, $\frac{\partial u_\alpha
}{\partial x_\beta }$, the partition function can be integrated
over the latter and the effective Hamiltonian for out-of-plane
deformations only can be exactly derived \cite{N89,NP87,LDR92,XMLR03}:
\begin{equation}
\mathcal{H}=\int d^2\mathbf{r}\left[ \frac \kappa 2\left( \nabla
^2h\right)
^2+\frac{K_0}8\left( P_{\alpha \beta }^T\frac{\partial h}{\partial x_\alpha }%
\frac{\partial h}{\partial x_\beta }\right) ^2\right],
\label{Ebeneff}
\end{equation}
where $P_{\alpha \beta }^T=\delta _{\alpha \beta }-\left( \partial
^2/\partial x_\alpha \partial x_\beta \right) /\nabla ^2$ is the
operator separating transverse components and
\begin{equation}
K_0=\frac{4\mu \left( \mu +\lambda \right) }{2\mu +\lambda }
\label{modul1}
\end{equation}

The Hamiltonian (\ref{Ebeneff}) describes interactions of soft
(long-wavelength) fluctuations, similar to the
Ginzburg-Landau-Wilson Hamiltonian for the critical point
\cite{WK74,M76,BDFN92}. The difference is that two-dimensional
systems are ``critical'' at \textit{any} finite temperatures. In
analogy with the theory of critical phenomena, one can introduce
\cite{NP87,N89} scaling hypothesis for correlation functions and
the corresponding exponents $\eta ,\eta ^{\prime },\zeta $ related
with the behavior of effective bending rigidity, $\kappa
_{eff}\left( q\right) \propto q^{-\eta },$ elastic moduli, $\mu
_{eff}\left( q\right) ,\lambda _{eff}\left( q\right) \propto
q^{\eta ^{\prime }}$ and characteristic out-of-plane deformation,
$\sqrt{\left\langle h\right\rangle ^2}\propto
L^\zeta $. All these exponent can be expressed through the only exponent $%
\eta $:
\begin{equation}
\zeta =1-\eta /2,
\begin{array}{cc}
& \eta ^{\prime }=2\left( 1-\eta \right)
\end{array}
\label{exponn}
\end{equation}
The normal-normal correlation function, instead of
Eq.(\ref{corr112}), behaves as
\begin{equation}
G\left( q\right) \propto q^{-\left( 2-\eta \right) }
\label{corr113}
\end{equation}
at $q\rightarrow 0$, and the membrane is ``flat'' (in a sense that $%
\left\langle \vec{n}\left( \vec{R}\right) \vec{n}\left( 0%
\right) \right\rangle \rightarrow const$ at $R\rightarrow \infty $) for $%
0<\eta <1$. However, the amplitude of out-of-plane displacements
grows with the sample size which means that the membrane is
essentially corrugated, in
 contrast with the surface of three-dimensional crystals where $\sqrt{%
\left\langle h\right\rangle ^2}$ remains much smaller than the
interatomic distances. Note that the {\it in-plane} phonon modes
become even softer due to the coupling with the out-of-plane ones,
since $\mu _{eff}\left( q\right) ,\lambda _{eff}\left( q\right)
\rightarrow 0$ at $q \rightarrow 0$. In this sense, there is no real
crystalline order since there is no infinitely narrow
delta-functional Bragg peaks but, instead, some power-law
singularities at the reciprocal lattice points so the Mermin theorem
is, of course, not violated. This broadening of the Bragg peaks is
clearly seen both in experiments \cite{Metal07} and simulations
\cite{FLK07} for graphene.
\begin{figure}[!t]
\begin{center}
\includegraphics[height=4cm]{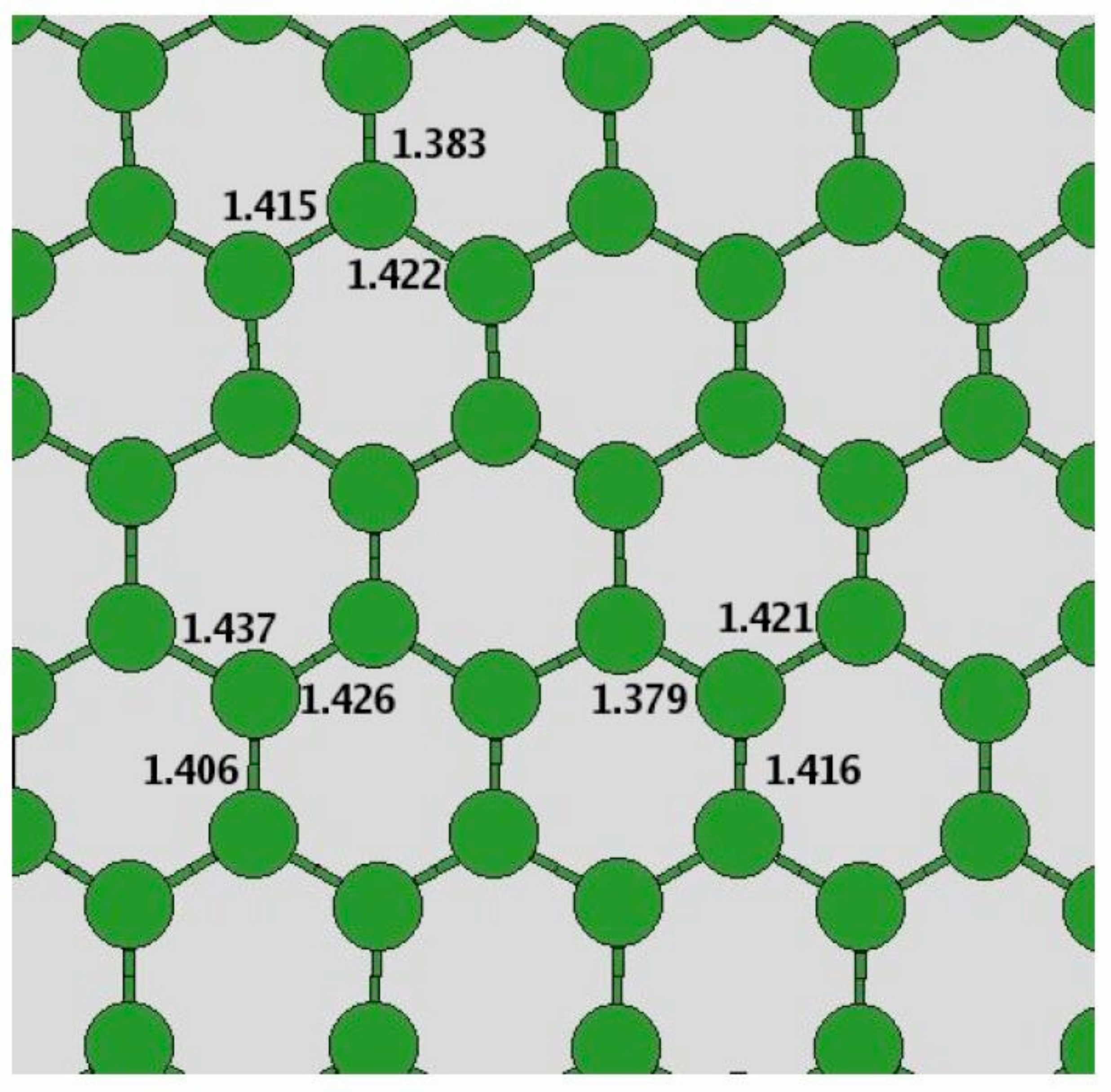}
\caption[fig]{Portion of one typical atomic configuration of graphene sample at
room temperature. The numbers indicate the bond length in \AA
(taken with permission from Ref. \cite{FJK07}).}
\label{rip2}
\end{center}
\end{figure}
The continuum model (\ref{Ebencrys}), which is called the model of
phantom membranes, has a transition to a crumpled phase at a
temperature of the order of $\kappa$. The term ``phantom'' means
that the model does not include self-avoidance, the natural
condition of true physical systems. It is assumed that
self-avoidance removes the phase transition to the high
temperature crumpled phase while the scaling properties of the
``flat'' phase remain the same as in phantom membranes
\cite{N89}.
\begin{figure}[!t]
\begin{center}
\includegraphics[width=8cm,angle=0]{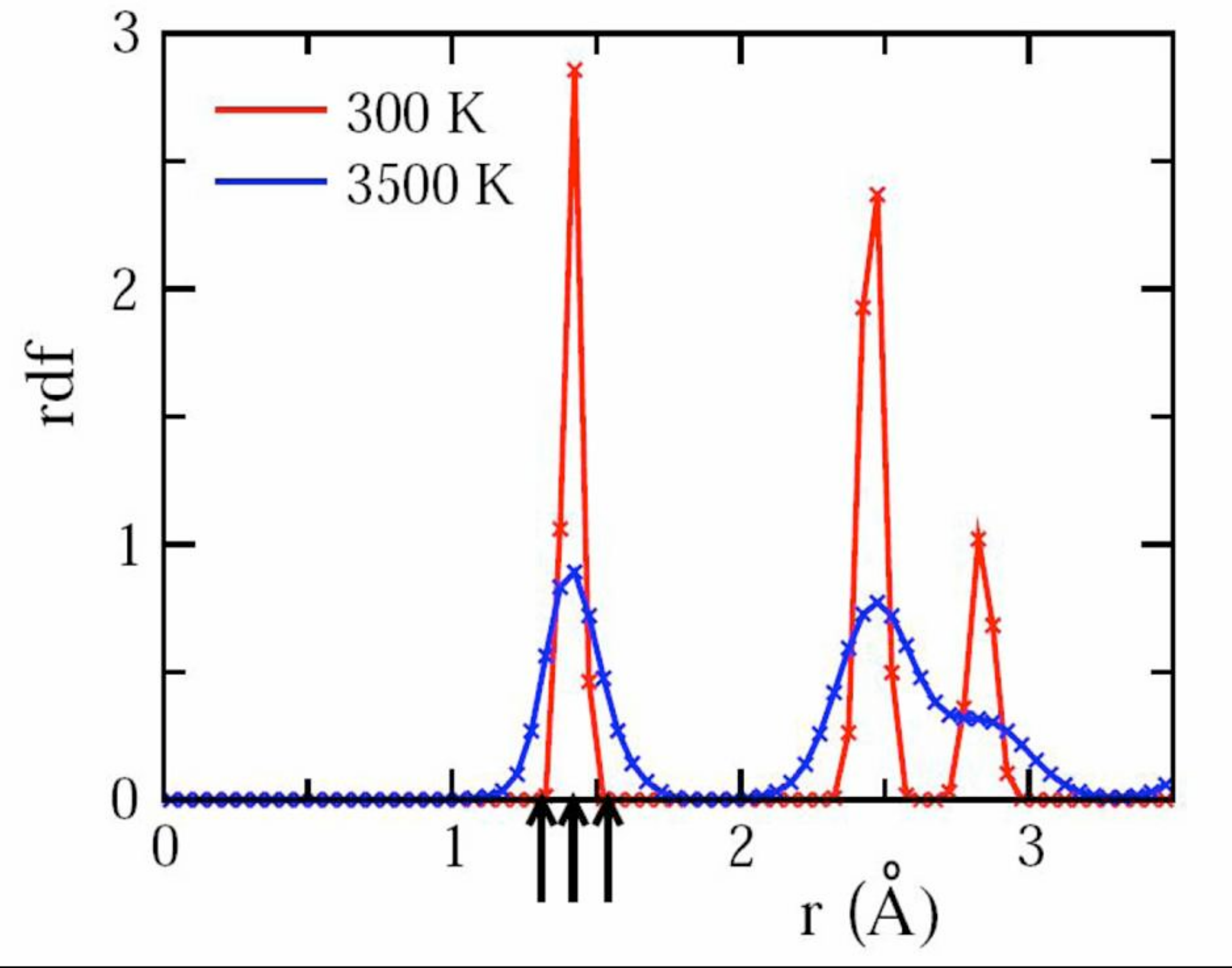}
\caption[fig]{Radial distribution function at T =300 K and
T =3,500 K as a function of interatomic distance. The arrows indicate the length
of double (r =1.31 \AA ), conjugated (r =1.42 \AA ) and single (r =1.54 \AA ) bonds
(taken with permission from Ref. \cite{FJK07}).}
\label{rip3}
\end{center}
\end{figure}
To calculate the exponent $\eta$ some approximations should be done. The $%
\epsilon$-expansion in $\epsilon = 4-d$, where $d$ is the space
dimensionality, so efficient in the theory of critical phenomena
\cite {WK74,M76,BDFN92} is practically useless for low-dimensional
systems. Instead, one can use an expansion in inverse number of
component $n$ of the vector of out-of-plane displacement $h$. Of
course, in real physical problem $n=1$. However, if one consider
$D$-dimensional membrane in $d$-dimensional space, $n=d-D$ can be
formally considered as a large parameter in the problem
\cite{NP87}, than, $\eta \approx 1/n$. More subtle use of this
parameter is the Self Consistent Screening Approximation
\cite{LDR92,XMLR03} which neglects vertex corrections when
consider diagrammatically the interaction of fluctuations; the
vertex corrections have formal smallness in $1/n$. This
approximation (for the real case $d=3,D=2$) yields $\eta= 2\left( \sqrt{15}%
-1 \right)/7 \approx 0.821$. A very recent non-perturbative
renormalization group approach \cite{KM09} yields a close value
$\eta \approx 0.849$. The discretized version of the model
(\ref{Ebencrys}) was investigated by Bowick
\textit{et al.} \cite{BDFN92} by means of Monte Carlo simulations giving $%
\eta\approx 0.72$.

For the case of crystalline membranes, contrary to the case of
liquid ones, the effective bending rigidity grows with the
temperature \cite{N89}, due to anharmonic coupling of
out-of-plane and in-plane displacements. This result has a very
simple physical meaning: it is known from the elasticity theory
that corrugations strengthen plates making them effectively
thicker \cite{LL70,TW59}, and amplitudes of corrugations increase
with the temperature increase.

The long-range orientational (``hexatic'') or translational
(crystalline) order can be in principle destroyed by  spontaneous
creation of topological defects, disclinations and dislocations,
respectively. Whereas for two-dimensional crystals in
two-dimensional space the disclination energy grows with the
sample size as $L^2$ (and the dislocation one as $\ln L$) for the
case under consideration (two-dimensional crystals in
three-dimensional space) the elastic energy strongly decreases due
to screening by bending deformations; as a result, the
disclination energy grows as $\ln L$ and the dislocation energy
remains constant in the thermodynamic limit $L\rightarrow \infty $
\cite{N89}. This means that, strictly speaking, the long-range
crystalline order is destroyed also via the creation of
dislocations by thermal fluctuations. However, for graphene with
its extremely strong chemical bonding (e.g., vacancy formation
energy about 7 eV \cite{CS06}) the concentration of such thermal
dislocations should be very small; no evidences of their formation
were found in atomistic simulations for temperatures up to 3000 K
\cite{FJK07,ZKF09}.

A quantitative information about intrinsic corrugations in
graphene can be obtained using atomistic simulations with some
interatomic interaction potential for carbon (of course, it cannot
be just pairwise potential, it should be dependent on the angles
between bonds which is crucially important for a covalent chemical
bonding). The Monte Carlo simulations \cite{FJK07,ZKF09,LKYZF09}
used the so called bond-ordered potential LCBOPII \cite {LGMF05}.
This potential is based on a large database of experimental and
theoretical data for molecules and solids and has been proved to
describe very well thermodynamic and structural properties of all
phases of carbon and its phase diagram in a wide range of
temperatures and pressures \cite{LGMF05,GLMFF05}. Of particular
importance here, is that bond order potentials correlate
coordination to bond strength, allowing changes between single,
double and conjugated bonds with the correct energetics.

\begin{figure}[!t]
\begin{center}
\includegraphics[height=4cm]{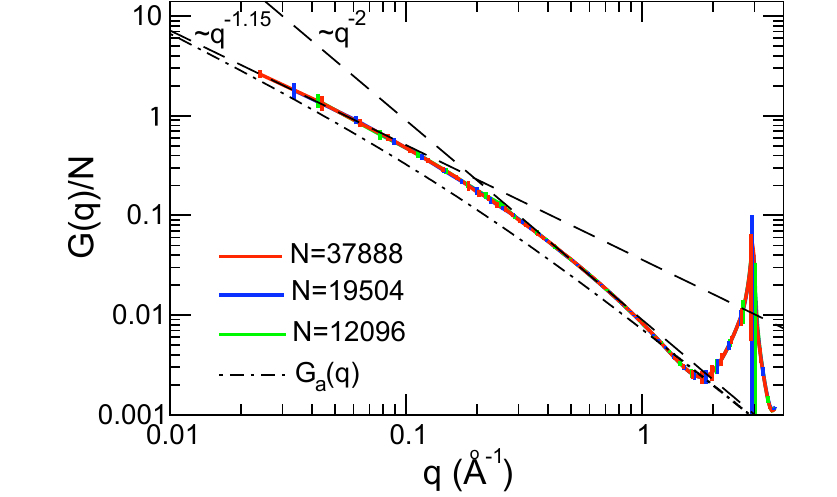}
\caption[fig]{Normal-normal correlation functions
calculated for three systems with N=12096, 19504, and N=37888
carbon atoms. The dashed lines show the asymptotic harmonic behavior
with power laws 
with 
between these limiting cases (taken with permission from Ref. \cite{LKYZF09}).
}
\label{rip4}
\end{center}
\end{figure}

A typical configuration of graphene at room temperature obtained
in the simulations is shown in Figure \ref{rip1}. For the crystallite of
about 10000 atoms a typical size of the height fluctuations was 0.07 nm
(a half of interatomic distance) with a typical spatial scale 5 to 10 nm.
As a result, the bond lengths are rather broadly distributed,
changing between lengths of double and single carbon-carbon bonds
(Figs. \ref{rip2} and \ref{rip3}). The
normal-normal correlation function shown in Fig. \ref{rip4} ( Fig. 2
from \cite{LKYZF09}) demonstrates a crossover from the harmonic
behavior (\ref{corr112}) to anharmonic one, (\ref{corr113}), at
$q^{*} \approx 2$ nm$^{-1}$, with $\eta \approx 0.85$. This value
is in a very good agreement with the predictions of the continuum
model \cite{LDR92,KM09}.

\section{Observable consequences of the gauge fields:
microscopic effects}\label{sec_obs}

\subsection{Index theorems and zero energy states} \label{sec_zero}

The energy spectrum of massless Dirac fermions coupled to gauge fields has
a very special property, the existence of zero-energy chiral modes
equally shared by electrons and holes. This is a consequence of
one of the most important theorems of modern mathematics, the Atiyah-Singer index theorem \cite{AS68,AS84}%
. This theorem has important applications in quantum field and
superstring theories \cite{kaku,nakahara}. It was mentioned
already in one of the first works on graphene \cite{Netal05} (see
also Ref. \cite{K07}) that the anomalous (``half-integer'')
quantum Hall effect in single-layer graphene is a consequence of
the Atiyah-Singer theorem (later \cite{KP08} a similar statement
has been also proven for the anomalous quantum Hall effect in
graphene bilayer \cite{Netal06}).  The
existence of the zero-energy states for massless Dirac fermions in
inhomogeneous magnetic fields in two dimensions was
demonstrated explicitly long ago by Aharonov and Casher \cite{AC79}. Here
we present their result and discuss its relevance for graphene
physics.

Let $\mathbf{A}\left( x,y\right) $ is an arbitrary two-dimensional
vector potential satisfying the condition $\nabla \mathbf{A}=0.$
Thus, one can introduce a scalar ``potential'' $\phi \left(
x,y\right) $ such that, that
\begin{equation}
A_x=-\partial _y\phi ,A_y=\partial _x\phi
\end{equation}
and
\begin{equation}
\nabla ^2\phi =B  \label{potphi}
\end{equation}
is the magnetic field. For zero-energy the Dirac equation is split
into two independent equations for the spinor components $\psi
_\sigma $ $\left(
\sigma =\pm \right) $%
\begin{equation}
\left( \partial _x+i\sigma \partial _y-iA_x+\sigma A_y\right) \psi
_\sigma =0 \label{dirzero}
\end{equation}
The potential $\phi $ can be excluded by the substitution
\begin{equation}
\psi _\sigma =e_{}^{-\sigma \phi }f_\sigma
\end{equation}
where
\begin{equation}
\left( \partial _x+i\sigma \partial _y\right) f_\sigma =0
\label{dirzero1}
\end{equation}
\[
\]
and, thus, $f_{+}$ and $f_{-}$ are analytic and complex conjugated
analytic entire functions of $z=x+iy$, respectively.

Since the Green's function of the Laplace operator in two
dimensions
\begin{equation}
\left\langle \mathbf{r}\left| \frac 1{\nabla ^2}\right|
\mathbf{r}^{\prime }\right\rangle =\frac 1{2\pi }\ln \frac{\left|
\mathbf{r-r}^{\prime }\right| }{r_0}  \label{greens}
\end{equation}
the solution of Eq.(\ref{potphi}) at large distances has the
asymptotics), with
\begin{equation}
\phi \left( r\right) \simeq \frac \Phi {2\pi }\ln \frac r{r_0}
\label{asy1}
\end{equation}
where $\Phi =\int dxdyB$ is the total magnetic flux through the
system.
Thus, at large $r$%
\begin{equation}
e_{}^{-\sigma \phi }\simeq \left( \frac{r_0}r\right) ^{\sigma \Phi
/2\pi } \label{asy2}
\end{equation}
Since the entire function cannot go to zero in all directions at infinity, $%
\psi _\sigma $ can be normalizable only assuming that $\sigma \Phi
>0,$ that is, zero-energy solutions can exist only for one
(pseudo)spin directions, depending on the sign of the total
magnetic flux.

Let us count how many independent solutions of Eq.
(\ref{dirzero1}) we have. As a basis, we can choose just polynoms
searching the solutions of the form
\begin{equation}
\psi _{+}=z^je_{}^{-\phi }
\end{equation}
(to be specific, we consider the case $\Phi >0,$ where
$j=0,1,2...$ One can easily see from Eq.(\ref{asy2}) that the
solution is integrable with the square only assuming that $j\leq
N$, where $N$ is the integer part of $\Phi /2\pi $.

According to the Atiyah-Singer theorem, the index of Dirac
operator in our situation, that is, the difference between numbers
of solutions with zero energy and positive and negative sigmas is
equal to $N$. The Aharonov-Casher procedure gives us an explicit
shape of these solutions and proves that, actually, there are
solutions only for one spin projection. These solutions are robust
(topologically protected) in a sense that they are not shifted
from zero energy of broadened by any inhomogeneities of the
magnetic field.

It has been proven in Ref. \cite{KP08} that the Atiyah-Singer
theorem can be applied also for the case of bilayer graphene; in
this situation the number of zero-energy solutions is twice the
number of the solutions for single layer with the same magnetic
flux. Explicitly, these solutions have been constructed, as a
generalization of the Aharonov-Casher solutions, in Ref.
\cite{K09}. Thus, the zero-energy states in the case of bilayer
are also topologically protected and cannot be broadened by any
inhomogeneities of the magnetic field.

This conclusion leads to an important consequence for the quantum
Hall effect in single- and bilayer graphene. Whereas smooth
inhomogeneities of electrostatic potential lead to a broadening of
all Landau levels, like in conventional homogeneous two
dimensional electron gas, and zero-energy Landau level is not an
exception \cite{KN07}, random vector potential due to ripples
cannot broaden the zero-energy Landau level which, thus, should be
essentially narrower than the other Landau levels. This conclusion
seems to be in an agreement with experimental data on quantum Hall
activation gaps in single-layer graphene \cite{Getal07}. It would
be very interesting to check it also for the case of bilayer.
\begin{figure}{t}
\begin{center}
\includegraphics*[width=4.5cm]{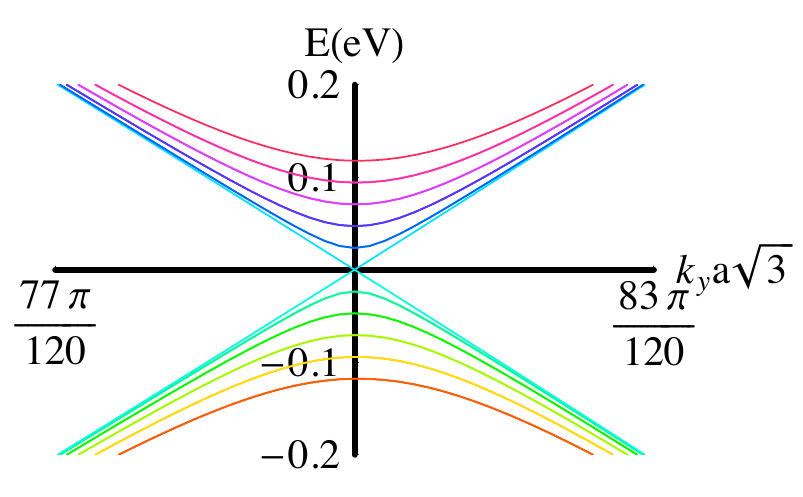}
\includegraphics*[width=4.5cm]{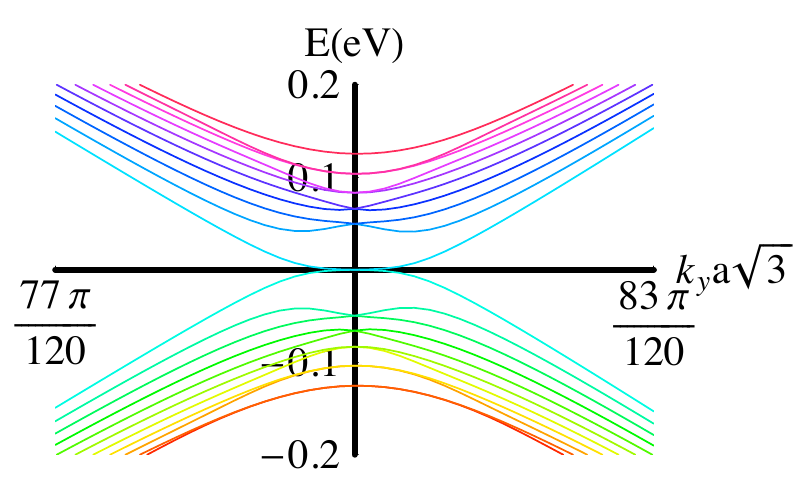}
\includegraphics*[width=4.5cm]{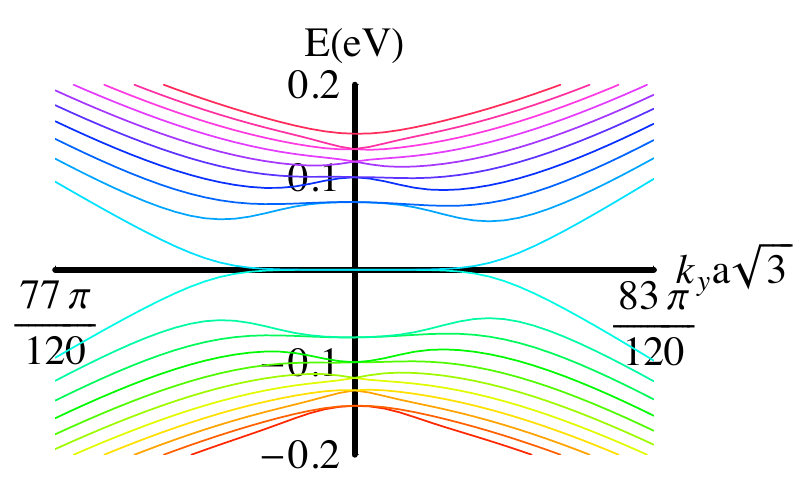}
\end{center}
\caption{Low energy states induced by a one dimensional ripple
modelled by a sine modulation of the atomic displacements. The
average hopping is $t_\parallel = 3$eV. The width of the
  ripple is $1200 a = 168$nm. The modulations of the hoppings are: Left,
  $\delta t / t = 0$, center, $\delta t / t = 0.02$, right, $\delta t / t = 0.04$ (taken with permission from Ref. \cite{GKV07}).}
\label{bands1}
\end{figure}
If the pseudomagnetic field created by ripples is strong enough it
can result in the appearance of the mid-gap states \cite{GKV07}.
Figure \ref{bands1}\cite{GKV07} shows energy spectrum of a
graphene ribbon with the simplest one-dimensional sine modulation
of the atomic displacements and, thus, vector potential. One can
clearly see an appearance of the midgap states for strong enough
modulation. If the modulated scalar potential is also present the
Atiyah-Singer theorem is not applicable and, in general, a gap can
open. General conditions for opening the gap in oscillating
electric and magnetic fields are discussed in Ref.\cite{S09}.
The existence of the midgap states for sine modulation has
been confirmed by ab initio calculations \cite {Wetal08}.

The midgap states are  probably  essential in the chemistry of
graphene. It was demonstrated by a straightforward ab initio
calculation \cite{BK09} that the energy of chemisorption of
hydrogen on rippled graphene is much lower than that on the flat samples
indicating that  the ripples can be centers of chemical activity. This
effect is especially strong if the corrugations are large enough
to create the midgap states: hydrogen leads to the splitting of the
peak at zero energy and thus to an essential total energy decrease.
This splitting results from a resonance between the midgap states
created by hydrogen itself and those created by the ripples.

\subsection{Inhomogeneities in the electronic density} \label{inhom}
\begin{figure}
  \begin{center}
    \includegraphics[height=4cm]{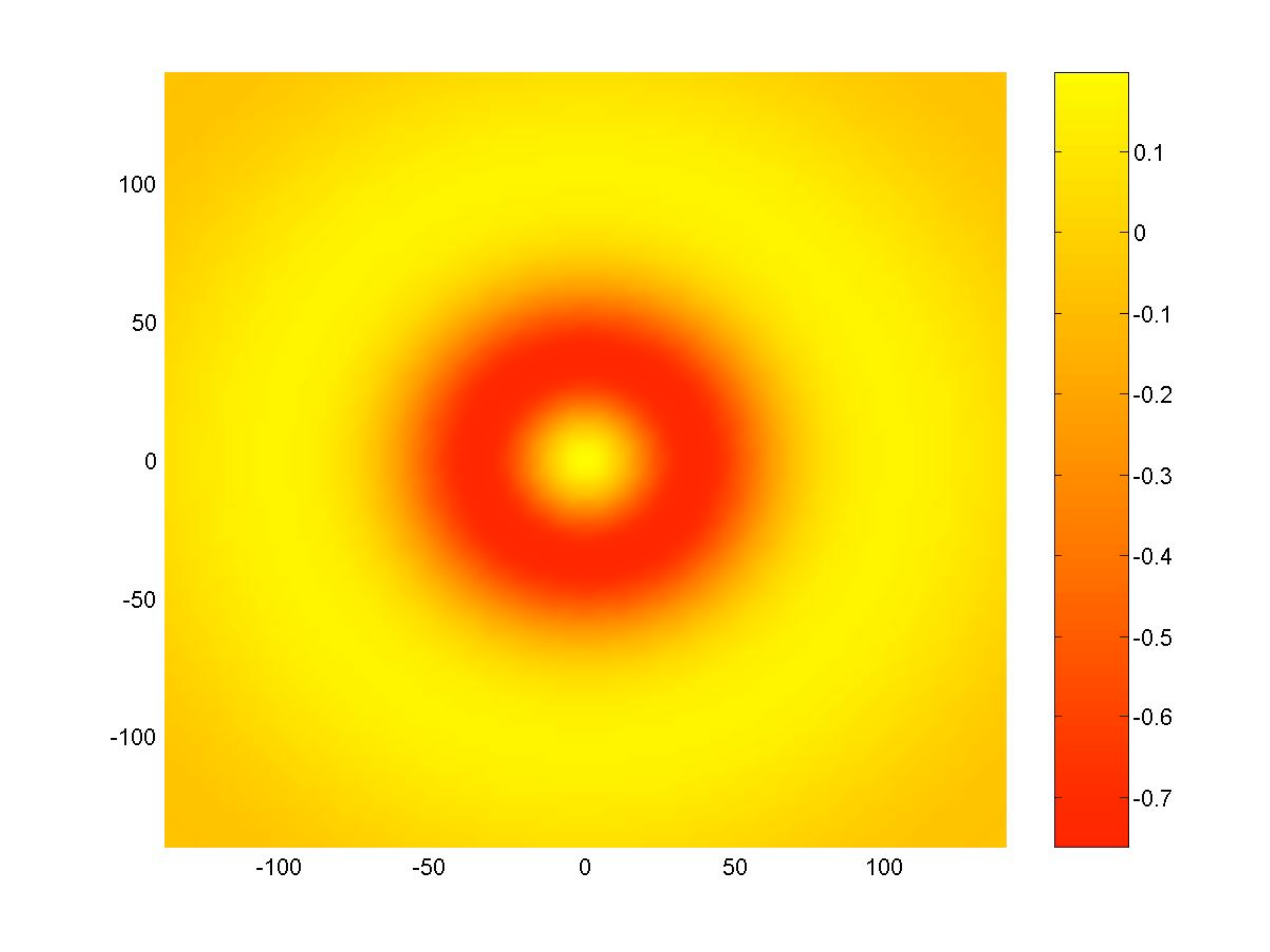}
    \hspace{1cm}
   \includegraphics[height=4cm]{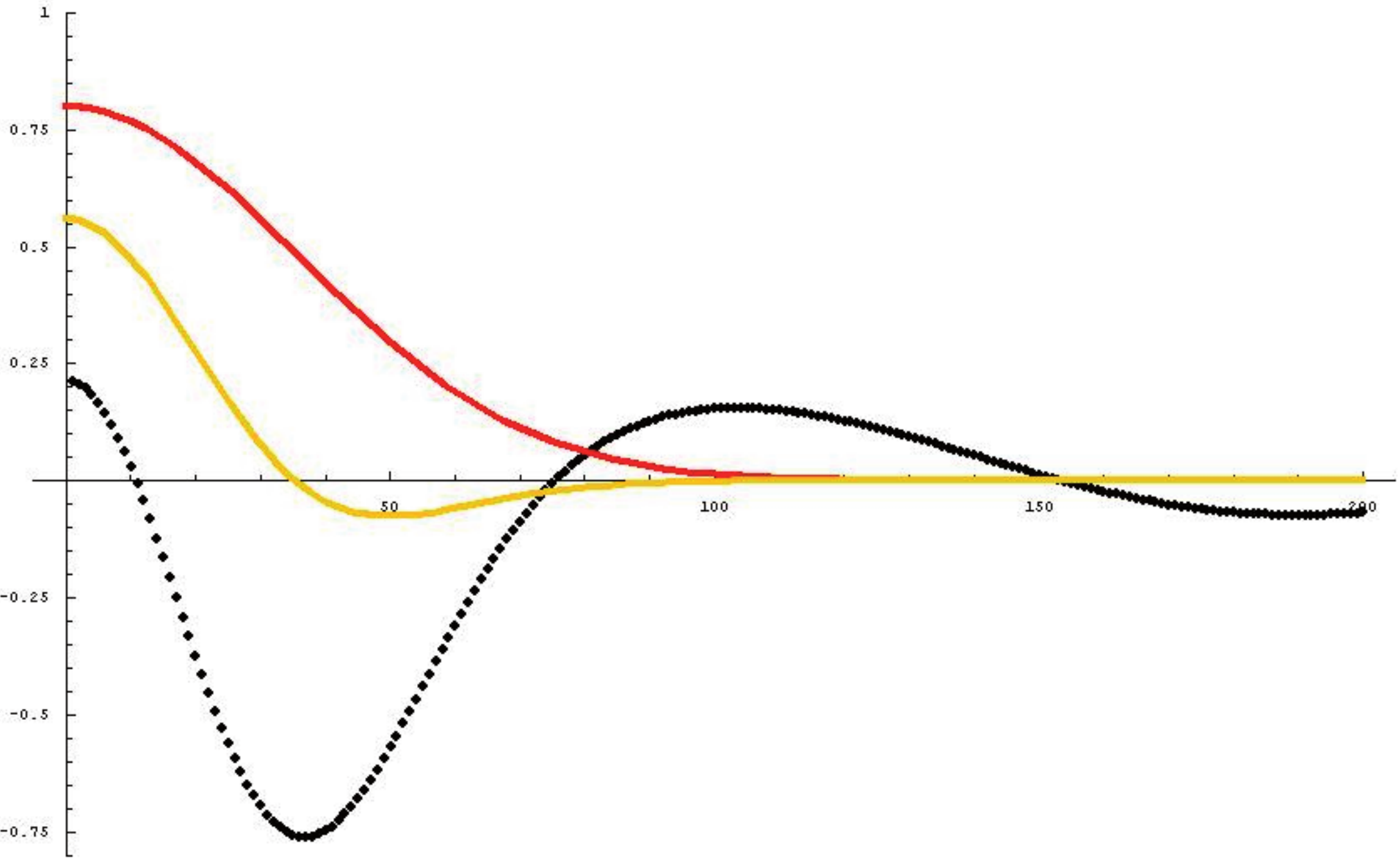}
    \caption{(Color online)Left: Effect of the curved bump of fig. \ref{gaussian}
    on the local density
    of states of the graphene sheet. The color code is indicated in the figure.
    Darker (lighter) areas represent
   negative (positive) corrections to the density of states of the flat graphene sheet.
   Right: Correction to the density of states
    (dotted line) in arbitrary units, versus the
    shape of defect (red -- upper -- line) and curvature of the defect (yellow --middle-- line)
    for a gaussian bump of an average width of 50 $\AA$ (taken with permission from Ref. \cite{JCV07}). }
    \label{bollo}
\end{center}
\end{figure}
One of the effects of having samples locally curved or rippled is
to induce charge inhomogeneities in the samples. This effects were
studied within the geometric (covariant) approach  in Refs.
\cite{CV07a,JCV07} by computing the local density of states
through the Green's function. In a curved surface, the spinor
Green's function obeys the equation
\begin{equation}
i\gamma^{\alpha}e_{\alpha}^{\
\mu}\left(\partial_{\mu}-\Omega_{\mu}\right)\
G({\bf r},{\bf r'})=\delta({\bf r-r'})(-g)^{-\frac{1}{2}},
\end{equation}
where the geometric factors are described in Appendix
\ref{sec_curvedspace}. From this equation the local density of
states can be obtained through the expression:
\begin{equation}
\rho(E,\textbf{r})=-\frac{1}{\pi}Im Tr
[G(E,\textbf{r},\textbf{r})\gamma^{0}].
\end{equation}
The gaussian shape of Fig. \ref{gaussian}  was shown in
\cite{JCV07} to induce oscillations in the density of states of
the type depicted in Fig. \ref{bollo} a). It is interesting to
note that the local maximum of correction to the flat DOS
corresponds to the regions in space where the curvature changes
sign as it can be seen in Fig. \ref{bollo} b).
An interesting prediction of this model is a space variation of
the Fermi velocity $\tilde{v}_{r}$ in the radial direction given
by
\begin{equation}
\tilde{v}_r(r,\theta)=\frac{v_F}{\sqrt{1+\alpha f(r)}},
\end{equation}
where $f(r)$ is the equation of the surface (\ref{surface}). It is
worth mentioning that irrespective of the shape of the
corrugations, the effective Fermi velocity in the presence of
intrinsic curvature will always be less than the flat value $v_F$.

The same formalism applied to a given distribution of topological
defects gives rise also to inhomogeneities of the local DOS where
a local charge density is accumulated around the pentagons and
repelled near the heptagons \cite{CV07a}.

\subsection{Strains, gauge fields, and weak (anti)localization}
In normal metals, the wavevector of the electrons at the Fermi
energy is comparable to the lattice spacing. This allows us to
define wavepackets which are much smaller than the distance over
which the electrons move between collisions, the mean free path. The
trajectories of these wavepackets in an applied electric field are
described by the classical equations of motion.

\begin{figure}
  \begin{center}
    \includegraphics[width=7cm]{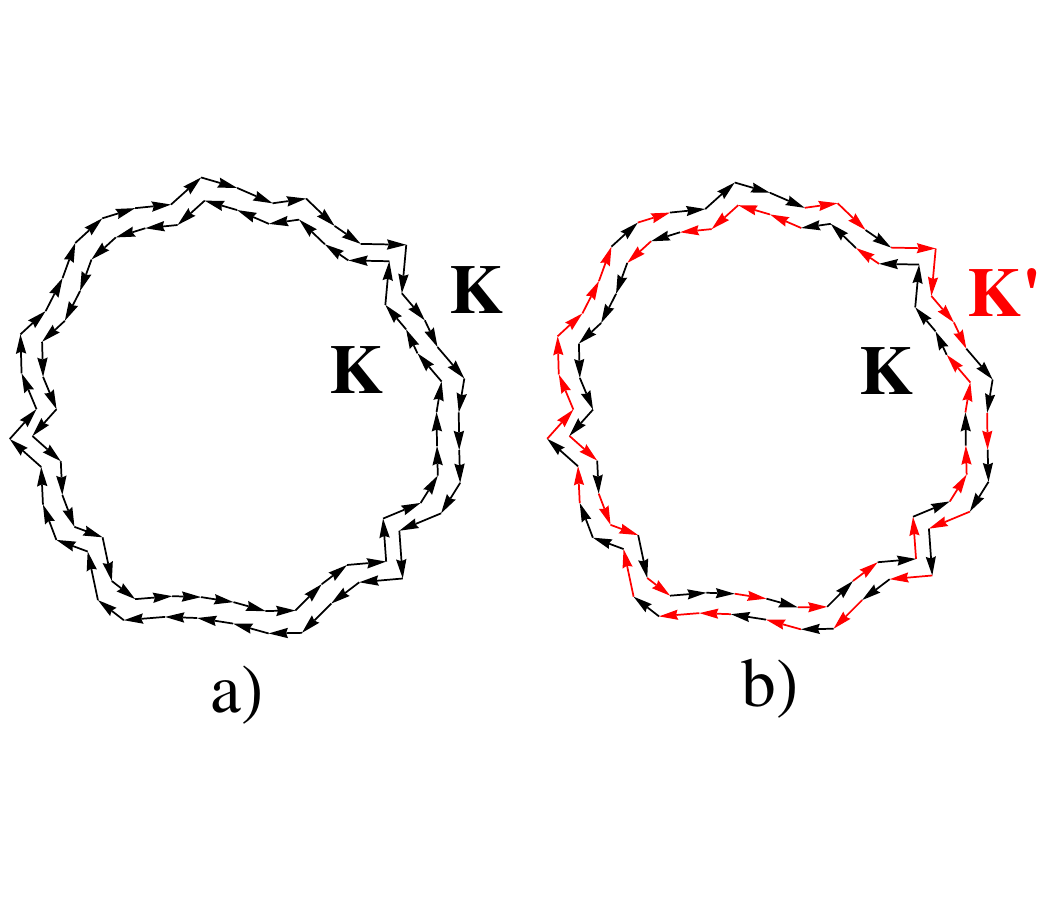}
    \caption{(Color online) Counterpropagating paths which contribute to the magnetoresistance in
    weak magnetic fields.  a) Paths where the wavefunctions  belong to the same valley. Although these paths are not related by time reversal symmetry, near the Dirac point are sufficiently similar
    to lead to interference effects.  b) Paths related by time reversal symmetry. Interference
    between such paths is the leading contribution to the magnetoresistance if intervalley scattering
    is sufficiently strong. }
    \label{WL_interference}
\end{center}
\end{figure}

A full analysis of the conductance of a metal requires the analysis
of the effects associated to the wave nature of the electrons
\cite{B84,LR85}. These corrections to the classical description can
be studied as a series in powers of $k_F \ell$, where $k_F$ is the
inverse Fermi wavelength, and $\ell$ is the mean free path. The
classical description is obtained in the limit $k_F \ell \rightarrow
\infty$. The leading quantum effects arise from the interference
between two counterpropagating paths, as schematically depicted in
Fig.~\ref{WL_interference}. In a normal metal, such paths show a
constructive interference. As a result, the probability that the
electron is scattered backwards is enhanced with respect to the
classical case. This interference is suppressed in the presence of a
magnetic field, and this leads to an increase of the conductivity in
weak magnetic fields, the so called weak localization effect. When
$k_F \ell \rightarrow 1$, interference effects become strong, and
they lead to the formation of localized
wavefunctions\cite{A58,AALR79}, and to insulating behavior. In
systems with large spin-orbit coupling, the spin  is pinned to the
momentum the electron. The rotation of the spin when the electron
moves around a closed loop adds a phase of $\pi$ to the
wavefunction. As a result, closed loops show negative interference,
and the magnetoresistance is negative, the so called weak
antilocalization.

Electrons in graphene have the sublattice and the valley degrees of
freedom, which can play a role similar to the spin in systems with a
strong spin-orbit coupling. The situation is schematically shown in
Fig.~\ref{WL_interference}. Counterpropagating paths can exist where
the electron resides always in the same valley, as shown in
Fig.~\ref{WL_interference} a). The two paths are not exactly the
same, because they are not related by time reversal symmetry, which
exchanges the valleys. The equivalence between the two paths arising
from the same valley becomes exact at the Dirac point. As a result,
these paths can contribute to the magnetoresistance for weak
magnetic fields, provided that intervalley scattering can be
neglected. The contribution of these paths leads to weak
antilocalization, because of the contribution to the total phase of
the sublattice polarization, the pseudospin. This situation can be
also described as the result of the existence of an approximate
symmetry which plays the role of time reversal symmetry within each
valley.

The contribution of the paths in Fig.~\ref{WL_interference} a) is
suppressed by the gauge fields due to lattice deformations
\cite{Metal06,MG06}. The interference between these paths is also
reduced by the inequivalence of the wavefunctions from the same
valley with opposite momenta, which increases away from the Dirac
energy, the so called trigonal warping \cite{MKFSAA06}. As a result,
there is a regime in graphene, when intervalley scattering is small,
but gauge fields are significant, where there are neither weak
localization nor weak antilocalization effects. This regime
describes well the experimental situation \cite{Metal06,THGS08}. The
absence of any magnetoresistance in the most recent experiments
\cite{THGS08} is remarkable, because the intravalley scattering
length associated with gauge fields or other perturbations which
break the effective time reversal symmetry at each valley is much
shorter than other scales, and it is difficult to explain by the
height of the ripples observed by AFM measurements.

In the presence of strong intravalley scattering, the leading paths
which contribute to the quantum corrections of the conductivity are
shown in Fig.~\ref{WL_interference} b). These are the same processes
which describe the magnetoresistance in ordinary metals, and they
lead to weak localization, and a negative magnetoresistance. Bilayer
graphene, were smaller corrugations are expected, shows weak
localization effects at low fields \cite{Getal07b}.

\subsection{Modelling disorder in graphene with random gauge fields}
\label{random} The influence of ripples and other types of
disorder on the transport properties of a system requires  to
assume a density of defects with some statistical distribution and
to average over defects. The standard techniques of disordered
electrons \cite{AS06} can be applied to rippled graphene by
averaging over the random effective gauge fields induced by
curvature or elastic deformations. Modelling disorder with random
magnetic fields was first proposed in the context of disordered
degenerate semiconductors \cite{Fradkin1,Fradkin2}. Dirac fermions
in random magnetic fields were used to study critical point
transitions between integer and fractional fillings in the Quantum
Hall Effect \cite{LFSG94,YS98}. Most of these works involved
disordered systems without interactions what reduces effectively
the dimensionality of the problem to two dimensions allowing the
use of very powerful tools as bosonization or conformal field
theory techniques \cite{CMW96}. A complete classification of
disordered graphene without interaction in the random matrices
language was worked out in \cite{OGM06,AE06,MKetal06}. Combined
disorder and interactions was first addressed in the context of
critical points between integer and fractional fillings in the
Quantum Hall Effect \cite{LFSG94,YS98,Y99} and later applied to
the graphene problem in \cite{GGV01,SGV05,VC08}. A recent complete
survey can be found in \cite{FA08} and the included references. A
computation of the effect of the curvature induced by topological
defects on the conductivity of neutral graphene using the random
field approach was done in \cite{CV09}.

The low energy excitations in graphene combining the two Dirac
points can be described by the four-dimensional Hamiltonian:
\begin{eqnarray}
H_0&=iv_F\int d^2x \overline\Psi(\vec x)\vec\gamma
\cdot\vec\nabla\Psi(\vec x) \label{4Dhamil}
\end{eqnarray}
where $\overline\Psi\equiv \Psi\dag\gamma_0$ with the $4\times4$
matrix $\gamma_0\equiv\sigma_3\otimes\sigma_3$ (in this section we
put $\hbar =1$). We further have
$\vec\gamma\equiv(\gamma_1,\gamma_2)=(-i\sigma_2,i\sigma_1)\otimes\sigma_3$.
The $\sigma_\mu$ denote the usual Pauli matrices such that
$\{\gamma_\mu,\gamma_\nu\}=2g_{\mu,\nu}{\bf{1}}_{4\times4}$,
$g_{\mu,\nu}$ denoting the Minkowski tensor where $g_{0,0}=1$,
$g_{i,i}=-1$ with $i=1,2$, and zero otherwise.

The long range Coulomb interaction in terms of the Dirac spinors
reads
\begin{align}
H_{ee}&=\frac{v_F}{4\pi}\int d^2x d^2x'\overline\Psi(\vec
x)\gamma_0\Psi(\vec x)\frac{g}{|\vec x-\vec x'|}
\overline\Psi(\vec x')\gamma_0\Psi(\vec x')
\end{align}
where $g=e^2/v_F$ is the dimensionless coupling constant.

In order to describe disorder effects, the Dirac spinors are
coupled to a gauge field $A(\vec x)$,
\begin{align}
\label{Hdisorder} H_{disorder}=\frac{v_\Gamma}{4}\int d^2x
\overline\Psi(\vec x)\Gamma\Psi(\vec x)A(\vec x)
\end{align}
where $v_\Gamma$ characterizes the strength and the $4\times4$
matrix $\Gamma$ the type of the vertex. In general, $A ( \vec x )$
is a quenched, Gaussian variable with the dimensionless variance
$\Delta$, i.e.,
\begin{align}
\label{Gauss} \langle A(\vec x)\rangle=0\quad,\quad\langle A(\vec
x)A(\vec x')\rangle=\Delta_\Gamma\delta^2(\vec x-\vec x')\quad.
\end{align}
Inclusion of long ranged correlated disorder is a much more difficult task that has been addressed in \cite{NM06,Kh07,CV09}.

A perturbative renormalization group (RG) approach to the problem as the one performed in \cite{SGV05,VC08,FA08} allows to study the phase diagram of the system as a function of the Coulomb interaction parameter and the strength of the various disorder couplings. It is known that  the (unscreened) Coulomb interactions in graphene induce an upward flow of the Fermi velocity that grows at low energies \cite{GGV94,GGV99}. That makes the effective coupling $g\sim\frac{e^2}{v_F}$ to flow to zero. Typically the inclusion of disorder changes the flow of the Fermi velocity due to wave function renormalization, i.e.,
$Z_{v_F}\to Z_{v_F}/Z_\Psi$. In the simplest one loop analysis done in \cite{SGV05},
from the $\beta$-function
$\beta_{v_F}=\Lambda\partial_\Lambda Z_{v_F}\tilde v_F$, one  obtains
flow equations for the effective Fermi velocity $v_F^{eff}$ of the form:

\begin{align}
\label{FlowLutt}
\frac{d}{d\ell}\frac{v_F^{eff}}{\tilde v_F}=
\frac{1}{16\pi}\left[\frac{e^2}{v_F^{eff}}-
\frac{\Delta}{2}\left(\frac{v_\Gamma^{eff}}{v_F^{eff}}\right)^2\right],
\end{align}
where $\ell=\ln \Lambda/\Lambda_0\sim1/\varepsilon$.

The phase diagrams obtained in \cite{SGV05} for the various types of disorder  considered can be summarized as follows:

\vspace{0.5cm}

i) For a random chemical potential $(\Gamma=\gamma_0)$,
$v_\Gamma=v_1$ remains constant under
renormalization group transformation. There is  an unstable fixed
line at $v_F^*=v_1^2\Delta/(2e^2)$. In the $(g,\Delta)$-plane,
the strong-coupling and the weak-coupling phases are separated
by a hyperbola, with the critical electron interaction
$g^*=e^2/v_F^*=2e^4/(v_1^2\Delta)$.

\vspace{0.5cm}

ii) A random gauge potential involves the vertices $\Gamma=i\gamma_1,i\gamma_2$.
The vertex strength renormalizes as $v_\Gamma=v_F$.  There is  an
attractive Luttinger-like fixed point for each disorder correlation strength
$\Delta$ given by $v_F^*=2e^2/\Delta$ or $g^*=\Delta/2$.

\vspace{0.5cm}

iii) For a random mass term $\Gamma={\bf{1}}_{4\times4}$, topological disorder
$\Gamma=i\gamma_5$, and $\Gamma=i\tilde\gamma_5$, we have $v_\Gamma=v_F^2/v_3$.
There is thus again an attractive Luttinger-like fixed point for each
disorder correlation strength $\Delta$ given by $v_F^*=\root 3\of{2v_3^2e^2/\Delta}$
or $g^*=\root3\of{\Delta e^4/(2v_3^2)}$.

\vspace{0.5cm}

More sophisticated calculations do not introduce essential changes in this simple description.

\subsection{Scattering of charge carriers by gauge fields}

High electron mobility found already in the first work on graphene
\cite {Netal04} is one of its most attractive features, in view of
potential applications. Submicron mean free path is routinely
achievable; even order of magnitude higher mobility $\mu \simeq
$10$^5$ cm$^2$/Vs was reached recently for freely suspended
graphene membranes \cite{Betal08,DSBA08}. Despite numerous
experimental and theoretical efforts, the physics of this high
mobility is not well understood yet and there are still
controversial views on the main mechanism limiting the electron
mobility. Following a general line of our review, we discuss here
the role of gauge fields created by corrugations of graphene in
its electron transport properties.

We proceed with a simple and physically transparent semiclassical
picture based on the Boltzmann kinetic equation. Its applicability
to graphene is not a priori obvious, due to potential role of
quantum relativistic effects known as the Zitterbewegung (in terms
of solid state physics - interband processes mixing electrons and
holes) \cite{Kat06b,AK07}. However, it has been formally
demonstrated in Ref. \cite{AK07}, using short-range scatterers as
an example, that the interband scattering processes are
negligible, and, thus, the semiclassical Boltzmann equation is
justified, assuming that
\begin{equation}
\varepsilon _F\tau /\hbar \gg 1/\left| \ln \left( k_Fa\right)
\right| ,\label{zitter}
\end{equation}
where $\varepsilon _F$ and $k_F$ are the Fermi energy (counted
from the Dirac point) and wave vector, respectively, and $\tau $
is the electron mean free path related with the resistivity $\rho
$ by the Drude formula \cite {Z01}
\begin{equation}
\rho =\frac 2{e^2v_F^2N\left( \varepsilon _F\right) }\frac 1{\tau
\left( k_F\right) } . \label{drude}
\end{equation}
The condition (\ref{zitter}) excludes only a relatively small
doping interval where the conductivity is close to a minimal
metallic conductivity of order of $e^2/h.$ The expression for
$\tau $ depends on the scattering mechanism. For point defects
with the concentration $n_{imp}$ and angular dependent scattering
cross section $\sigma \left( \theta \right) $ one has
\cite{Z01,SA98}
\begin{equation}
\frac 1{\tau \left( k\right) }=n_{imp}v_F\sigma _{tr}\left(
k\right), \label{tau}
\end{equation}
where
\begin{equation}
\sigma _{tr}=\int\limits_0^{2\pi }d\theta \frac{d\sigma \left(
\theta \right) }{d\theta }\left( 1-\cos \theta \right),
\label{tau1}
\end{equation}
is the transport cross section.

For the case of radially symmetric potential the scattering cross
section can be expressed in terms of the scattering phases $\delta
_m$ as \cite {OGM06,KN07,HG07,N07,CPP07,SKL07a,SKL07b,G08}
\begin{equation}
\sigma _{tr}=\frac 4k\sum\limits_{m=0}^\infty \sin ^2\left( \delta
_m-\delta _{m+1}\right).  \label{sigtr}
\end{equation}

Consider first the case of short-range scatterers with the radius
of potential $R$ much smaller than the electron de Broglie
wavelength $\lambda _F=2\pi /k_F.$ Then only $s$-scattering
survives ($m=0$), with $\delta _0\left( k\right) \sim kR$ in a
generic case, with a negligible contribution to the resistivity,
\begin{equation}
\rho \simeq \frac h{4e^2}n_{imp}R^2 \label{rho1}
\end{equation}
(we keep here and further a factor 4 to remind that there are four
channels of conductivity, due to two valleys and two spin
projections). This is not surprising: the massless Dirac fermions
in graphene has the same dispersion law as light and, thus,
obstacles with sizes smaller than the wavelength are inefficient,
like in optics \cite{BW80}. On the contrary, for conventional
nonrelativistic two-dimensional electrons even weak scattering
leads to a formation of shallow bound states and to singularities
in the scattering matrix\cite{LL77}, namely, $\delta _0\left(
k\right) \sim 1/\ln \left( kR\right) .$ For the case of graphene,
such situation takes place for a special case of ``unitary'', or
``resonant'' scatterers with a quasibound state located near the
Dirac point \cite{OGM06,KN07,HG07}. In such a case, the
contribution to the resistivity is much stronger than (\ref{rho1})
\cite {KN07,SPG07,KG08},
\begin{equation}
\rho \simeq \frac h{4e^2}\frac{n_{imp}}{n\ln ^2\left( k_FR\right)
} \label{rho2}
\end{equation}
where $n$ is the charge carrier concentration.

For the case of long-range Coulomb scattering potential, the
scattering phases are energy independent \cite{N07,SKL07a,PNCN07}
which leads to a contribution to the resistivity inversely
proportional to $n$; taking into account the screening does not
change this result \cite{NM06,A06,AHGS07}. A rough estimation for
the resistivity reads \cite{AHGS07}
\begin{equation}
\rho \simeq 20\frac h{e^2}\frac{n_{imp}}n  \label{rho3}
\end{equation}

Experimentally \cite{Netal05,ZTSK05}, the resistivity is roughly
inversely proportional to the charge carrier concentration, or,
equivalently, the mobility is independent, or weakly dependent on
$n.$ Since the scattering by Coulomb charges is the simplest
mechanism explaining this behavior it is not surprising that this
was considered as an explanation ``by default''. This explanation
seems to be in agreement with experimental data on graphene
chemically doped by potassium \cite{Jetal08}. On the contrary,
experiments on the chemical doping by gaseous impurities such as
NO$_2$ show a rather weak dependence of the mobility on charge
carrier concentration \cite{Setal07a}. Recently, it was
demonstrated that the mobility is also weakly sensitive to
dielectric constant $\kappa $ of substrate or to coverage of
graphene by polar liquids with high $\kappa $, such as ethanol or
water \cite{Petal09}. Clusterization of the charge impurities due
to low diffusion barriers may be a possible explanation of their
relatively small contribution to the resistivity
\cite{KGG09,WKL09b}. Anyway, the experimental data \cite{Petal09}
probably mean that charge impurities cannot be the main factor
limiting electron mobility in graphene.

Resonant scatterers such as vacancies \cite{SPG07} or covalently
bond impurities \cite{WKL09b,WKL09a} might be another option.
However, vacancies in graphene have huge energy, about 7 eV, and
should be rather exotic (if not created intentionally as in
experiments \cite{CCJFW09}); as for the covalent impurities it is
not clear why they should \textit{always} have resonant levels
close to the Dirac point. For these reasons, an alternative
scenario \cite{KG08}, the scattering by gauge fields created by
frozen ripples, is worth to be considered seriously.

Let us consider a generic perturbation of the form
\begin{equation}
H^{\prime }=\sum\limits_{\mathbf{pp}^{\prime }}\Psi
_{\mathbf{p}}^{\dagger }V_{\mathbf{pp}^{\prime }}\Psi
_{\mathbf{p}^{\prime }},  \label{hamprime}
\end{equation}
where $\Psi _{\mathbf{p}}$ is the Dirac spinor dependent on quasimomentum $%
\mathbf{p}$ and
\begin{equation}
V_{\mathbf{pp}^{\prime }}=V_{\mathbf{pp}^{\prime }}^{\left( 0\right) }+A_{%
\mathbf{pp}^{\prime }}^{\left( x\right) }\sigma
_x+A_{\mathbf{pp}^{\prime }}^{\left( y\right) }\sigma _y
\label{Vscat}
\end{equation}
contains both scalar (electrostatic) and vector (gauge field)
random potentials. Following Ref. \cite{KG08} we will use Born
approximation in the perturbation $H^{\prime }$ and
Kubo-Nakano-Mori approach \cite{K57,N57,M65} (as a recent example
illustrating technical details see, e.g., Ref. \cite {IK02}). This
approach allows us to obtain, in a simple and straightforward way,
the results which are equivalent to solution of Boltzmann equation
by variational principle \cite{Z01}. The inverse mean free path
time is given by
\begin{equation}
\frac 1\tau =\frac 1{\hbar^2 \left\langle j_x^2\right\rangle
}\int\limits_{-\infty }^\infty dt\left\langle \left[ j_x\left(
t\right) ,H^{\prime }\left( t\right) \right] \left[ H^{\prime
},j_x\right] \right\rangle ,  \label{taub}
\end{equation}
where $j_x$ is the current operator in $x$ direction. Use of this
approach in graphene should be done carefully, namely, only
intraband (electron-electron or hole-hole, depending on the
positions of the Fermi energy) contributions to the operators
$j_x$ and $H^{\prime }$ should be taken into account. An accurate
justification of such procedure is done in Ref. \cite{AK07}. For
the case of static disorder, the result takes the form
\begin{equation}
\frac 1\tau =\frac{4\pi }{\hbar N\left( \varepsilon _F\right) }\sum\limits_{%
\mathbf{pp}^{\prime }}\delta \left( \varepsilon
_{\mathbf{p}}-\varepsilon _F\right) \delta \left( \varepsilon
_{\mathbf{p}^{\prime }}-\varepsilon _F\right) \left( \cos \theta
_{\mathbf{p}}-\cos \theta _{\mathbf{p}^{\prime }}\right) ^2\left|
W_{\mathbf{pp}^{\prime }}\right| ^2,  \label{tau2}
\end{equation}
where $\theta _{\mathbf{p}}$ is the polar angle of the vector
$\mathbf{p}$ and
\begin{align}
W_{\mathbf{pp}^{\prime }}&= {V_{\mathbf{pp}'}}^{\left( 0 \right)}
\frac{ 1+\exp \left[ -i\left( \theta _{\mathbf{p}}-\theta
_{\mathbf{p}'}\right) \right]}{2} + \nonumber \\
&+ \frac{1}{2} \left[ \left( A_{\mathbf{pp}^{\prime }}^{\left(
x\right) }+iA_{\mathbf{pp}^{\prime }}^{\left( y\right) }\right)
\exp \left( -i\theta _{\mathbf{p}}\right) +\left(
A_{\mathbf{pp}^{\prime }}^{\left( x\right)
}-iA_{\mathbf{pp}^{\prime }}^{\left( y\right) }\right) \exp \left(
i\theta _{\mathbf{p}^{\prime }}\right) \right].   \label{w}
\end{align}
\[
\]
Note that the electrostatic part of the scattering disappears for
the back scattering ($\theta _{\mathbf{p}}-\theta
_{\mathbf{p}^{\prime }}=\pi $) which is related to the ``Klein
tunneling''\cite{KNG06} but this is not the case for the
scattering by vector potential. For a rough estimations in order
of magnitude Eq. (\ref{tau2}) can be rewritten as
\begin{equation}
\frac 1\tau \simeq \frac{2\pi N\left( \varepsilon _F\right) }\hbar
\left( \left\langle V_{\mathbf{q}}^{\left( 0\right)
}V_{-\mathbf{q}}^{\left(
0\right) }\right\rangle +\left\langle \mathbf{A}_{\mathbf{q}}\mathbf{A}_{-%
\mathbf{q}}\right\rangle \right) _{q\approx k_F} , \label{tau3}
\end{equation}
where $\mathbf{q=p}^{\prime }\mathbf{-p}$ is the scattering vector. Since $%
N\left( \varepsilon _F\right) \propto k_F$, to have concentration
independent mobility one needs that one of the correlation functions in Eq. (%
\ref{tau3}) scales as $1/q^2$. Note that the second term, that is,
the scattering by gauge fields, is not sensitive to dielectric
screening.

Consider now the case of scattering by gauge fields created by
intrinsic ripples due to thermal fluctuations (see Section
\ref{thermal}). For not too small wave vectors, in the harmonic
regime, $q>q^{*}$ one can neglect a coupling between bending and
stretching modes. This case is relevant for electron transport
assuming that
\begin{equation}
k_F>q^{*},  \label{ineq99}
\end{equation}
which does not look too restrictive. Further we will assume this
condition
to be satisfied. Thus, the vector potential is quadratic in derivatives $%
\partial h/\partial x,\partial h/\partial y$ and the estimation of the
correlation function in (\ref{tau3}) is
\begin{equation}
\left\langle
\mathbf{A}_{\mathbf{q}}\mathbf{A}_{-\mathbf{q}}\right\rangle
\approx \left( \frac{\hbar v_F}a\right) ^2\sum\limits_{\mathbf{q}_1\mathbf{q}%
_2}\left\langle h_{\mathbf{q-q}_1}h_{\mathbf{q}_1}h_{-\mathbf{q+q}_2}h_{-%
\mathbf{q}_2}\right\rangle \left[ \left( \mathbf{q-q}_1\right) \cdot \mathbf{%
q}_1\right] \left[ \left( \mathbf{q-q}_2\right) \cdot
\mathbf{q}_2\right], \label{AqAq}
\end{equation}
and the correlator in right-hand side of Eq. (\ref{AqAq}) can be
decoupled by Wick theorem since the field $h\left( x,y\right) $
can be considered as a Gaussian in this regime. Thus, the
scattering rate under the condition (\ref
{ineq99}) is determined by the correlation function $\left\langle h_{\mathbf{%
q}}h_{-\mathbf{q}}\right\rangle ,$ that is, a Fourier component of
the correlator
\begin{equation}
\Gamma \left( r\right) =\left\langle \left[ h\left( r\right)
-h\left( 0\right) \right] ^2\right\rangle .  \label{corr99}
\end{equation}
For thermally excited ripples in the harmonic regime
\begin{equation}
\Gamma \left( r\right) \simeq \frac T\kappa  r^2, \label{corrharm}
\end{equation}
which is a Fourier transformation of the Eq. (\ref{corr112}).
However, we consider first a more general case $\Gamma \left(
r\right) \propto r^{2H}$ (for
instance, for the ripples due to roughness of substrate one could expect $%
H\approx 1/2,$ see Ref. \cite{Ietal07}).

For $2H<1,$ the correlation function (\ref{AqAq}) has a finite limit at $q=0$%
\begin{equation}
\left\langle
\mathbf{A}_{\mathbf{q}}\mathbf{A}_{-\mathbf{q}}\right\rangle
_{q=0}\approx \left( \frac{\hbar v_F}a\right) ^2\frac{z^4}{R^2},
\label{corr88}
\end{equation}
where $z$ and $R$ are the characteristic height and radius of the
ripples, respectively. The corresponding contribution to the
resistivity is
\begin{equation}
\delta \rho \simeq \frac h{4e^2}\frac{z^4}{R^2a^2}.
\end{equation}
For $2>2H>1$, the resistivity is proportional to $n^{1-2H}$ and,
for $2H=1,$ to $\left| \ln \left( k_Fa\right) \right| .$ In all
these cases the corresponding contributions are too small in
comparison with experimentally observed resistivity of graphene.

Let us now come back to the case of thermally excited ripples (\ref{corrharm}%
), $H=1$. The integral in Eq. (\ref{AqAq}) is logarithmically
divergent at small wave vectors, but this divergence should be cut
at $q_1\simeq q^{*}.$ The corresponding contribution to the
resistivity is estimated as
\begin{equation}
\delta \rho \simeq \frac h{4e^2}\left( \frac T{\kappa a}\right) ^2\frac{%
\left| \ln \left( q^{*}a\right) \right| }n . \label{res99}
\end{equation}
This contribution gives us correct concentration dependence of the
resistivity and, for room temperature $T=300K$, correct order of
magnitude of mobility, $\mu \simeq 10^4$ cm$^2$/Vs. However, it
predicts a strong temperature dependence of the resistivity,
whereas experimentally it is very weak for the case of graphene at
substrate \cite{Metal08b}. The authors of Ref. \cite {KG08}
postulated that the ripple structure for graphene on substrate is
frozen and becomes temperature independent below the room
temperature.

The status of this hypothesis is not clear yet, neither
theoretically nor experimentally. Whereas earlier STM studies of
the ripples of graphene on substrate \cite{Ietal07,Setal07} have
found corrugations more or less similar to those of the substrate,
recent work \cite{Getal09} claims that an ``intrinsic'' component
postulated in Ref.\cite{KG08} is also noticeable. Further
investigations of the temperature dependence of the ripple
structure are desirable to clarify the issue. From the theoretical
point of view, it was demonstrated in Ref. \cite{BK09}  that
ripples can attract adatoms and chemical groups (such as hydrogen
or hydroxyl) and can be stabilized by them. Again, the problem
requires more studies to clarify the situation.

Up to now, we consider out-of-plane deformations as a classical
static field. In terms of quantum mechanics, the corresponding
scattering mechanism is described as the two-phonon scattering
processes by flexural phonons \cite {Metal08b}. Introducing in a
standard way \cite{Z01} phonon creation and annihilation operators
$b_{\mathbf{q}}^{\dagger },b_{\mathbf{q}}$ one can represent the
out-of-plane displacement field as
\begin{equation}
h\left( \mathbf{r}\right) =\sum\limits_{\mathbf{q}}\sqrt{\frac
\hbar
{2M\omega _{\mathbf{q}}}}\left( b_{-\mathbf{q}}^{\dagger }+b_{\mathbf{q}%
}\right) e^{i\mathbf{qr}},  \label{phon1}
\end{equation}
where $M$ is the mass of carbon atoms, $\omega
_{\mathbf{q}}=\sqrt{\kappa/\rho_m} q^2$ is the frequency of flexural
phonons, $\rho_m$ is the mass density. Since the vector potential is
quadratic in $\partial h/\partial x,\partial h/\partial y$ the
interaction Hamiltonian has the form
\begin{equation}
H^{\prime }=\sum\limits_{\mathbf{pqk}}\mathcal{A}_{\mathbf{pqk}}c_{\mathbf{p}%
}^{\dagger }c_{\mathbf{p+k-q}}\left( b_{-\mathbf{q}}^{\dagger }+b_{\mathbf{q}%
}\right) \left( b_{\mathbf{k}}^{\dagger }+b_{\mathbf{-k}}\right),
\label{phon2}
\end{equation}
where we take into account only intraband scattering processes ($c_{\mathbf{p%
}}^{\dagger },c_{\mathbf{p}}$ are electron creation and
annihilation operators) and the amplitude $\mathcal{A}$ behaves at
small wave vectors as
\begin{equation}
\mathcal{A}_{\mathbf{pqk}}\propto \frac{\hbar v_F}{Ma\sqrt{\omega _{\mathbf{q%
}}\omega _{\mathbf{k}}}}\mathbf{qk}.  \label{phon3}
\end{equation}
Then, we simply substitute the Hamiltonian (\ref{phon2}) into Eq. (\ref{tau1}%
). The calculations are very similar to those for two-magnon
scattering processes in half-metallic ferromagnets \cite{IK02}.
The result (where we skip for simplicity numerical factors of
order one) reads
\begin{eqnarray}
\frac {1}{\tau} &\simeq \frac{\hbar}{T\left( v_F M\right) ^2N\left(
\varepsilon_F\right) }\sum\limits_{\mathbf{pp}^{\prime }\mathbf{q}}\frac{\left( \mathbf{%
v}_{\mathbf{p}}-\mathbf{v}_{\mathbf{p}^{\prime }}\right) ^2}{\omega _{%
\mathbf{q}}\omega _{\mathbf{q+p-p}^{\prime }}}\left|
\mathbf{q}\left(
\mathbf{q+p-p}^{\prime }\right) \right| ^2f_{\mathbf{p}}\left( 1-f_{\mathbf{p%
}^{\prime }}\right) \times \nonumber \\
&\times \left[ 2N_{\mathbf{q}}\left(
 1+N_{\mathbf{q+p-p}^{\prime }}\right)
 \delta \left( \varepsilon_{\mathbf{p}^{\prime}}-\varepsilon_{\mathbf{p}}-
 \hbar \omega_{\mathbf{q}}
 +\hbar \omega_{\bf{q+p-p}'}\right)
  + \right. \nonumber \\ &+ \left( 1+N_{\bf q}\right) \left( 1+N_{\mathbf{q+p-p}'}\right)
\delta \left( \varepsilon_{\mathbf{p}'} - \varepsilon_{\mathbf{p}}
 +\hbar \omega_{\mathbf{q}}+\hbar
\omega_{\mathbf{q+p-p}'} \right) + \nonumber \\ &+ N_{\mathbf{q}}
N_{\mathbf{q+p-p}'}
\left. \delta \left( \varepsilon_{\mathbf{p}%
^{\prime }}-\varepsilon_{\mathbf{p}}-\hbar \omega_{\mathbf{q}}-\hbar \omega_{\mathbf{%
q+p-p}^{\prime }}\right) \right] ,  \label{phon4}
\end{eqnarray}
where $f_{\mathbf{p}}=f\left( \varepsilon_{\mathbf{p}}\right) $ and $N_{%
\mathbf{q}}=N_B\left( \omega_{\mathbf{q}}\right) $ are the Fermi
and Bose distribution functions, respectively. For
temperatures $T>\hbar \omega _{2k_F}^2 \simeq 1$ K
phonons can be considered as classical neglecting the phonon frequencies
in the conservation laws and replacing $N_{\mathbf{q}}$ by
$T/\hbar \omega _{\mathbf{q}}$ in Eq. (\ref {phon4}). Then, the
result coincides with Eqs. (\ref{tau3}), (\ref{res99}). The case
of very low temperatures was considered in Ref. \cite{MO08}. The
minimal conductivity from a geometric model of topological lattice
defects has been computed in \cite{CV09}.

\section{Observable consequences of the gauge fields: mesoscopic effects, strains in suspended samples}
\label{sec_mesos}


\subsection{Aharanov-Bohm phases in suspended samples}

Mesoscopic deformations have been observed in graphene, and they
can be induced in a controlled way. A particularly appropriate
setup is a suspended graphene flake
\cite{Metal07,Ietal07,DSBA08,Betal08a,Betal08,
GLetal08,Betal08b,BSHSK08,BRBH09,Betal09,VLBL09,Cetal09}. These
systems are being very actively investigated, as the carrier
mobility can be substantially larger than that of graphene on a
substrate. The conductance minimum at the charge neutrality point
is also sharper than in non-suspended samples. It has been
observed that these samples are under tension, which can be
modulated by the electric force between the flake and the gate.

Typical sizes, $L$, of suspended samples are of the order of a few
microns. Away from very close to the neutrality point these
dimensions are much larger than the Fermi wavelength and the
screening length, which is proportional to it. Hence, the long
wavelength scalar potentials, $V_{0} ( \vec{r} )$, which may be
generated by compression \cite{SA02b,M07} are screened. A simple
Thomas-Fermi approximation leads to $V_{scr} ( \vec{r} ) \approx
V_0 ( \vec{r} ) / ( k_{FT} L )$ where $k_{FT} = e^2 k_F / ( \pi
v_F )$ is the screening length. Gauge fields, on the other hand,
remain unscreened.

\subsubsection{Electronic transport in ballistic samples} We first
assume that the samples are free of disorder and the only source
of electron scattering is the gauge fields induced by the
deformations. The simplest geometry which can be studied is a
rectangular flake, as sketched in Fig. \ref{sketch}. If the
bending energy is neglected, a complete analytical form can be
found for the deformation\cite{FGK08}:
\begin{align}
h ( x ) &= \frac{h_0}{L^2} \left( x^2 - \frac{L^2}{4} \right) \nonumber \\
h_0^3 &= \frac{3 \pi}{64} \frac{e^2}{E} \left( n L^2 \right)^2
\end{align} \label{height_1}
where $E=4(\lambda+\mu)\mu/(\lambda + 2\mu) \approx 24$ eV
\AA$^{-2}$ is the Young's modulus of graphene. The in-plane
stresses are:
\begin{equation}
u_x ( x ) = - \frac{2 x^3 h_0^2}{3 L^4} + \frac{x h_0^2}{6 L^2}
\end{equation}
where the boundary condition $u_x ( \pm L / 2 ) = 0$ has been
used. As a result, the only non zero component of the strain
tensor is:
\begin{equation}
u_{xx} = \frac{h_0^2}{6 L^2}
\end{equation}

The effect of the uniform strain is to induce a constant gauge
field in the suspended region, $A_y = \beta u_{xx} / a$. The
effect of this field is to shift the momentum parallel to the
interface between the suspended and clamped region, $k_y$. An
electron outside the suspended region with momentum  $\vec{k} = (
k_x , k_y )$ and energy $\epsilon_{\vec{k}} = v_F \sqrt{k_x^2 +
k_y^2}$ can be either reflected at the boundary, with momentum
$\vec{k}' = ( - k_x , k_y )$ or transmitted, with momentum
$\vec{k}'' = ( k_x' , k_y + A_y )$ with $k_x'$ such that ${k_x'}^2
+ ( k_y + A_y )^2 = k_x^2 + k_y^2$. For incident momenta in the
range $k_x^2 < 2 A_y k_y + A_y^2$ there are no transmitted states,
and the incident electron is reflected. The transmission
amplitude, $T$,  is determined by the equations:
\begin{align}
1 - R &= T \nonumber \\
\frac{k_x + i k_y}{k} + \frac{-k_x + i k_y}{k} R &= \frac{k_x' + i
( k_y + A_y )}{k} T \label{transmission}
\end{align}
and the transmission coefficient is ${\cal T} = |T|^2 ( k_x' / k_x
)$. At normal incidence, $k_y = 0$, the transmission coefficient
can be expanded, ${\cal T} \approx 1 - A_y^2 / ( 4 k_x^2 )$. The
full expression for the transmission coefficient in the setup
shown in Fig. \ref{sketch}, with two boundaries, can be obtained
analytically \cite{FGK08}. The presence of two barriers leads to
Fabry-Perot interferences as function of the incident angle,
$\theta = \arctan ( k_y / k_x )$. The angular dependence of the
transmission is shown in Fig. \ref{transmission_angle}. Note the
skew scattering in the two valleys.

The above analysis is only valid for clean, ballistic systems. A
finite concentration of impurities breaks the conservation of
parallel momentum, which leads to the suppression of the
transmission through each interface, Eq.~(\ref{transmission}). The
lack of momentum conservation gives a finite transmission when
strains in a ballistic sample completely suppresses it
\cite{FGK08}. Evanescent waves allow carriers to penetrate in the
suspended region. These carriers can be scattered into propagating
modes by the impurities. A perturbative calculation shows that the
induced transmission grows linearly with the concentration of
impurities. The effect is proportional to the scattering strength,
which can be parametrized in terms of the mean free path induced
by the impurities.

\subsection{Interferences between strains and real magnetic fields}
The gauge fields associated to strains interfere with real
magnetic fields applied to the system. When the strain can be
treated as a small perturbation, the resulting effect is the
existence of an inhomogeneous effective field distribution, given
by the sum of the constant applied field and the effective field
induced by the strains. Electronic states in the vicinity of the
inhomogeneities are deformed, and can lead to localized ``snake
states'', where current flows along the perturbed region, as in
the presence of a nonuniform magnetic field \cite{KROC08,MVetal08,MVP09}.
\begin{figure}[!t]
\begin{center}
\includegraphics[width=10cm,angle=0]{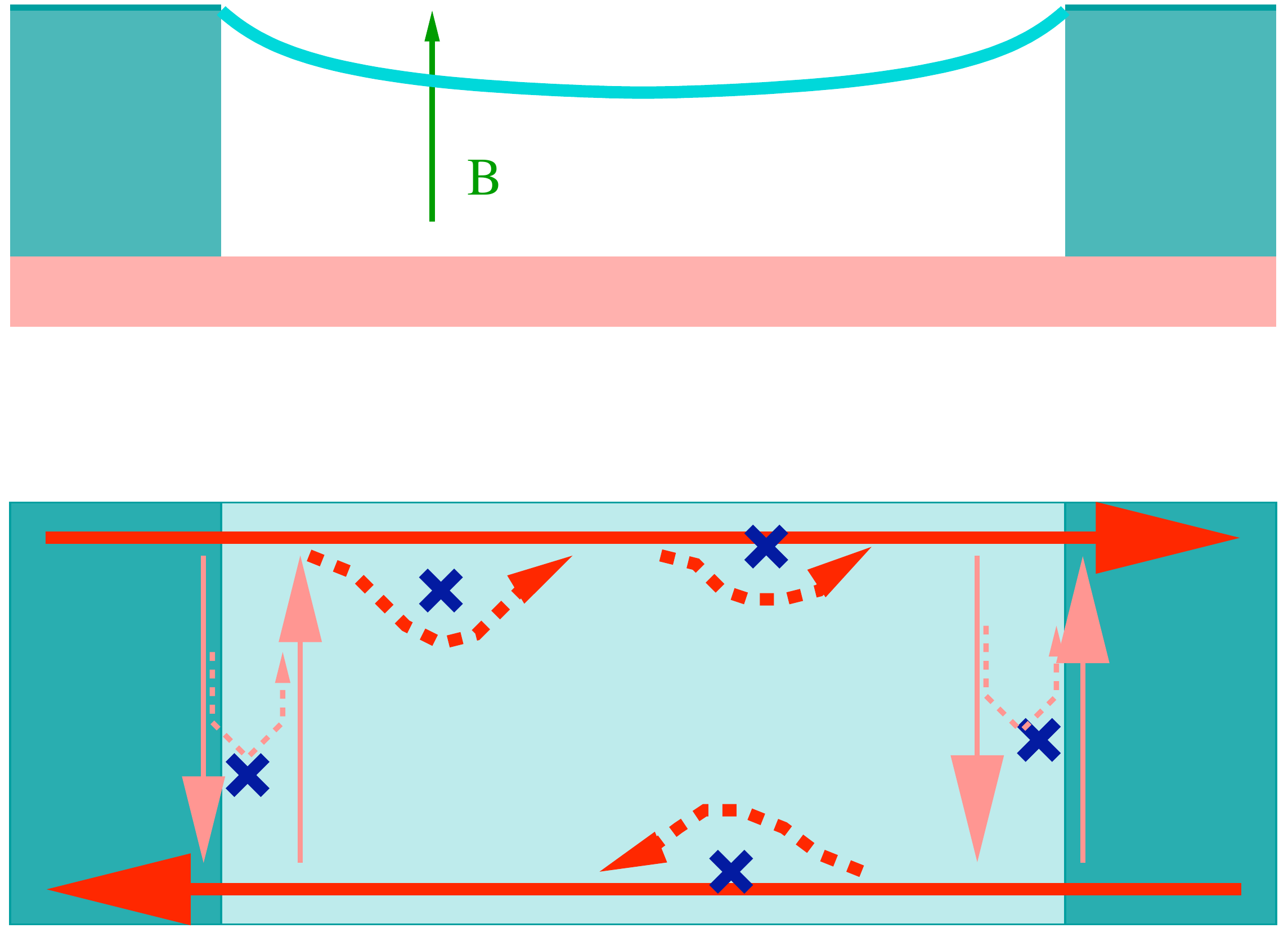}
\caption[fig]{Sketch of the expected effect in the presence of
strains and a constant magnetic field. The geometry is the same as
in Fig.[\ref{sketch}]. The strains induce currents which lead to
backscattering between the chiral edge states. A small
concentration of impurities suppresses the effect.}
\label{disorder_magnetic}
\end{center}
\end{figure}

The qualitative effect of strains in the geometry in Fig.
\ref{sketch} is shown in Fig. \ref{disorder_magnetic}. The strains
induce currents parallel to the interface between the suspended
and the non suspended regions, which allow for backscattering
between the chiral edge states created by the magnetic field
\cite{LPSFG09}. The states along the interface are not chiral, and
impurities induce backscattering between them. If this effect is
sufficiently strong, these states will become localized, and the
coupling between the chiral edge channels will be suppressed.

The hamiltonian which describes the system, in the absence of
impurities, and neglecting the edges, is:
\begin{equation}
{\cal H} = v_F \left( \begin{array}{cc} 0 &i \partial_x + i k_y
\mp i A_y^{strain} \theta ( x ) - i \frac{x}{l_B^2} \\ i
\partial_x - i k_y \pm i A_y^{strain} \theta ( x ) + i
\frac{x}{l_B^2} \end{array} \right) \label{hamil_field}
\end{equation},
where $l_B$ is the magnetic length associated to the external
field, and we assume that, in a system infinite in the $y$
direction the momentum $k_y$ is conserved. The two signs in
Eq.~(\ref{hamil_field}) refer to the two valleys in the Brillouin
Zone. By squaring the hamiltonian, one obtains the effective
Schr\"odinger equation:
\begin{equation}
\left[ - v_F^2 \partial_x^2 + \frac{v_F^2 ( x - x_\pm )^2}{l_B^4}
\mp v_F A_y^{strain} \delta ( x ) \right] \Psi ( x ) =
\epsilon_{k_y}^2 \Psi ( x )
\end{equation}
where $x_- = k_y l_B^2$ and $x_+ = x_- + A_y^{strain}$. For a
given value of $k_y$, the eigenvalue $\epsilon_{k_y}$ depends only
on $v_F$ and $A_y^{strain} l_B$. The strain induces two effects:
i) it shifts the centers of the Landau levels, and ii) a delta
function potential appears at the interface. At distances from the
interface greater than $l_B$, the Landau levels are not perturbed,
and $\epsilon_{k_y} \rightarrow \pm v_F \sqrt{n} / l_B$, where $n$
is the Landau level index. For $A_y^{strain} l_B \ll 1$ we can use
perturbation theory to analyze the shift in the energies of the
Landau levels. The leading effect is due to the delta function
potential. The maximum change in the energy is:
\begin{equation}
\epsilon_{k_y}^n \approx \pm \frac{v_F \sqrt{n}}{l_B} \left( 1 +
\pm c_n \frac{ A_y l_B}{2 n} \right)
\end{equation}
where $c_n$ is a numerical constant. The Landau levels become
dispersive, with a velocity $v^* = \partial \epsilon_{k_y}^n /
\partial k_y \approx \pm v_F A_y^{strain} l_B$. The Landau level
energy has a maximum, or a minimum, for a value of $k_y$ such that
the center is close to the interface. Hence, there are two counter
propagating states per Landau level and per valley. These modes
have a width $\sim l_B$. Using the Born approximation, it can be
shown that the reflection coefficient induced by a single charged
impurity is of order $e^2 / v_F$ for $A_y^{strain} l_B \sim 1$.

\begin{figure}[!t]
\begin{center}
\includegraphics[width=10cm,angle=0]{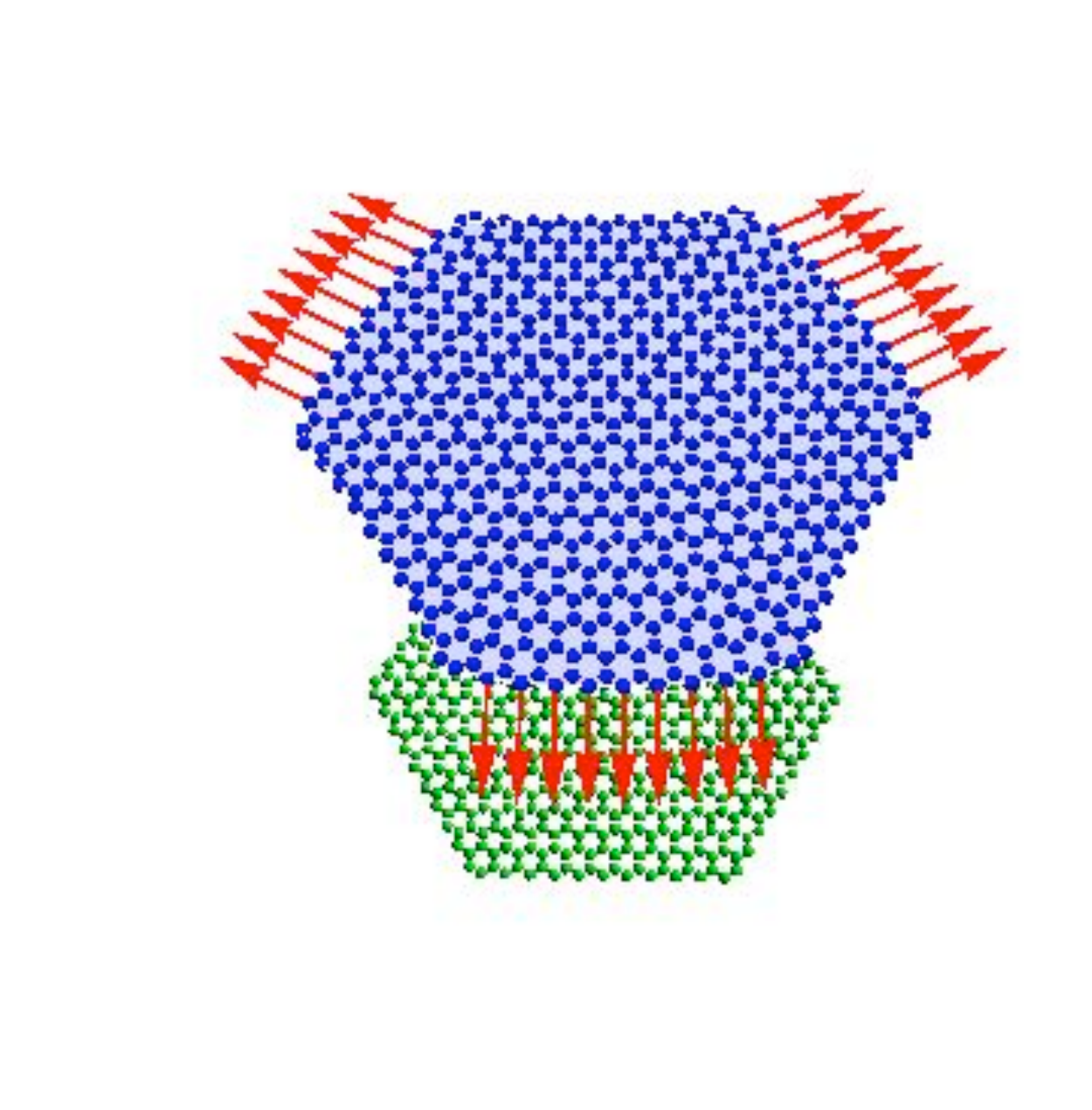}
\caption[fig]{Sketch of applied strains which lead to an almost
constant effective magnetic field inside the sample.}
\label{qdot_strain}
\end{center}
\end{figure}

\subsection{Generation of effective magnetic fields. Strain
engineering} The previous analysis suggests that strains can
generate a wide variety of effective magnetic fields. This has
open the possibility of using ``strain engineering'' in order to
modify the electronic properties of graphene samples
\cite{PN08,AV08,PNP09,GKG09,FR09,VPP09,CJS10,MPMT09,RUC09,GGKN09,BK09,Retal09,CJS10b}.
Changes in the strain can enhance the chemical activity of
graphene, or confine the massless carriers, avoiding the Klein
paradox \cite{KNG06}. Experimentally, strains can be induced in
graphene flakes by bending the substrate
\cite{Metal09,Hetal09,Tetal09b}, and the amount of strain can be
measured by Raman spectroscopy. Strains can be due to the
difference in expansion coefficients between graphene and the
substrate, leading to wrinkles \cite{Betal09}. There is
experimental evidence that the electronic properties are modified
in strained samples \cite{Tetal09}.

A particularly interesting possibility is the generation of an
effective {\em constant} magnetic field. The results in the
Appendix  \ref{sec_radial} show that a strain distribution with
trigonal symmetry induces such a field. An experimental setup
which leads to an approximate uniform field is shown in
Fig.~\ref{qdot_strain}, see \cite{GKG09,GGKN09}. Such situation
will change drastically the electronic properties of the system.
The electronic spectrum will be split into Landau levels, with
counterpropagating chiral modes at the edges. The combination of
insulating behavior in he bulk and gapless modes at the edges is
reminiscent of a topological insulator \cite{H88,KM05}.

\subsection{Effective electric fields}
A time-independent strain gives rise to static gauge fields.
Hence, a time-dependent strain will induce effective electric
fields \cite{OGM09}. These fields appear in oscillating graphene
sheets \cite{Betal08,Getal08}. As in the static case, the fields
acting on different valleys have opposite sign, so that there is
no net charge transport and the field is not screened. The time
dependence of the scalar potential also induced by the strains
\cite{SA02b} also leads to an effective electric field, although
with the same sign for the two valleys. This potential can be
screened. The effective electric fields induce electronic
currents, which dissipate energy. This ohmic damping reduces the
quality factor of the oscillator.

The description of the decay of long wavelength phonons into
electron-hole pairs by means of an effective electric field acting
on the electrons has also been used to understand phonon damping
in carbon nanotubes \cite{Setal08b} (for a review of the extensive
literature on nano electromechanical systems, NEMS, based on
carbon nanotubes, see \cite{HPWZ08}). In a clean system, the
losses due to currents induced by the effective electric fields
are equivalent to the phonon lifetime which can be calculated
using second order perturbation theory. A formulation using
effective fields can easily be extended to diffusive systems,
where the electronic wave functions can only be estimated using
semiclassical arguments. Standard perturbation theory is adequate
for clean, ballistic carbon nanotubes, but a description using a
finite mean free path and local conductivities is more appropriate
for large graphene samples, as the elastic mean free path is
typically much smaller than the dimensions of the sample. It is
worth noting that the existence of two counter-propagating valley
currents imply that the ohmic dissipation in a graphene resonator
is limited, at high temperatures, by valley drag, which, in turn,
is dependent on the strength of the electron-electron interaction.

\section{Conclusions } \label{sec_conclusions}

To summarize, we have presented an overview of the role of gauge
fields in  graphene and other carbon-based materials. The geometric
structure of honeycomb lattice naturally leads, in the continuum
limit, to massless Dirac fermions as elementary excitations. The
internal degree of freedom which plays the same role as spin in
quantum electrodynamics has here a very simple geometric meaning, it
is just the sublattice label (pseudospin). Additional internal
quantum numbers are the valley label (isospin) and real spin. If we
neglect spin-orbit coupling (which is very small in graphene) the
latter is split from other degrees of freedom. We distinguish two
types of gauge fields: topological, associated to lattice defects,
such as disclinations and dislocations, and related to modulations
of the hoppings, induced by elastic deformations.

The coupling of the topological fields to the electrons does not
depend on the parameters of the material. Topological fields come in
two mutually non commuting varieties, depending on whether they mix
valleys or not. The gauge fields associated to long wavelength
strains do not mix valleys, and their coupling to the electrons
depend on the value of the strength of the electron-phonon
interaction.

Being exactly a two-dimensional system, graphene shows a thermally
induced bending instability which leads to intrinsic ripples
(corrugations) even in a free sample. For the case of exfoliated
graphene on a substrate, the ripples will arise also due to
interaction with the substrate. In both cases, the ripples are
sources of a random pseudomagnetic gauge field which modifies the
electronic transport in the  weak (anti)localization regime, and it
also induces essential contributions to the resistivity. The ripples
may be one of the main limiting factors restricting mobility of
charge carriers in graphene.

Regular gauge fields, or even artificially designed fields can exist
in suspended samples, where they can lead to a number of novel
features unique to graphene, such as the confinement of electrons,
or  the formation of localized orbitals similar to the Landau levels
which exist in real magnetic fields.

The study of gauge fields in graphene establishes important
relations with contemporary mathematical physics. For example, due
to the  Atiyah-Singer index theorem the gauge fields in both
single-layer and bilayer graphene can create  states with zero
energy, the states being chiral (which means that for a given valley
they belong to only one sublattice). The inhomogeneity of
pseudomagnetic field created by the ripples should broaden all
Landau levels in the quantum Hall regime except the zero-energy one.
This prediction of the index theorem can be checked experimentally
and seems to be in an agreement with the available data. It is
interesting to note that the gauge potentials which arise from
lattice deformations have intrinsic meaning, although their effects
on the electrons are invariant under gauge transformations, so that
different physical deformations can lead to the same electronic
structure.

We have not explored here extensions to other systems, such as
multilayered graphene, or the topological insulators. The same
arguments used to define gauge fields in single layer graphene also
apply to multilayered samples. The effect of a given deformation, or
gauge field, on the electrons will depend on the number of layers
and in the way they are connected. In topological insulators with
surface states described by the Dirac equations, gauge fields can
also be defined. Because of the different symmetries of these
compounds, such fields should arise from perturbations which break
time reversal symmetry.

\section{Acknowledgments}
\label{thank}

MAHV thanks A. Cortijo and F. de Juan  for many
illuminating discussions. FG and MAHV acknowledge support from MEC (Spain) through
grant FIS2005-05478-C02-01 and CONSOLIDER CSD2007-00010, and by
the Comunidad de Madrid, through CITECNOMIK,
CM2006-S-0505-ESP-0337. MK acknowledges support from Stichting
voor Fundamenteel Onderzoek der Materie (FOM), the Netherlands.

\appendix

\section{The Dirac equation in curved space}\label{sec_curvedspace}

In general when trying to formulate a mathematical relation
defined in a flat space to a curved space, one uses a general
covariance principle that amounts to substitute any given
magnitude transforming as a tensor in the flat space by the
corresponding magnitude transforming as a tensor under general
transformations in the curved manifold. This simple substitution
is complicated for spinors because there are no spinor
representations in the group of general transformations. This
makes necessary to introduce an alternative formalism based on
tetrads \cite{BD82,W96}. Instead of the usual metric $g_{\mu\nu}$
we must introduce at each point $X$ described in arbitrary
coordinates, a set of locally inertial coordinates $\xi_X^a$ and
the fielbein fields  $e_\mu^a(x)$, a set of orthonormal vectors
labelled by $a$ that fixes the transformation between the local
(Latin indices) and the general (Greek indices) coordinates:
\begin{equation}
e_\mu^a(X)\equiv\frac{\partial \xi_X^a(x)}{\partial
x^\mu}\vert_{x=X}.
\end{equation}
We can now list the various geometric objects needed to derive the
Dirac physics in the curved space:
The metric tensor of the curved manifold $g_{\mu\nu}(x)$ is
related to the flat, constant  metric $\eta_{ab}$ by the equation
\beq
g_{\mu\nu}(x)=e_\mu^a (x) e_\nu^b (x) \eta_{ab};
\eeq
its determinant, needed to define a scalar density lagrangian is
given by
\begin{equation}
\sqrt{-g}=[\det(g_{\mu\nu})]^{1/2}\;=\;\det[e_\mu^a(x)].
\label{jacobian}
\end{equation}
The curved space gamma matrices $\gamma^\mu(x)$ satisfying the
commutation relations
\begin{equation}
\{\gamma^\mu(x)\gamma^\nu(x)\}=2g^{\mu\nu}(x),
\label{curvedgammas}
\end{equation}
are related with the constant, flat space matrices $\gamma^a$ by
\begin{equation}
\gamma^\mu(x)=e^\mu_a(x)\gamma^a.
\end{equation}
The most complicated object needed to complete the analysis is the
spin connection $\Omega_\mu(x)$ associated to the spinor covariant
derivative. This is an important object for the physics since it
acts as a gauge field. The construction of the spin connection is
done by observing that the derivative of the spinor does not
transform as a vector under a coordinate transformation in the
flat tangent space. It does if we  introduce for the spinors a
covariant derivative of the form
\begin{equation}
\mathcal{D}_{a}=e^{\mu}_{a}\left[\frac{\partial}{\partial
x^{\mu}}+\Omega_{\mu}\right].\label{covariantderivative}
\end{equation}
which has a standard transformation under a change of coordinates
in both the vector and the spinor indices.
To get the structure of the $\Omega_{\mu}(x)$ matrices it suffices
to consider their transformation properties in flat space. For the
spin one half representation they take the form
\begin{equation}
\Omega_\mu(x)=\frac{1}{4}\gamma_a\gamma_b
e^a_\lambda(x)g^{\lambda\sigma}(x) \nabla_\mu e^b_\sigma(x),
\end{equation}
 with
\begin{equation}
\nabla_\mu e^a_\sigma=\partial_\mu
e_\sigma^a-\Gamma_{\mu\sigma}^\lambda e_\lambda^a
\end{equation}
 where
$\Gamma_{\mu\sigma}^\lambda$ is the usual affine connection which
is related to the metric tensor by
\begin{equation}
\Gamma_{\mu\sigma}^\lambda=\frac{1}{2}g^{\nu\lambda}\{\frac{\partial
g_{\sigma\nu}}{\partial x^\mu}+\frac{\partial g_{\mu\nu}}{\partial
x^\sigma}-\frac{\partial g_{\mu\sigma}}{\partial x^\nu}\}.
\label{christoffel}
\end{equation}

Finally the Dirac equation is given by
\begin{equation}
i\gamma^{\mu}(x)[\partial_{\mu}+\Omega_\mu(x)]\psi(x)=0
.
\end{equation}

We will exemplify the formalism by giving the details of the
computation  of the various geometrical factors for the two main
problems described in Section \ref{sec_defects}: the topological
defects and the smooth gaussian deformation.

We begin by describing the formalism for a smooth deformation
where the curvature is non singular everywhere. We start by
embedding a two-dimensional surface with polar symmetry (this is
only for simplicity and can be easily extended to any shape) in
three-dimensional space (described in cylindrical coordinates).
The surface is defined by a function $z(r)$ giving the height with
respect to the flat surface z=0, and parametrized by the polar
coordinates of its projection onto the z=0 plane. The metric for
this surface is obtained as follows: We compute
\begin{equation}
dz^{2}=(\frac{dz}{dr})^{2}dr^{2}\equiv \alpha f(r)dr^{2},
\label{surfacegauss}
\end{equation}
 and substitute for the line element:
\begin{equation}
ds^{2}=dr^{2}+r^{2}d\theta^{2}+dz^{2}=\left(1+\alpha
f(r)\right)dr^{2}+r^{2}d\theta^{2}. \label{generalmetric}
\end{equation}
In particular the the  gaussian bump shown in Fig. \ref{gaussian}
is defined by:
\begin{equation}
z=A\exp(-r^{2}/b^{2}), \label{gaussianformula}
\end{equation}
so that
\begin{equation}
dz^{2}=\frac{A^{2}}{b^{4}}4r^{2}\exp(-2r^{2}/b^{2})dr^{2},
\end{equation}
which corresponds to Eq. (\ref{generalmetric}) with
$$\alpha=(A/b)^{2}\;,\; f(r)=4(r/b)^{2}\exp(-2r^{2}/b^{2}).$$
From the line element  we can write  the metric in a more usual
form:
\begin{equation}
g_{\mu\nu}=
\left(%
\begin{array}{cc}

   -(1+\alpha f(r)) & 0 \\
   0 & -r^2 \\
\end{array}%
\right), \label{metric}
\end{equation}
where we have omitted the time coordinate which is trivial in the
case of a static background.
The affine connection  $\Gamma^{\lambda}_{\mu\nu}$ which only depends on the metric is for the metric (\ref{metric}):
\begin{eqnarray}
\Gamma^{r}_{rr}=\frac{\alpha f'}{2(1+\alpha f)}, &
\Gamma^{r}_{\theta\theta}= - \frac{r}{1+\alpha f}, &
\Gamma^{\theta}_{r\theta}=\frac{1}{r} \;\;, \label{connections}
\end{eqnarray}
where $f'=df/dr$, and the rest of the elements are zero or related
by symmetry.

The geometrical (gaussian) curvature $K$ is
 \begin{equation}
 K(r)=\frac{\alpha f'(r)}{2r(1+\alpha f(r))^2  }.
 \label{curvature}
 \end{equation}

 The fielbein fields $e^{a}_{\; \mu}$ satisfy:
\begin{equation}
\label{flatfielbein1} g_{\mu\nu}=e^a_{\; \mu}e^b_{\;
\nu}\eta_{ab},
\end{equation}
where
$\eta_{ab}$ is the identity matrix in two dimensions (note that
this relation does not fix $e^a_{\; \mu}$ uniquely). We choose the
$e^a_{\; \mu}$ to be
\begin{eqnarray}
 e^{1}_{\; r}=(1+\alpha f)^{1/2}\cos \theta \qquad & e^{1}_{\; \theta}=-r\sin \theta  \nonumber\\
e^{2}_{\; r}=(1+\alpha f)^{1/2}\sin \theta \qquad &
e^{2}_{\; \theta}=r \cos \theta;\nonumber\\
\end{eqnarray}

that reduce to the flat set  when $\alpha=0$. Now we can compute
the spin connection coefficients,
\begin{equation}
\omega_{\mu}^{\; ab}=e^a_{\; \nu} \left(\partial _{\mu} +
\Gamma^{\nu}_{\mu\lambda} \right)e^{b\lambda},
\end{equation}
which are found to be:
\begin{eqnarray}
\omega_{\theta}^{\; 12}=1-(1+\alpha f)^{-1/2}, \label{spinconcef}
\end{eqnarray}
the rest being zero or related by symmetry (the spin connection
$\omega$ is antisymmetric in the upper indices \cite{W72}).

The spin connection
\begin{equation}
\Omega_{\mu}=\frac{1}{8}\omega_{\mu}^{\;
ab}\left[\gamma_a,\gamma_b\right],
\end{equation}
turns out to be
\begin{eqnarray}
\Omega_{r}=0 ,& \Omega_{\theta}= \frac{1-(1+\alpha
f)^{-1/2}}{2}\gamma^{1}\gamma^{2}. \label{spincon}
\end{eqnarray}
\vspace{0.5cm}

In the case of topological defects, for the metric defined by
(\ref{genmetric}):
\begin{equation}
g_{\mu\nu}=
\left(%
\begin{array}{ccc}
  -1 & 0 & 0 \\
  0 & e^{-2\Lambda} & 0 \\
  0 & 0 & e^{-2\Lambda} \\
\end{array}%
\right),
\end{equation}
the gamma matrices and the spinor connection in the curved
background   are immediately found to be
$$
\gamma^{0}(\textbf{r})=\gamma^{0}\;,\;
\gamma^{i}(\textbf{r})=e^{\Lambda({\bf r})}\gamma^{i}\;\; (i=1,2)
$$
$$
\Omega_{1}(\textbf{r})=-\frac{1}{2}\gamma^{1}\gamma^{2}\partial_{y}\Lambda\;\;,\;\;
\Omega_{2}(\textbf{r})=-\frac{1}{2}\gamma^{2}\gamma^{1}\partial_{x}\Lambda\;,
$$
and the determinant of the metric tensor is
$$
\sqrt{-g}=e^{-2\Lambda}.
$$
The special feature of the conical defects is that the Riemann
curvature tensor is zero everywhere but for the apex of the cone
where it has a delta function singularity.

\section{In plane strains and effective fields in radial coordinates}
\label{sec_radial}
We analyze a circular graphene quantum dot. We
first analyze the distribution of stresses. We assume that the dot
is free of lattice defects. The stresses are due to forces applied
at the boundaries.

The elastic free energy in circular coordinates is:
\begin{align}
{\cal F} &= \frac{\lambda}{2} \int  2 \pi r d r d \theta \left(
\partial_r u_r + \frac{u_r}{r} + \frac{\partial_\theta u_\theta}{r}
\right)^2 + \nonumber \\ &+ \mu \int 2 \pi r  d r d \theta \left[
\left( \partial_ r u_ r \right)^2 + \left( \frac{\partial_\theta
u_\theta}{r} + \frac{u_r}{r} \right)^2 + \right. \nonumber
\\ &+ \left. \frac{1}{2} \left(
\partial_r u_\theta + \frac{\partial_\theta u_r}{r} -
\frac{u_\theta}{r} \right)^2 \right]
\end{align}
where $u_r$ and $u_\theta$ are the two components of the
displacement vector in circular coordinates. The stress tensor is:
\begin{align}
\sigma_{rr} &= \lambda \left( \partial_r u_r +
\frac{\partial_\theta
u_\theta}{r} + \frac{u_r}{r} \right) + 2 \mu \left( \partial_r u_r \right) \nonumber \\
\sigma_{\theta \theta} &= \lambda \left( \partial_r u_r +
\frac{\partial_\theta u_\theta}{r} + \frac{u_r}{r} \right) + 2 \mu
\left( \frac{\partial_\theta u_\theta}{r}
+ \frac{u_r}{r} \right) \nonumber \\
\sigma_{r \theta} &= 2 \mu \left( \frac{\partial_\theta u_r}{r} +
\partial_r u_\theta - \frac{u_\theta}{r} \right)
\end{align}

 The resulting equilibrium equations for $u_r$ and $u_\theta$ are:
 \begin{align}
 &\lambda
 \left(
 - \partial^2_r u_r - \frac{\partial_r u_r}{r} +
 \frac{u_r}{r^2} - \frac{\partial_r \partial_\theta u_\theta}{r} +
 \frac{\partial_\theta u_\theta}{r^2}
 \right) + \nonumber \\
 &+ 2 \mu \left( -
 \frac{\partial_r u_r}{r} - \partial^2_r u_r + \frac{u_r}{r^2} +
 \frac{3 \partial_\theta u_\theta}{2 r^2} - \frac{\partial^2_\theta
 u_r}{2 r^2} - \frac{\partial_r \partial_\theta u_\theta}{2 r}
 \right) = \nonumber \\ &= 0 \nonumber \\
 &\lambda \left( - \frac{\partial_r \partial_\theta u_r}{r} -
 \frac{\partial_\theta u_r}{r^2} - \frac{\partial_\theta^2
 u_\theta}{r^2} \right) + \nonumber \\ &+ 2 \mu \left(
 -\frac{3 \partial_\theta u_r}{2 r^2} - \frac{\partial_\theta^2
 u_\theta}{r^2} - \frac{\partial_r u_\theta}{2 r} +
 \frac{u_\theta}{2 r^2} - \frac{\partial_r \partial_\theta u_r}{2r} -
 \frac{\partial_r^2 u_\theta}{2} \right) = \nonumber \\ & = 0
 \label{equilibrium}
 \end{align}
We look for solutions of the type:
\begin{align}
u_r &= a_r r^m e^{i n \theta} \nonumber \\
u_\theta &= a_\theta r^m e^{i n \theta} \label{ansatz_0}
\end{align}
Inserting these expressions in Eqs.~\ref{equilibrium}, we find:
\begin{align}
\left[ \lambda \left( - m^2 + 1 \right) + \mu \left ( - 2 m^2 + 2
+
n^2 \right) \right] a_r &+ \nonumber \\
+ (i n ) \left[ \lambda ( - m + 1 ) + \mu (
3 - m ) \right] a_\theta &= 0 \nonumber \\
(  - i n ) \left[ \lambda ( m + 1 )  + \mu ( 3 + m ) \right] a_r
&+
\nonumber \\
\left[ \lambda n^2 + \mu ( 2 n^2 - m^2 + 1) \right] a_\theta &= 0
\label{harmonics}
\end{align}
These equations admit solutions for $n = \pm ( m \pm 1 )$. For
$m=0$ there are only two solutions, with $n=\pm 1$. For $m=0$, a
finite value of $a_\theta$ leads to a rotation of the disk, and
does not change the physical properties.

The induced gauge field on the Dirac electrons of the graphene
layer depends on the relative orientation of the graphene axes.
Choosing the direction $\theta = 0$ as one of the lattice axis,
the gauge potential can be written as:
\begin{align}
A_r &= \frac{\beta}{a} \left[ \left( \partial_r u_r -
\frac{\partial_\theta u_\theta}{r} - \frac{u_r}{r} \right) \cos (
3 \theta ) - \left( \frac{\partial_\theta u_r}{r} + \partial_r
u_\theta - \frac{u_\theta}{r}
\right) \sin ( 3 \theta )    \right] \nonumber \\
A_\theta &= \frac{\beta}{a} \left[ - \left( \partial_r u_r  -
\frac{\partial_\theta u_\theta }{r} - \frac{u_r}{r} \right) \sin (
3 \theta ) - \left( \frac{\partial_\theta u_r}{r} + \partial_r
u_\theta - \frac{u_\theta}{r} \right) \cos ( 3 \theta )  \right]
\end{align}
where $\beta = \partial \log ( t ) / \partial \log ( a ) \approx 2
- 3$ is the logarithmic derivative of the nearest neighbor hopping
$t$ with respect to the nearest neighbor distance, $a$. The
effective magnetic field is:
\begin{equation}
B = \frac{\partial_\theta A_r}{r} - \partial_r A_\theta -
\frac{A_\theta}{r}
\end{equation}
\begin{table}
\begin{tabular}{||c|c|c||}
\hline\hline $n$ &$a_\theta / a_r$&$B( r , \theta )$ \\ \hline
\hline  $- m - 1$ &$-i$ &$4 i m ( m-1) e^{-i(m-2) \theta} r^{m-2}$ \\
\hline $-m + 1$&$i \frac{\lambda ( m+1)+\mu (
m+3)}{\lambda(m-1)+\mu(m-3)}$ &$4
i m (m-1) ( m-2 ) \frac{\lambda + \mu}{\lambda ( m-1) + \mu ( m-3 )}e^{-i(m-4)\theta} r^{m-2}$\\
\hline $m - 1$ &$-i \frac{\lambda ( m+1)+\mu (
m+3)}{\lambda(m-1)+\mu(m-3)}$ &$-4
i m (m-1) ( m-2 ) \frac{\lambda +  \mu}{\lambda ( m-1) + \mu ( m-3 )}e^{i(m-4)\theta} r^{m-2}$\\
\hline $m+1$&$i$ &$-4 i m ( m-1) e^{i(m-2)\theta}r^{m-2}$ \\
\hline \hline \end{tabular} \caption{Displacements and induced
effective magnetic fields for solutions of the equations of
elasticity with well defined symmetries under rotations, see
Eq.~(\ref{ansatz_0}).}\label{table}\end{table} The effective
magnetic fields induced by solutions of equilibrium elasticity of
the type described in Eq.~(\ref{ansatz_0}) are given in
Table[\ref{table}]. Non trivial solutions which induce a constant
field can be obtained by setting $m=2$ and combining the first and
second solutions in Table[\ref{table}].

\bibliography{gauge_12}

\end{document}